\documentclass[cernpreprint,atlasdraft=false, LANGEDIT=true, USenglish, texmf, orcidlogo]{atlasdoc}
\usepackage{atlaspackage}
\usepackage{atlasbiblatex}
\usepackage{subfig}
\usepackage[tableposition=top]{caption}

\usepackage{atlasphysics}
\usepackage{atlasheavyion}
\usepackage{placeins}
\graphicspath{{logos/}{figures/}}

\usepackage{ANA-HION-2022-15-PAPER-defs}

\LEcontact{Bob Kowalewski bob.kowalewski@cern.ch}

\AtlasTitle{Measurement of photonuclear jet production in ultra-peripheral Pb+Pb collisions at $\sqrt{s_{\text{NN}}} = 5.02$ TeV with the ATLAS detector}

\AtlasAbstract{%
In ultra-relativistic heavy ion collisions at the LHC, each nucleus acts
a sources of high-energy real photons that can scatter off the opposing
nucleus in ultra-peripheral photonuclear ($\gamma+A$) collisions. Hard
scattering processes initiated by the photons in such collisions provide a
novel method for probing nuclear parton distributions in a kinematic region
not easily accessible to other measurements. ATLAS has measured production
of dijet and multi-jet final states in ultra-peripheral Pb+Pb collisions
at $\sqrt{s_{\text{NN}}} = 5.02$ TeV using a data set recorded in 2018 with
an integrated luminosity of 1.72 $\text{nb}^{-1}$.
Photonuclear final states are selected by requiring a rapidity gap
in the photon direction; this selects events where one of the outgoing nuclei remains intact.
Jets are reconstructed using the anti-$k_\text{t}$ algorithm with radius parameter, $R = 0.4$.
Triple-differential cross-sections, unfolded for
detector response, are measured and presented using two sets of
kinematic variables. The first set consists of the total transverse momentum ($H_\text{T}$),
rapidity, and mass of the jet system. The second set uses $H_\text{T}$ and particle-level nuclear and
photon parton momentum fractions, $x_\text{A}$ and $z_{\gamma}$, respectively.
The results are compared with leading-order (LO) perturbative QCD
calculations of photonuclear jet production cross-sections, where
all LO predictions using existing fits fall below the data in the shadowing region.
More detailed theoretical comparisons will allow these results to strongly constrain nuclear parton distributions, and these data provide results from the
LHC directly comparable to early physics results at the planned Electron-Ion Collider.
}

\AtlasRefCode{HION-2022-15}

\PreprintIdNumber{CERN-EP-2024-221}

\AtlasJournalRef{\PRD 111 (2025) 052006}
\AtlasDOI{10.1103/PhysRevD.111.052006}

\hypersetup{pdftitle={ATLAS document},pdfauthor={The ATLAS Collaboration}}

\begin{document}

\maketitle

\tableofcontents

\section{Introduction}
%

% The next lines are included from the .//tex/intro.tex input file
Studies of hard-scattering processes are a crucial component of the
ultra-relativistic heavy-ion physics programs at the Large Hadron
Collider (LHC) and the Relativistic Heavy Ion Collider (RHIC). Such studies
will soon reach or, in some cases, have
already reached \cite{Paukkunen:2010qg,Eskola:2019bgf,Eskola:2019dui,}
a level of precision
where measurements are sensitive to nuclear modifications of the
parton distribution functions (PDFs) in the colliding nuclei. The modification of
nuclear PDFs (nPDFs), relative to the PDFs of free nucleons, has been a
subject of extensive study since the first measurements by the EMC
experiment \cite{EuropeanMuon:1983wih} decades ago. However, the scarcity of available
data has hindered the interpretation of the observed nuclear
modifications, limiting the precision of global fits compared to those performed
for the proton PDFs (see,
e.g. Refs. \cite{Pumplin:2002vw,Martin:2009iq,Harland-Lang:2014zoa,NNPDF:2014otw,}).

Recent advances in the methods used to perform global fits to
nuclear PDFs \cite{Walt:2019slu,} and extensions to next-to-next-to-leading-order (NNLO) \cite{Khanpour:2020zyu} in
perturbative QCD, combined with the
inclusion \cite{Eskola:2019dui,AbdulKhalek:2020yuc, Helenius:2021tof,Eskola:2021nhw,} of recent data from the LHC and
experiments at Jefferson Laboratory, have significantly reduced
uncertainties in the extracted nuclear PDFs. However, global fits remain
limited by the precision %
and kinematic coverage of experimental data,
requiring
new measurements that cover a large kinematic range in
Bjorken $x$ and $Q^2$ for a heavy nucleus. %
Measurements are particularly important in the region of intermediate $Q^2$ ($100$~$\GeV^2\lesssim Q^2 \lesssim 10^{4}$~$\GeV^2$), where
data are lacking for a wide range of Bjorken $x$ values.
Such data will be provided by
deep-inelastic scattering measurements at the Electron-Ion Collider
(EIC) \cite{Accardi:2012qut} when it becomes operational, but those data will also have
limited $Q^2$ coverage at low $x$ ($x \lesssim 10^{-2}$), where ``shadowing'', the suppression of the
nuclear PDFs at low $x$, is observed \cite{Klasen:2023uqj}.

\begin{figure}[!b]
\centerline{
\hspace{\fill}
\subfloat[]{\includegraphics[width=0.3\textwidth]{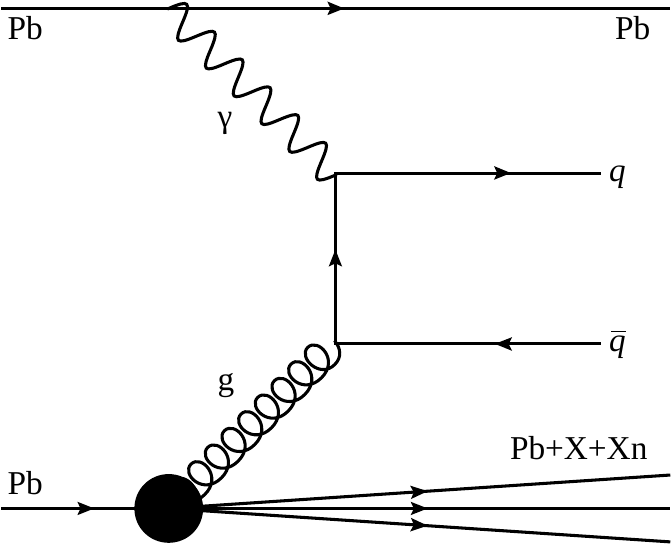}}
\hspace{\fill}
\subfloat[]{\includegraphics[width=0.3\textwidth]{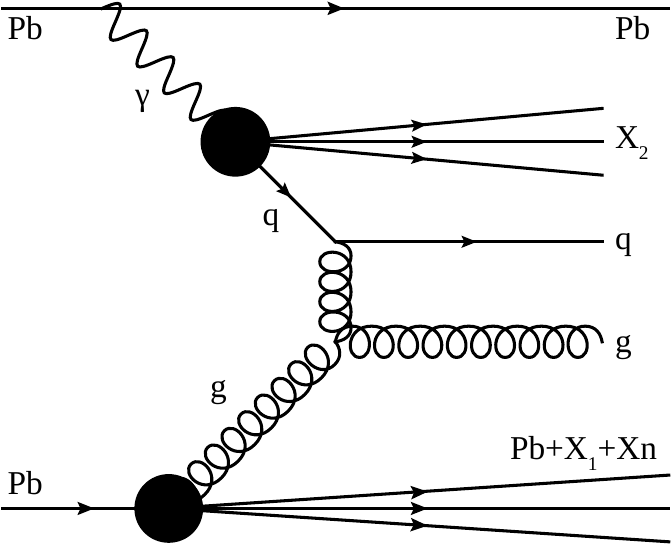}}
\hspace{\fill}
}
\caption{
Diagrams representing different types of leading-order contributions to dijet
production in high-energy photon-nucleus collisions. Diagram (a)
represents the direct contribution in which the photon itself
participates in the hard scattering. Diagram (b)
represents the ``resolved'' contribution in which a virtual excitation
of the photon, into a state involving at least a $q\bar{q}$ pair and
possibly multiple gluons, participates in the hard scattering.
The black circles represent hadronic processes where
hard partons are contributed from the initial state. Additional hadronic
particle production occurs in the final state resulting from the breakup
of the struck nucleus ($X$, $X_1$) or the resolved photon remnant ($X_2$).
}
\label{fig:photonuc}
\end{figure}

Ultra-peripheral nuclear collisions
(UPCs) \cite{Bertulani:2005ru,Baltz:2007kq,Klein:2020fmr} provide an opportunity to
study nuclear PDFs using quasi-real photons associated with one of the
colliding nuclei as an electromagnetic (EM) probe of the other
nucleus \cite{Strikman:2005yv,Guzey:2018dlm,Guzey:2019kik,}. In a leading-order (LO)
description of such
``photonuclear'' (\gamA) collisions, the photon can either scatter
directly off a quark or gluon from the opposing nucleus or fluctuate, virtually, into a hadronic state
that undergoes a hard-scattering process with that nucleus. The two
different processes, often referred to as ``direct'' and ``resolved,''
respectively, are illustrated in Figure~\ref{fig:photonuc}.

At the LHC, the incident nuclei generate large fluxes of
coherent photons with energies up to around $80$ GeV. Thus, it is
possible to study hard-scattering processes with $Q^2$ values well in
excess of 100 $\GeV^2$  initiated by the photons. Such
photonuclear collisions typically produce two or more detected jets that can be
used to reconstruct the kinematics of the incoming
particles. Therefore, UPC photonuclear production of dijet or multi-jet
final states in \PbPb\ collisions provides a direct probe of
lead nPDFs. These measurements cover a unique range of parton kinematics due to the
photonuclear collision system, allowing them to probe nPDFs at higher $Q^2$ than fixed-target
experiments, over a broad range in $x$. These UPC measurements are complementary to other
measurements \cite{CMS-HIN-16-003,HION-2023-05} at the LHC in
\pPb\ collisions that also provide useful constraints on the
lead PDFs. Since \pPb\ collisions have a larger underlying event and
more energetic hadronic probe, they have wider coverage in Bjorken $x$
but do not extend as low in $Q^2$. The kinematic coverage from photonuclear processes provides
access to both the shadowing region at low $x$ and the
anti-shadowing region, where the nPDFs rise at intermediate $x$ ($10^{-2} \lesssim x \lesssim 10^{-1}$).
The use of an EM probe in UPCs also reduces the theoretical uncertainty associated with
hadronic physics when extracting nPDFs.

The cross-section for \gamA\ processes and the
final-state jet kinematics depend
crucially on the photon flux provided by the incident nuclei. Thus,
the recent incorporation of nuclear photon fluxes \cite{Jackson:1999} into
the \pythia \cite{Sjostrand:2014zea}
event generator has greatly facilitated UPC measurements at the
LHC by allowing for more accurate modeling of the relevant initial state.
Recent measurements, such as those of dilepton production in ultra-peripheral
\gamgam\ scattering \cite{HION-2016-02,HION-2021-16,CMS-HIN-19-014,STDM-2019-19}, have provided  stringent tests of
the current theoretical understanding of the flux generated by a
nucleus. However, there remain some open issues \cite{Harland-Lang:2020veo,Harland_Lang_2021,Harland-Lang:2023ohq}
associated with the flux calculation, particularly involving the treatment of
higher-energy photons produced inside the nucleus, which
are relevant for measuring photonuclear
jet production.

Additional uncertainty on the photon energy modeling arises from sensitivity to
the photon fragmentation functions \cite{Cor03} introduced by resolved processes.
To reduce the impact of  %
photon flux and fragmentation uncertainties, measurements of photoproduction
cross-sections should be performed in intervals of photon energy. Resolved processes, however,
do not allow for a straightforward extraction of the photon
energy, so the hard-scattering kinematics are best characterized by the quantity $\zg \equiv \yg\xg$, where \yg\ is
the fractional photon momentum, $\yg \equiv E_\gam/\left(\sqn/2\right)$.
$\xg$ is the fraction of the photon's momentum carried by the parton
entering the hard scattering, where $\xg = 1$ for direct processes. These variables are further
described
and
defined in terms of final-state jets in Section \ref{sec:KinVarDefs}.

UPC photonuclear scattering can be distinguished from
non-UPC hard-scattering processes by requiring the
photon-emitting nucleus to remain intact. Experimentally, this is accomplished by
using the zero-degree calorimeters (ZDCs), which detect the beam-energy neutrons
emitted in most hadronic nuclear interactions. The condition that no neutrons (\zn) are observed in one
direction, combined with a requirement for gaps in the particle
rapidity distribution on that side of the event, is effective at identifying
photonuclear collisions \cite{HION-2018-27}. A requirement that at
least one neutron (\xn) is observed in the other direction distinguishes
photonuclear events from, for example, \gamgam scattering processes, and suppresses these backgrounds.

Although the emission of a coherent photon by an entire nucleus typically does not lead to nuclear
breakup, it is well-known \cite{Baur:1998ay,BAUR2002359,PhysRevC.57.1920,}
that additional soft EM interactions between passing
nuclei -- so-called Coulomb excitations -- can cause the photo-emitting
nucleus to break up. This process typically occurs via the excitation of the
nucleus to a giant dipole resonance \cite{VEYSSIERE1970561}, which subsequently decays,
emitting several neutrons. The probability for such neutron emission is
sensitive to the inter-nuclear impact parameter \cite{Baltz:2002pp}. The sampled impact
parameters are smaller for \gamA\ processes producing jets than, for
example, \gamgam\ production of dileptons because of the higher required
photon energies.  Consequently, it is expected that nuclear breakup probabilities will be larger in \gamA\ processes
than, for example, $\gamgam \rightarrow \mu^+\mu^-$. To
reduce the sensitivity to theoretical uncertainties in modeling nuclear breakup,
measurements of the breakup probability as a function
of kinematic variables, in particular \zg, is essential.

This paper presents measurements of dijet and multi-jet cross-sections
from photoproduction processes in ultra-peripheral \PbPb\ collisions at
$\sqn = 5.02$~\TeV, using data recorded in 2018 with the ATLAS detector at the LHC,
corresponding to an integrated luminosity of 1.72~\inb. The jets
produced in \gamA collisions are reconstructed using the anti-\kt
algorithm \cite{Cacciari:2008gp,Fastjet} with a radius parameter $R = 0.4$. Triple-differential
cross-sections, unfolded for detector response, are measured using two
sets of kinematic variables. The first set is based on the kinematics of
the jet system: the jet system rapidity, \yjets, the jet system
mass, \mjets, and \HT, the scalar sum of the transverse momenta (\pt) of the
jets. The second set of kinematic variables more directly characterizes
the hard-scattering process and parton kinematics. It consists of \zg,
the parton momentum fraction in the photon direction; \xA, the nuclear parton momentum fraction; and \HT, which can be viewed
as a proxy for the momentum transfer, $Q$. Each of these variables is defined in coordinates oriented with
respect to the direction (referred to hereafter as ``photon-going'') of the photon-emitting nucleus, where the positive $z$-axis
points along the momentum direction of this nucleus. Table~\ref{tbl:kinranges} lists the ranges
covered by the measurement for each of the quantities in the two
different sets of variables.  The same ranges are used for both the reconstructed and unfolded results. For the differential cross-sections that are the
primary result of this measurement, between 8 and 12 intervals are defined
covering each of the kinematic variables using either uniform linear or logarithmic divisions, as indicated in the
table. %
\begin{table}[tb]
\centering
\caption{Ranges in the two sets of kinematic variables covered by the measurement
and the intervals used in each variable for the measured cross-sections.
}
\begin{tabular}{l|l|l|c|c|l}\hline
Kinematic set & Variable & Minimum               & Maximum       & Interval & Number of  \\
&          & value                 & value         & type     & intervals  \\\hline\hline
\kinsetjet & \yjets      & $-4$                  &   $2$         & linear   &  12        \\
& \mjets      & $35$  GeV             &   $335$ GeV   & log      &  8          \\
& \HT         & $35$  GeV             &   $275$ GeV   & log      &  8         \\\hline
\kinsetpart& \xA         & $2.0\times10^{-3}$    &   $0.5$       & log      &  10       \\
& \zg         &  $3.7\times10^{-4}$   &   $0.027$     & log      &  7        \\
& \HT         &  $35$  GeV            &   $275$ GeV   & log      &  8        \\ \hline
\end{tabular}
\label{tbl:kinranges}
\end{table}

The remainder of this paper is structured as follows:
Section~\ref{sec:detector} describes the ATLAS detector;
Section~\ref{sec:samples} details the data and simulated event samples used in the
measurement; Section~\ref{sec:objevent} discusses the reconstruction
of physics objects and event selections applied in the analysis;
Section~\ref{sec:analysis} presents the procedure for performing the nominal analysis and cross-section measurement;
Section~\ref{sec:syst} discusses systematic uncertainties;
Section~\ref{sec:breakup} describes the measurement of nuclear break-up effects;
Section~\ref{sec:results} presents the results;
and Section~\ref{sec:conclusion} summarizes the paper and presents
conclusions.

% End of text imported from the .//tex/intro.tex input file

%

%
\section{The ATLAS detector}
\label{sec:detector}

% The next lines are included from the .//tex/detector.tex input file
%
%
%
%

%
%
%
%
%
%
%
%
%
%
%
%
%
%
%

%
%
%
%

%
%
%
%
%
%
%
%

%
%
%
%
%
%
%

%
%
%
%
%
%

%

The ATLAS detector~\cite{PERF-2007-01} at the LHC covers nearly the entire solid angle around the collision point.\footnote{ATLAS uses a right-handed coordinate system with its origin at the nominal interaction point (IP)
in the centre of the detector and the \(z\)-axis along the beam pipe.
The \(x\)-axis points from the IP to the centre of the LHC ring,
and the \(y\)-axis points upwards.
Polar coordinates \((r,\phi)\) are used in the transverse plane,
\(\phi\) being the azimuthal angle around the \(z\)-axis.
The pseudorapidity is defined in terms of the polar angle \(\theta\) as \(\eta = -\ln \tan(\theta/2)\) and is equal to the rapidity
$ y = \frac{1}{2} \ln \left( \frac{E + p_z c}{E - p_z c} \right) $ in the relativistic limit.
Angular distance is measured in units of \(\Delta R \equiv \sqrt{(\Delta y)^{2} + (\Delta\phi)^{2}}\).
}
It consists of an inner tracking detector surrounded by a thin superconducting solenoid, electromagnetic and hadronic calorimeters,
and a muon spectrometer incorporating three large superconducting air-core toroidal magnets.

The inner-detector system (ID) is immersed in a \qty{2}{\tesla} axial magnetic field
and provides charged-particle tracking in the range \(|\eta| < 2.5\).
The high-granularity silicon pixel detector covers the vertex region and typically provides four measurements per track,
the first hit generally being in the insertable B-layer (IBL) installed before Run~2~\cite{ATLAS-TDR-19,PIX-2018-001}.
It is followed by the SemiConductor Tracker (SCT), which usually provides eight measurements per track.
These silicon detectors are complemented by the transition radiation tracker (TRT),
which enables radially extended track reconstruction up to \(|\eta| = 2.0\).
The TRT also provides electron identification information
based on the fraction of hits (typically 30 in total) above a higher energy-deposit threshold corresponding to transition radiation.

The calorimeter system covers the pseudorapidity range \(|\eta| < 4.9\).
Within the region \(|\eta|< 3.2\), electromagnetic calorimetry is provided by barrel and
endcap high-granularity lead/liquid-argon (LAr) calorimeters,
with an additional thin LAr presampler covering \(|\eta| < 1.8\)
to correct for energy loss in material upstream of the calorimeters.
Hadronic calorimetry is provided by the steel/scintillator-tile calorimeter,
segmented into three barrel structures within \(|\eta| < 1.7\), and two copper/LAr hadronic endcap calorimeters.
The solid angle coverage is completed with forward copper/LAr and tungsten/LAr calorimeter modules
optimised for electromagnetic and hadronic energy measurements respectively.

The muon spectrometer (MS) comprises separate trigger and
high-precision tracking chambers measuring the deflection of muons in a magnetic field generated by the superconducting air-core toroidal magnets.
The field integral of the toroids ranges between \num{2.0} and \qty{6.0}{\tesla\metre}
across most of the detector.
Three layers of precision chambers, each consisting of layers of monitored drift tubes, cover the region \(|\eta| < 2.7\),
complemented by cathode-strip chambers in the forward region, where the background is highest.
The muon trigger system covers the range \(|\eta| < 2.4\) with resistive-plate chambers in the barrel, and thin-gap chambers in the endcap regions.

The luminosity is measured mainly by the LUCID--2~\cite{LUCID2} detector that records Cherenkov light produced in the quartz windows of photomultipliers located close to the beampipe.

Two ZDCs, which measure neutrons emitted at small rapidity
separation from the incident nuclei, are used for triggering and for offline
event selection. The ZDCs are located symmetrically at a distance of
$\pm 140$~m from the nominal IP and cover $|\eta|>8.3$ along the beam axis.
Each calorimeter consists of four modules, each containing slightly more than one interaction
length of tungsten absorber.

Events are selected by the first-level (L1) trigger system implemented in custom hardware,
followed by selections made by algorithms implemented in software in the high-level trigger (HLT)~\cite{TRIG-2016-01}.
The first-level trigger accepts events from the \qty{40}{\MHz} bunch crossings at a rate below \qty{100}{\kHz},
which the high-level trigger further reduces in order to record complete events to disk at about \qty{1}{\kHz}.

A software suite~\cite{SOFT-2022-02} is used in data simulation, in the reconstruction
and analysis of real and simulated data, in detector operations, and in the trigger and data acquisition
systems of the experiment.

% End of text imported from the .//tex/detector.tex input file

%

%
\section{Data and simulated event samples}
\label{sec:samples}

% The next lines are included from the .//tex/samples.tex input file
\subsection{Data sample}

The measurements presented in this paper were performed using
1.72~\inb\ of \PbPb\ collision data recorded by ATLAS during
2018. During that data-taking period, the typical number of hadronic
\PbPb\ collisions per bunch crossing was $\mu \lesssim  5 \times 10^{-3}$. Although
the likelihood of multiple hadronic collisions in a single bunch crossing is small, pile-up may be induced
by EM dissociation processes that have much larger
cross-sections \cite{PhysRevLett.109.252302,PhysRevSTAB.17.021006,Klusek-Gawenda:2013ema,}.
This form of EM pile-up is accounted for using methods described in Section \ref{sec:corrs}.

The primary triggers used for the measurement involve a combination of
\znxn and transverse energy requirements at L1 and a requirement
of at least one jet above a given \pT threshold
in the HLT. In particular, the L1 triggers required that the total
transverse energy measured in the ATLAS calorimeters, \sumET, satisfies
$5 < \sumET < 200$~GeV. The HLT jet triggers are based on anti-\kt
jets with radius $R = 0.4$ reconstructed using topological clusters \cite{PERF-2014-07}
formed from energy deposits in the calorimeter. These triggers
were implemented with nominal jet-\pt\ thresholds of 10 and 20~\GeV. For 40\%
of the sampled luminosity, the jet triggers were applied only over
the pseudorapidity range $\etajet < 3.2$, while for the rest of the data
jets were triggered over the range $\etajet < 4.9$.
A separate set of events collected using only the L1
\znxn and \sumET conditions is used to evaluate the HLT jet-trigger
efficiency.

In order to measure \gamA collisions in which the photon-emitting
nucleus breaks up due to additional soft-photon exchanges, a
separate sample of \xnxn events was recorded. This sample used two
triggers that require neutrons in both ZDCs.  One of these selects events with a
maximum \sumET of 50~\GeV and at least one track in the ID, while
the other selects events with a minimum \sumET of 50~GeV. The sampled luminosities
were $33.2$ $\mu$b$^{-1}$ and $32.7$ $\mu$b$^{-1}$ for the first
and second trigger, respectively. A sample of minimum-bias $5.02$~\TeV \pp collisions is also used for background studies.
This \pp data sample was collected in 2017 using a trigger that required
at least one track in the ID and collected $2.66$ nb$^{-1}$ of integrated luminosity.

\subsection{Monte Carlo simulated samples}

The primary Monte Carlo (MC) sample used for this analysis was created %
using the \pythia event generator,
which can simulate photon-induced hard-scattering processes
\cite{Helenius:2018bai} and allows a user-defined photon flux, $F_{\gamma/A}$, to be specified.
This sample was produced using the photon flux associated with a
classical point charge integrated over the transverse dimensions above
a minimum cutoff, $b_{\text{min}} \approx 6.62$
fm~\cite{Jackson:1999}, the nominal radius of a lead nucleus, which is given in
Eq.~(\ref{eqn:PhotonFlux}) by the equivalent photon approximation (EPA):

\begin{equation}
F_{\gamma/A} (E_{\gamma}) = \frac{2}{\pi} \frac{\alpha Z^2 }{\beta_{\text{B}}^2}\frac{1}{E_{\gamma}}
\left[
u_{R} K_1(u_R)K_0(u_R) - \frac{u_{R}^2 \beta_{\text{B}}^2}{2}\,
\left(
K_1^2(u_R)-K_0^2(u_R)
\right)
\right]\,.
\label{eqn:PhotonFlux}
\end{equation}

\begin{figure}[!tb]
\centerline{
\includegraphics[width=0.5\textwidth]{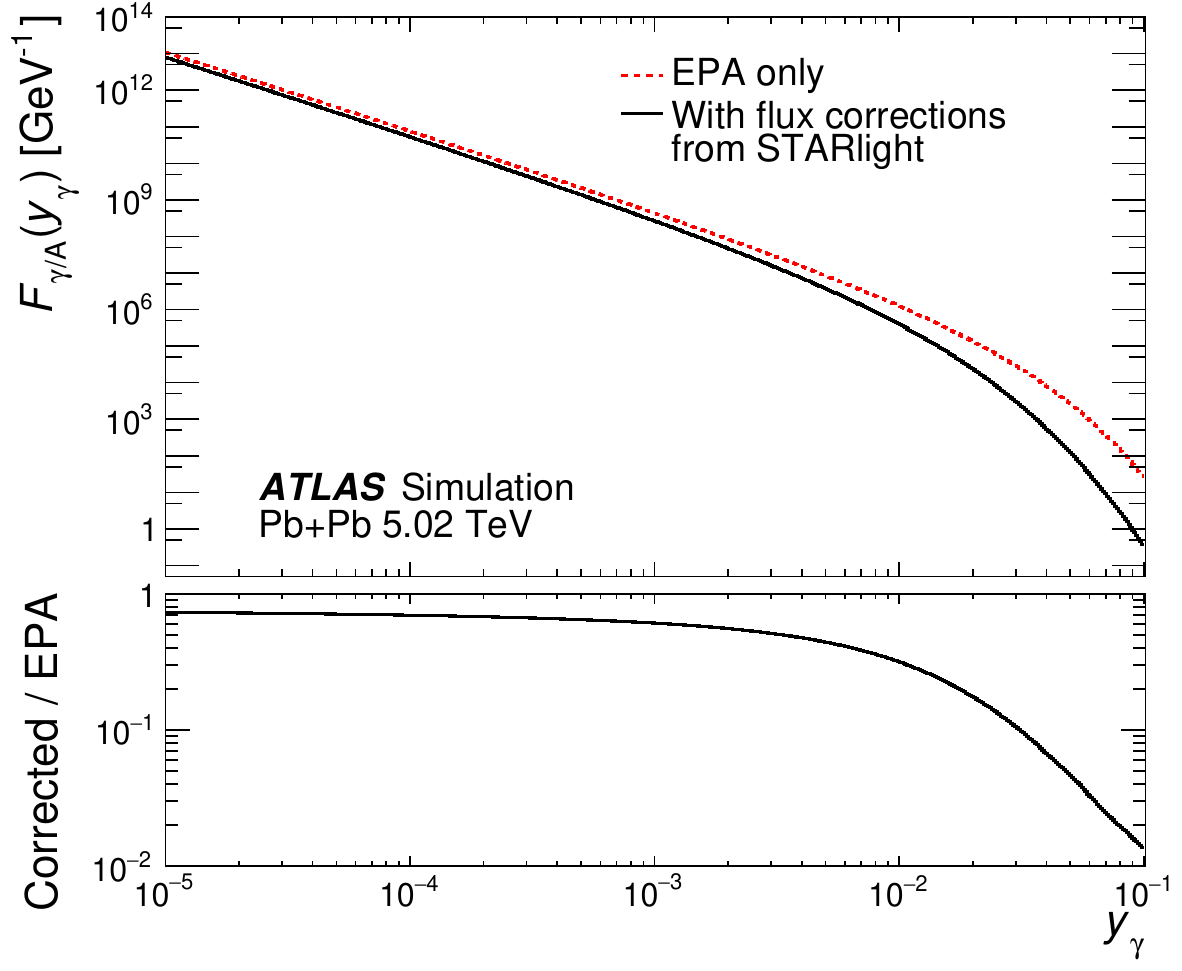}
}
\caption{
The photon flux generated coherently by a single nucleus in EPA with a hard cut-off
at the nuclear radius (red dashed), compared to a photon flux fully corrected using the
\textsc{Starlight} \cite{Klein:2016yzr} event generator (solid black). The ratio
shown in the bottom panel is used to correct the \pythia MC sample used in this analysis.
}
\label{fig:PhotonFluxMod}
\end{figure}

Here $u_R = E_{\gamma}b_{\text{min}}/\gamma_{\text{B}}\beta_{\text{B}} \hbar c$,
$\alpha$ is the fine-structure constant, $E_\gamma$ is the emitted photon energy,
and $\beta_\text{B}$ and $\gamma_\text{B}$ are the Lorentz boost parameters
associated with the photon-emitting nucleus.
$K_0$ and $K_1$ are modified Bessel functions
of the second kind. This description of the flux, however, is incomplete because it neglects
effects arising from the nuclear size. First, the hard cutoff at the nuclear
radius imperfectly approximates the requirement that no hadronic interactions occur
in the collision. This requirement can be simulated with the
\textsc{Starlight} event generator \cite{Klein:2016yzr}, which is used in this analysis to re-weight the
photon spectrum in the \pythia sample. Additionally, Eq.~(\ref{eqn:PhotonFlux}) neglects the size of the
struck nucleus, which is accounted for by modelling the nuclear thickness
with a Woods-Saxon distribution \cite{Klein:2016yzr} and integrating over the full range of impact
parameters. The corrected and un-corrected photon fluxes are compared  as a function of
$y_\gamma$ in Figure \ref{fig:PhotonFluxMod}, which demonstrates that these effects
are small for collisions with small $y_\gamma$ (large $b$) but important for large
$y_\gamma$ (small $b$), since collisions with smaller impact parameters are more
sensitive to details of the collision geometry. The bottom panel of Figure \ref{fig:PhotonFluxMod}
shows the ratio of the corrected and un-corrected fluxes, which is the combined re-weighting
factor applied to the \pythia sample as a function of $y_\gamma$.

The simulated signal events include
both direct and resolved photon processes; the latter require additional
modeling using the CJKL photon PDF \cite{Cor03} set. The events were generated
using the nCTEQ15 \cite{Kovarik:2015cma} nuclear parton distribution
functions and the A14 set of tuned parameters (``tune'')~\cite{ATL-PHYS-PUB-2014-021}. Final-state
stable particles, defined as those with $c\tau > 10$~mm, were then
passed to a \textsc{Geant4}-based simulation of the ATLAS detector
\cite{Agostinelli:2002hh,SOFT-2010-01}. Samples with different
selections on the hard-scattering kinematics were generated to ensure
full coverage of the kinematic range considered in this measurement.
Samples with equal numbers of events were generated for photons propagating in the
positive and negative $z$ directions.

Two other samples of events were generated to simulate potential
backgrounds in the measurement. A set of diffractive photo-production
events was generated using \pythia, %
but with a modified pomeron flux to account for coherent emission by the
Pb nuclei, which yields kinematics similar to those of coherent photon emission.
Since the sample of peripheral
\PbPb\ events that satisfy the event-selection criteria is overwhelmingly
dominated ($>99\%$) by events with a single binary nucleon-nucleon collision,
a sample of \pp collisions was
generated using \pythia, requiring a hard-scattering process
producing jets with $\pT > 15$~GeV, to simulate this possible hadronic background.

% End of text imported from the .//tex/samples.tex input file

%

%
\section{Event reconstruction and selection}
\label{sec:objevent}

% The next lines are included from the .//tex/objects.tex input file
\subsection{Event reconstruction}

Charged particle tracks in the ID are reconstructed over
the interval $|\eta^{\text{tr}}|<2.5$ using
the same methods and selections as applied in minimum-bias
\pp\ measurements \cite{STDM-2015-02}. Topological clusters are
reconstructed over the interval $|\eta^{\text{cl}}|<4.9$ from energy deposits
in the calorimeter \cite{PERF-2014-07}.

\begin{figure}[!tb]
\centerline{
\includegraphics[width=0.65\textwidth]{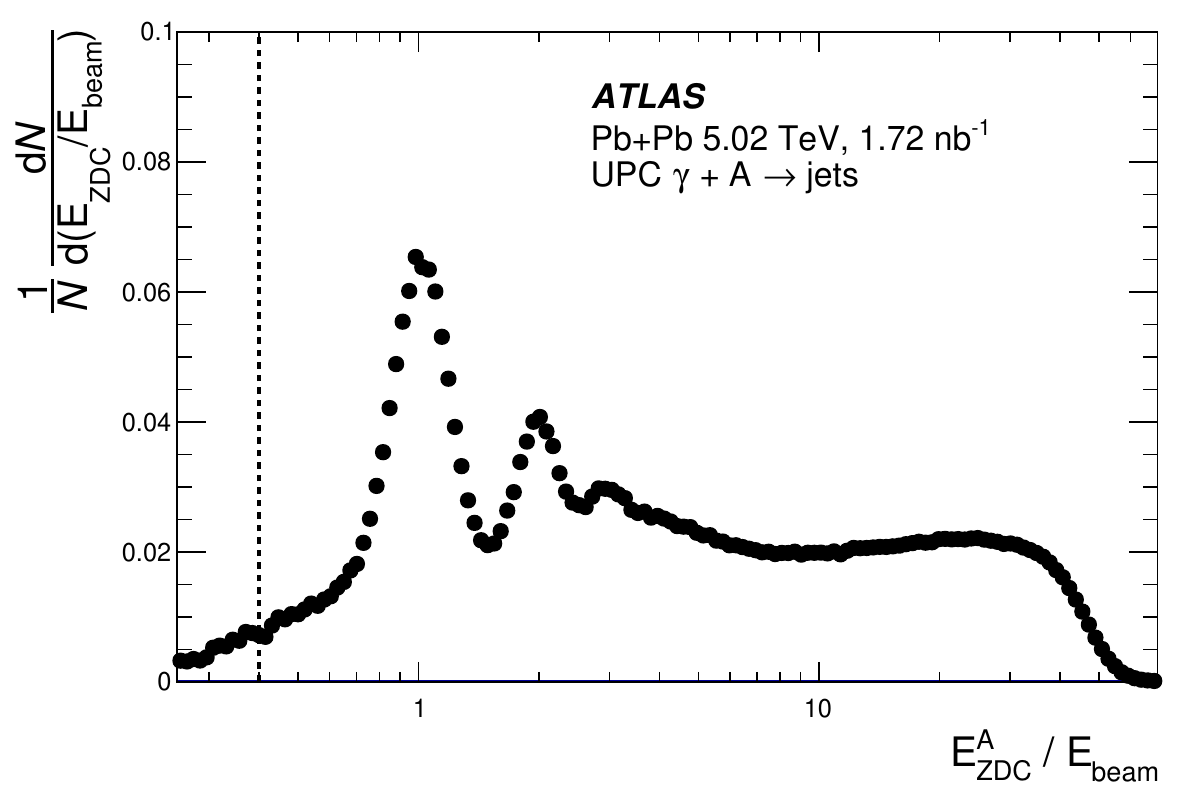}
}
\caption{
Distribution of energies measured in the ATLAS ZDCs
expressed in terms of the ratio of the measured energy
to the beam energy, $E_\mathrm{beam} = 2510$ \GeV. The energies are measured in the calorimeter
opposite the \zn side in events satisfying all photonuclear dijet
event selections except the \xn energy requirement which selects
events to the right of the dotted line.
}
\label{fig:ZDCEnergyXN}
\end{figure}
The jets used for this analysis are obtained from particle-flow inputs
derived from tracks and clusters as described in
Ref. \cite{PERF-2015-09}. They are reconstructed using the anti-\kt
algorithm with radius parameter $R = 0.4$. The jets are calibrated
to the hadronic scale using scale factors obtained from MC simulations
specifically derived for low-$\mu$ \pp data. A separate MC-based
correction derived from the \pythia\ photonuclear sample is applied to
account for differences in the hard-scattering physics and kinematics
between \gamA and \pp events. Separate \textit{in situ}
energy scale corrections \cite{JETM-2018-05} are applied that account for
differences in the jet response between data and simulation. These
corrections are derived in low-$\mu$ 13~\TeV\ \pp\ data using jets
recoiling against another object produced in the collision, such as another jet
in a different $\eta$ region or a \Zzero\ boson. Details on the calibration
are provided in Appendix~\ref{sec:JetResp}. The calibrations
are derived such that they are valid for calibrated jet \pT values
$\pTjet > 15$~\GeV; thus, only jets that satisfy this condition are included in
the analysis. To ensure that the jets are fully contained within the
acceptance of the detector, jets are also required to satisfy $\etajet < 4.4$.

The total energies of neutrons within the acceptance of the ZDCs are
obtained by summing individual energy measurements from the four longitudinal
segments of each calorimeter. The energy scale is calibrated using the
single neutron peak ($E = 2.51$~\TeV), for which an energy resolution of 17\% is
obtained. The primary analysis requires events to satisfy a \znxn\ requirement,
which is imposed by requiring the energy in one of the ZDCs -- the
``\zn'' direction -- to be less than 1~\TeV, while the energy in the
other -- the ``\xn'' direction -- is greater than 1~\TeV. The photon is
assumed to be traveling in the $0n$ direction and the struck nucleus in the
\xn direction. The distribution of ZDC energies
on the struck-nucleus side %
is shown in Figure~\ref{fig:ZDCEnergyXN} for events that satisfy the
trigger requirements and all the event selections described in Section~\ref{sec:eventsel}
except the ZDC energy requirement on the \xn side.  It is observed
that the single-neutron peak has a tail to small energies resulting from showers that start
deep in the calorimeter and are not fully contained. The fraction
of valid \xn\ events that would fall below the \xn\ energy
threshold, indicated in the figure with the dashed line, is estimated to be $\lesssim0.1~\%$.

\begin{figure}[!t]
\centerline{
\includegraphics[width=0.75\textwidth]{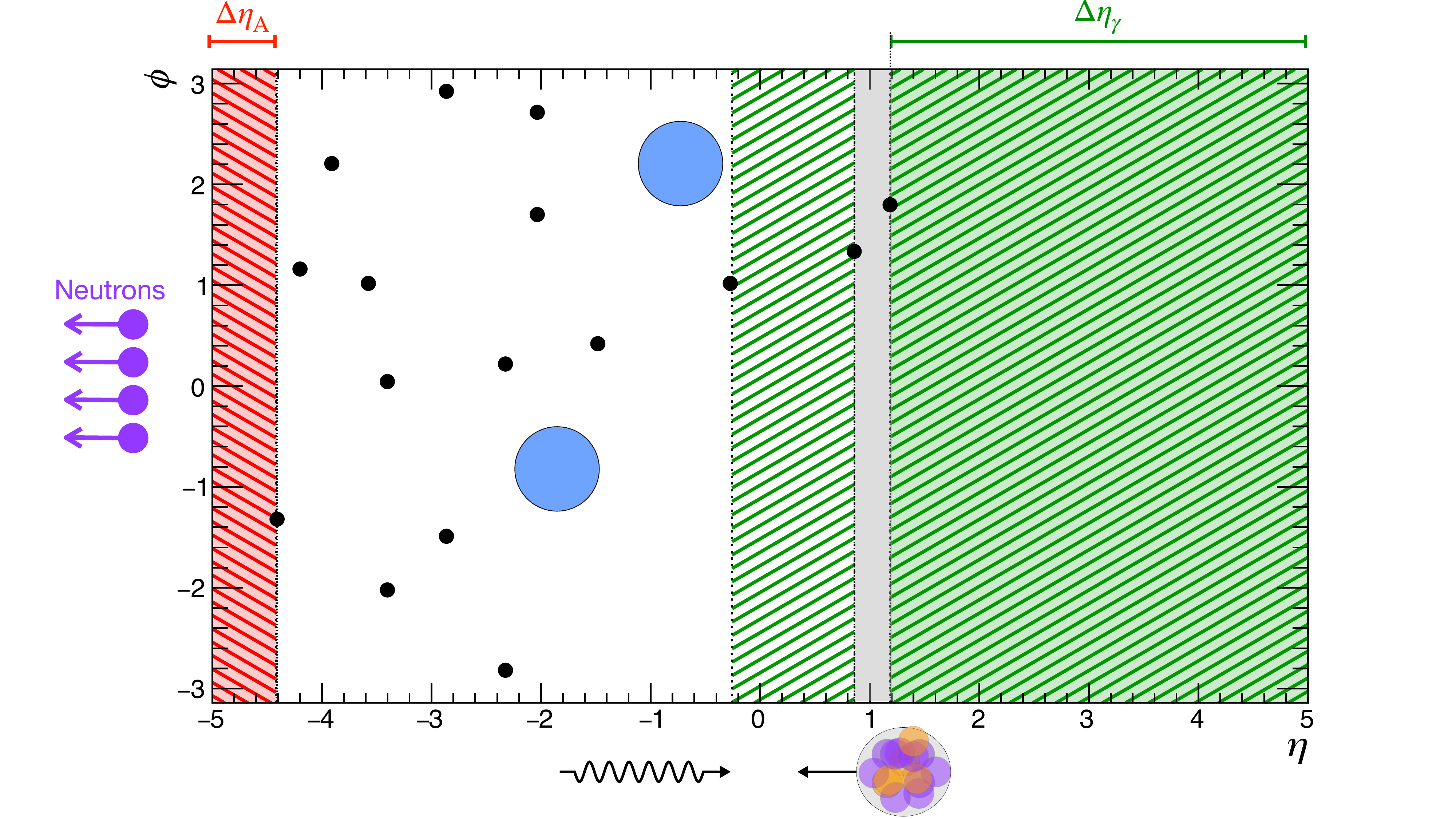}
}
\caption{Diagram illustrating how the various rapidity gap quantities
are computed in a typical event with photonuclear jet
production. Tracks and clusters are
indicated by black points while the jets are shown with blue
circles. The lack of neutrons in the positive $\eta$ direction
defines that as the photon-going direction. The \edgeGapGam\ and
\edgeGapA\ are indicated by the green and red solid shaded areas,
respectively. The regions contributing to the \sumGapGam\ are shown
with the hatched green shading, with the solid grey shaded region
indicating a $\Delta \eta$ that is smaller than 0.5 and excluded
from the \sumGapGam\ calculation. All hatched regions contribute to \sumGap.
}
\label{fig:GapCalculationDiagram}
\end{figure}

Another characteristic signature of photoproduction events is that they have large
gaps in the rapidities of produced particles.  This feature allows them to be separated
from hadronic backgrounds. These rapidity gaps are determined
using a combination of reconstructed tracks and clusters, both of which are
required to have $\pT > 200$~\MeV. Rapidity-dependent requirements are imposed on the significance
of cluster energies relative to noise levels to suppress contributions from electronic noise
\cite{STDM-2011-01}. So-called edge gaps, \edgeGapGam\ and \edgeGapA\,
are
defined %
as the interval between the edge of the detector ($\eta =\pm 4.9$) and the nearest track or cluster in the photon-going and
nuclear-going directions, respectively.

To prevent the rejection of
resolved photon processes, a second gap definition is constructed
by summing all intervals, $\Delta \eta$,  between $\eta$-adjacent tracks or clusters
with $\Delta \eta > 0.5$. This sum includes all intervals from the most forward jet in the photon-going
direction to the photon-going edge of the detector at $\eta = 4.9$, and the resulting quantity is
denoted \sumGapGam. A separate sum of all intervals satisfying
$\Delta\eta > 0.5$, regardless of orientation with respect to the jets,
is denoted as \sumGap and is used to reject $\gamma\gamma$ backgrounds.
Figure~\ref{fig:GapCalculationDiagram} illustrates how the
various rapidity gap quantities are calculated.

\subsection{Jet performance}

\begin{figure}[!tb]
\centering
\subfloat[]{\includegraphics[width=0.475\textwidth]{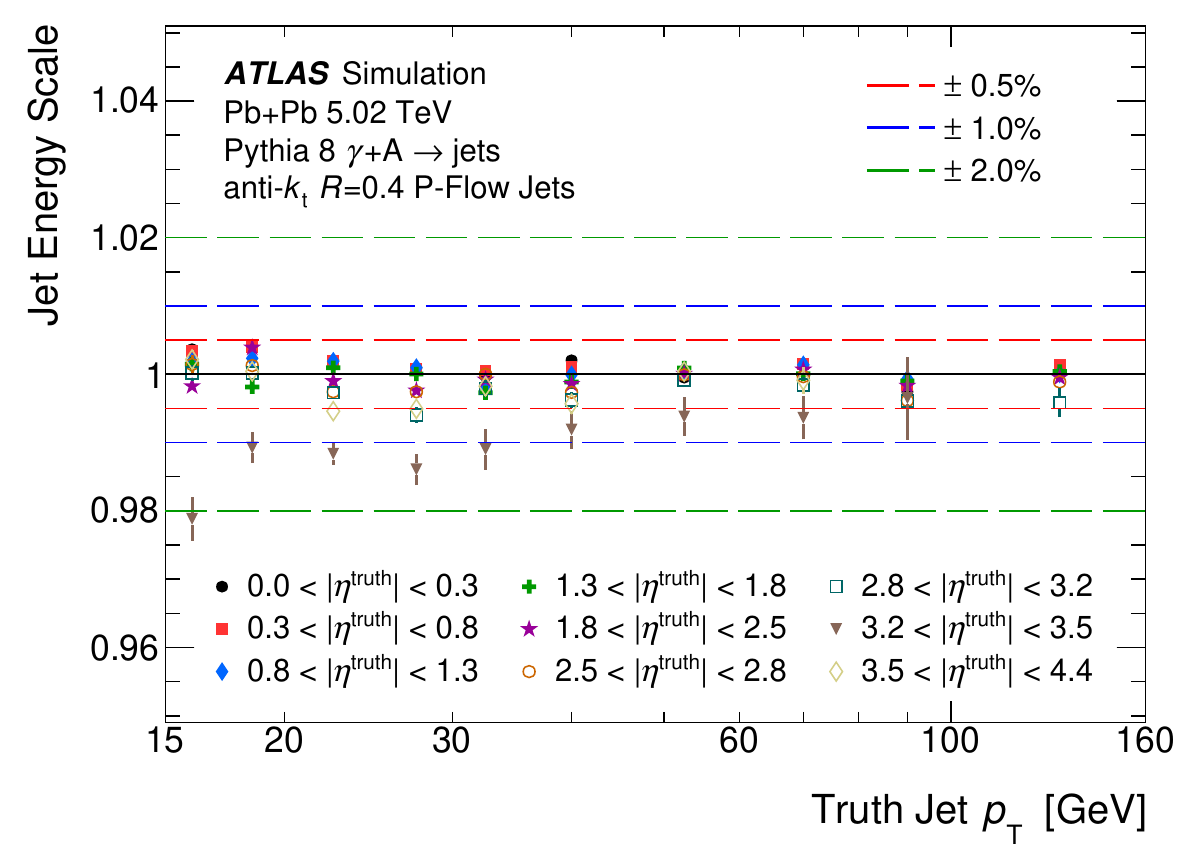}}
\subfloat[]{\includegraphics[width=0.475\textwidth]{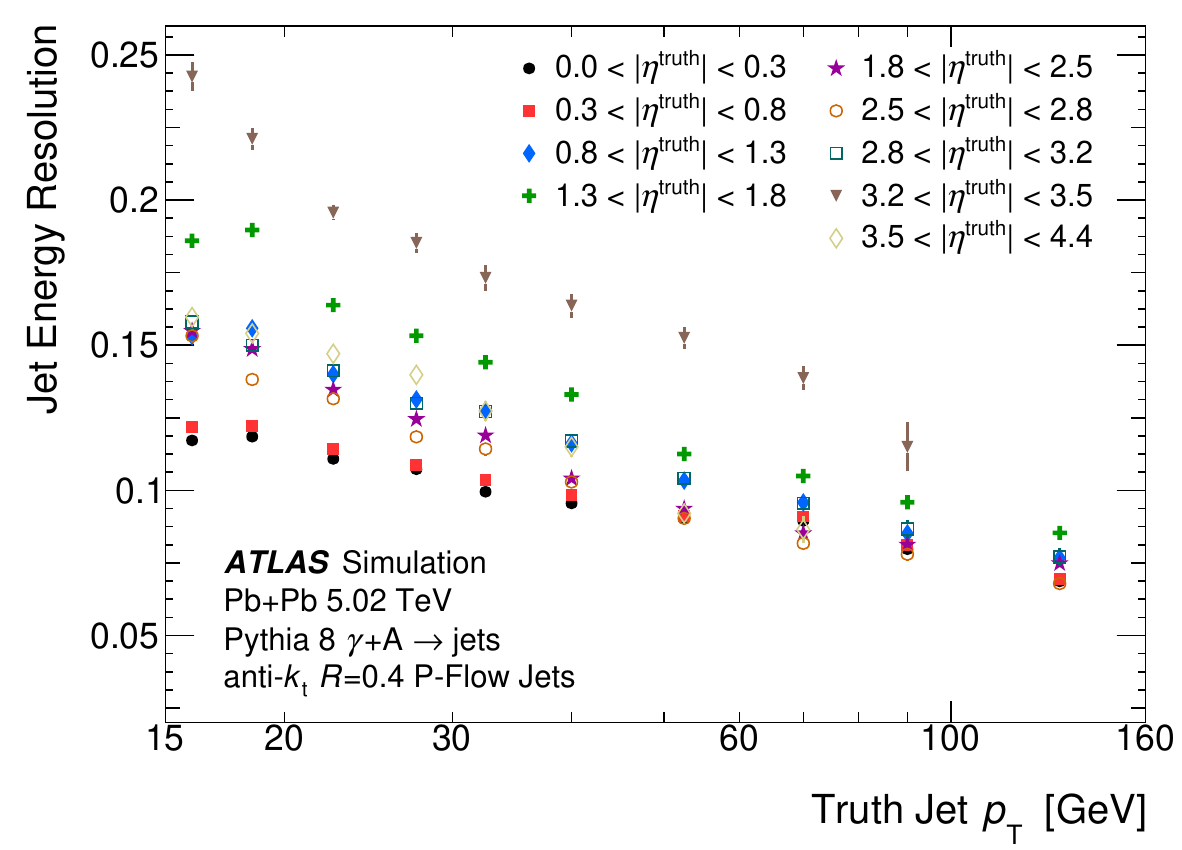}}
\caption{The (a) mean and (b) standard deviation of jet response distributions as
a function of \pTtruth\ for different $|\etaTruth|$ intervals. Results are restricted to
jets with $15 < \pTtruth < 160~\GeV$. The red and blue dashed lines
in (a) mark the $0.5\%$ and $1\%$ levels of closure, respectively.}
\label{fig:JetPerf}
\end{figure}

The jet reconstruction capabilities are evaluated in the \pythia \gamAjets MC sample
by matching truth jets
to the nearest reconstructed
jet within $\Delta \text{R} < 0.3$. Response distributions, $\pTreco / \pTtruth$,
are built in intervals of \pTtruth and $\etaTruth$.  Each distribution
is fit to a Gaussian function whose mean and standard deviation are referred to as the
jet energy scale (JES) and jet energy resolution (JER), respectively.
The JES and JER are shown in Figure~\ref{fig:JetPerf}
as functions of \pTtruth\ for different intervals of $|\etaTruth|$. The mean response
is generally within a half percent of unity over most of the \pTtruth\
and $\etaTruth$ range, which keeps the required unfolding corrections small.\par

The variation of the resolution with \pTjet\ and \etajet\ is consistent with what is observed in other ATLAS studies
of jet response in \pp collisions \cite{JETM-2018-05}. In this analysis, the resolution
at the lowest \pt\ values
is substantially smaller than in previous studies.
This is because the pile-up, which is the dominant contribution in typical LHC \pp analyses,
is negligible in the \PbPb\ UPC dataset.
The combined impact of deviations from unity in the JES and
the overall effect of JER are corrected in the unfolding procedure.
Possible differences in these jet calibrations between data and MC are
included as sources of systematic uncertainty.

\subsection{Jet system kinematic variables}
\label{sec:KinVarDefs}
This analysis in this paper relies on the measurement of the kinematics
of the outgoing system of $N$ jets resulting from hard-scattering
processes initiated by photons. The scalar transverse momentum sum,
\HT, is defined as
\begin{equation}
\HT \equiv \sum_i p_{\mathrm{T}\,i} ,
\end{equation}
while the $N$-jet system mass and rapidity are calculated as
\begin{align}
\mjets & \equiv\left[\left(\sum_i E_{i}\right)^2-\left|\sum_i
\vec{p}_{i}\right|^2\right]^{1/2},\\
\yjets & \equiv\frac{1}{2} \ln{\left(\dfrac{ \sum_i E_{i} + \sum_i
{p^*_{z}}_i}{\sum_i E_{i} - \sum_i
{p^*_{z}}_i}\right)}\, .
\end{align}
In the equations above, $i$ runs over all measured jets in an event that satisfy $\pTjet > 15$~\GeV and $\etajet < 4.4$. $E$ and $\vec{p}$
represent jet energies and momentum vectors, respectively, and $p_{z}$
represents the longitudinal component of the jet
momentum. The signs of $p_z^*$ are chosen to be positive in
the photon-going direction. As a result, the \yjets\ values are signed
such that negative (positive) \yjets\ values correspond to the jet
system being  shifted towards the nuclear (photon) direction.

Neglecting effects from initial-state parton showers of incoming quarks
and gluons and final-state particles not included in the jet
reconstruction, the quantities,
\begin{align}
\zg & \equiv\frac{ \mjets }{\sqn} e^{+\yjets} \, ,\\
\xA & \equiv\frac{ \mjets}{\sqn} e^{-\yjets}\, ,
\end{align}
correspond to fractions of the beam momentum carried by the
partons in the emitted photon and struck nucleus, respectively. More generally, these
quantities provide physical observables that are strongly correlated with
the initial parton kinematics. In direct processes, in
which the photon participates directly in the hard
scattering, \zg\ corresponds to the deep-inelastic scattering variable $y$.

\subsection{Event selection}
\label{sec:eventsel}
Events used in the measurement were recorded during
stable running conditions of the LHC and meet standard data-quality
criteria \cite{DAPR-2018-01}. For the nominal analysis, events are
must satisfy the \znxn\ condition and contain at least two jets
with $\pTjet > 15$~\GeV and $\etajet < 4.4$. For events
that have at least one jet within the acceptance of the ID, a
reconstructed primary vertex \cite{PERF-2015-01} is required. To reduce
the contribution of events where jets produced in the hard scattering fail
event selections or where the hadronic underlying event contributes additional jets,
the mass of the jet system must satisfy $0.9 \HT < \mjets < 4\HT$. This selection
was optimized by studying the kinematic edges in \xA\ and \zg\ as a function of \HT and
choosing boundaries that approximately match those introduced by single-jet requirements.

To suppress hadronic \PbPb\ interactions
that pass the \znxn\ requirement, the sum of gaps in the photon-going direction,
\sumGapGam, must satisfy $\sumGapGam > 2.5$. To suppress
jet production from \gamgam or diffractive photoproduction processes,
an edge gap requirement is imposed in the nuclear-going direction:
$\edgeGapA < 3$. To suppress background from $\gamma+\gamma\rightarrow
e^+e^-$ pairs, the total sum of gaps, \sumGap, is
required to be less than 9. Additional event cleaning
requirements
are applied to remove
backgrounds \cite{DAPR-2017-01} that result from
upstream interactions of lead ions, from beam-halo muons passing
through the detector, or from events whose calorimeter measurements are
significantly distorted by collisions occurring in bunch crossings
preceding the crossing of interest. Backgrounds from upstream interactions
are removed by rejecting events with more than one vertex, and by requiring
consistency between the numbers of tracks and topo-clusters within the angular acceptance
of the ID. Beam-halo muon backgrounds are removed via vetos on
jets measured at large time delays relative to the bunch crossing. Events
distorted by prior collisions are removed by rejecting events where significant
regions of the calorimeter record negative energy signals, which are characteristic
of this distorted response.

The jet kinematic and event-level
selections applied in the analysis are summarized in
Table~\ref{tbl:select}. Events passing these selections
are referred to in the remainder of this paper
as UPC \gamAjets\ events. While events are selected
using a gap requirement, an efficiency correction is applied for this
selection, so the gap requirement is not part of the fiducial
definition and should not be applied for theoretical
comparisons to the measured cross-sections.

\begin{table}[tb]
\centering
\caption{Jet and event selections applied in the measurement. The
event cleaning requirements are discussed in Section \ref{sec:eventsel}.
$E_{0n}$ and $E_{Xn}$ are the ZDC energies on either side of the ATLAS
detector, which determine the photon-going and nucleus-going directions.
}
\begin{tabular}{l|l}\hline
Jet & Event \\ \hline
$\pTjet>15$~\GeV & ZDC \znxn: $E_{0n} < 1~\TeV$ and $E_{Xn} > 1~\text{TeV}$ \\
$\etajet < 4.4$  &  $N_{\text{jet}} \geq 2$\\
& $0.9\HT < \mjets < 4\HT$\\
& $\sumGapGam > 2.5$, $\edgeGapA < 3$, and $\sum \Delta \eta < 9$\\
\hline
\end{tabular}
\label{tbl:select}
\end{table}
\begin{figure}[!b]
\centerline{
\subfloat[]{\includegraphics[width=0.48\textwidth]{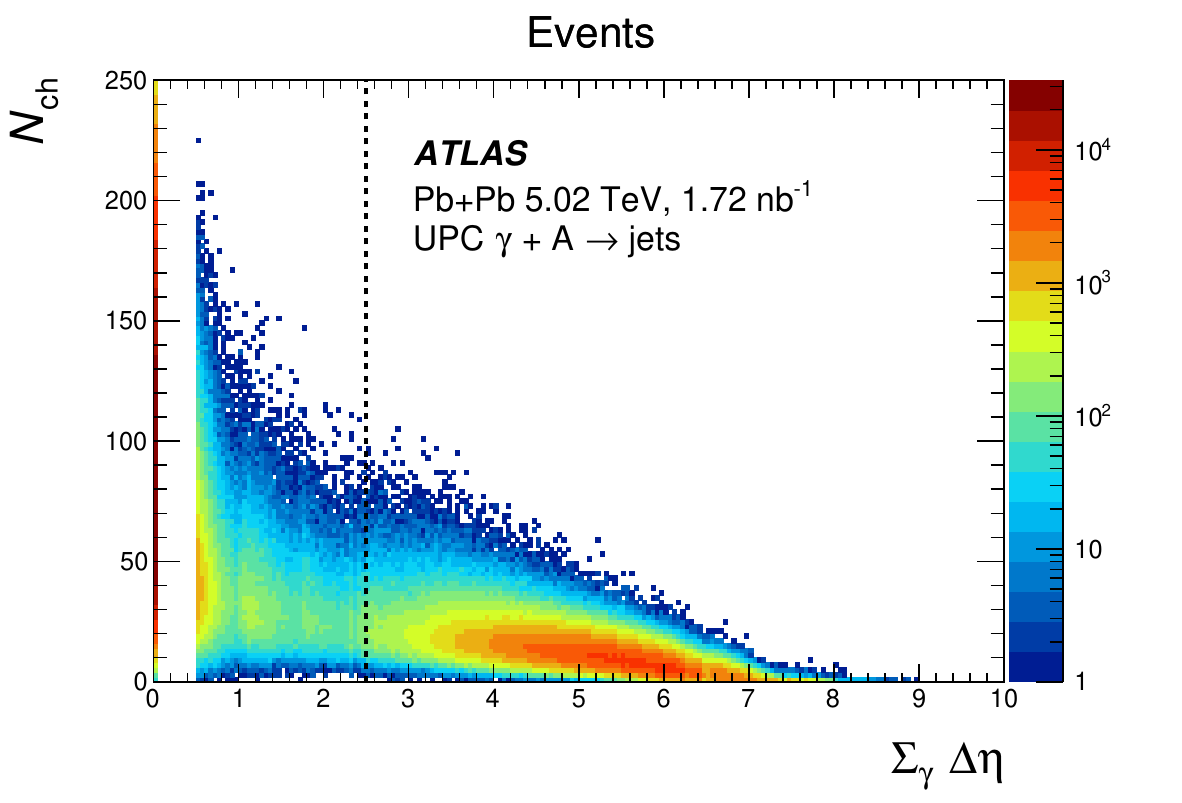}}
\subfloat[]{\includegraphics[width=0.48\textwidth]{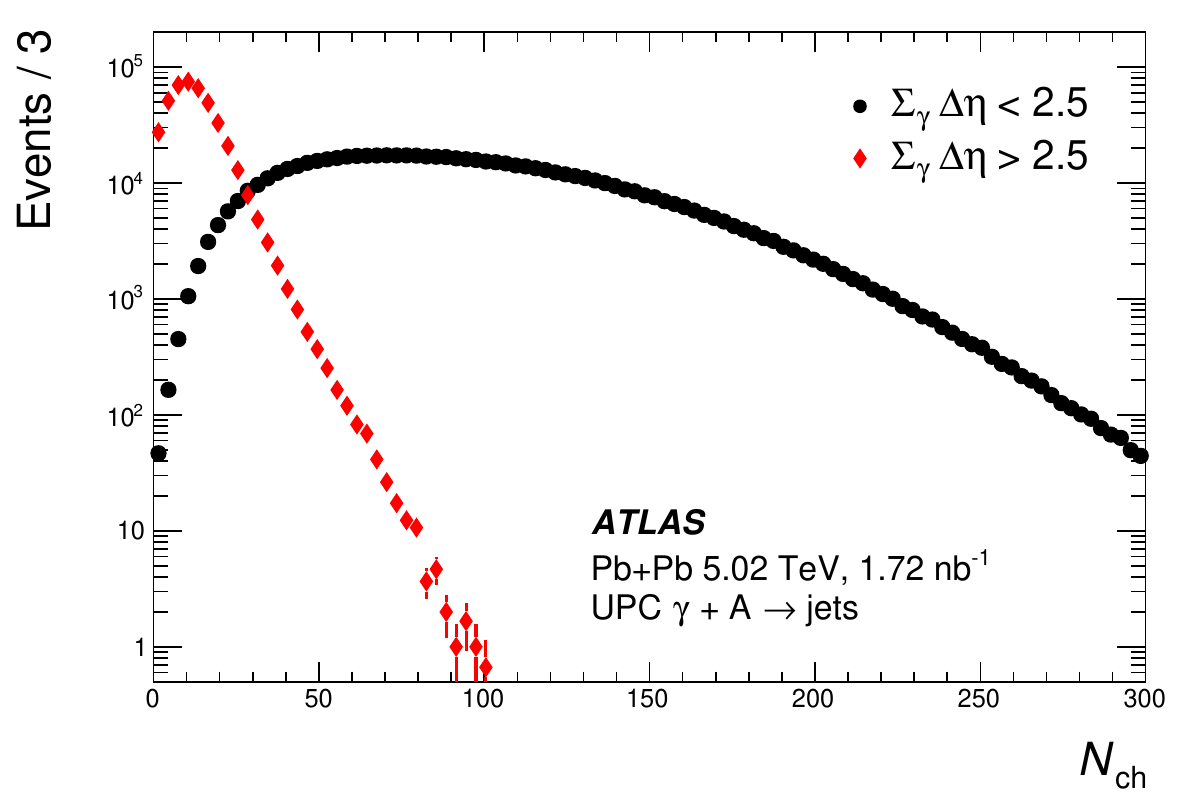}}
}
\caption{
(a) The distribution of charged track multiplicity, \Nch, versus the
sum of gaps in the photon-going direction, \sumGapGam. (b) The
distribution of \Nch for events passing and failing the $\sumGapGam >
2.5$ requirement. The distributions are obtained by applying all
event selections detailed in
Table~\ref{tbl:select} except the requirement on \sumGapGam.
}
\label{fig:gapSelect}
\end{figure}

The impact of the gap selection on the analysis is demonstrated in
Figure~\ref{fig:gapSelect}, which shows the correlation between the
multiplicity of charged tracks, \Nch, and \sumGapGam. \Nch is
defined as the number of reconstructed tracks with $|\eta^\text{tr}| < 2.5$
and $\pT^\text{tr} > 0.2~\GeV$. In
particular, events with small or zero \sumGapGam have a very broad
\Nch distribution as these events result primarily from hadronic
interactions in which the particle production spans the full ID
acceptance. In contrast, events with larger \sumGapGam have much
smaller multiplicities (even after accounting for the reduced geometrical acceptance),
which is consistent with the expected behavior for
photonuclear interactions, where hadronic production is
primarily backward and partially or fully outside the acceptance of the ID.

%

% End of text imported from the .//tex/objects.tex input file

%

%
%
%
%
%

%
%
%
%
%

%
%
%
%
%

%
\section{Analysis}
\label{sec:analysis}

% The next lines are included from the .//tex/analysis.tex input file
This section presents the methods used to derive
UPC \gamAjets cross-sections from the analysis components described
in Sections \ref{sec:samples} and \ref{sec:objevent}. First,
the inefficiencies arising from the application of sample selection requirements are
corrected for.
Then, residual background contamination is constrained via
detailed studies of the rapidity gap distributions, and the
resulting jet kinematics are validated by comparison with expectations from MC simulations.
Finally, three-dimensional cross-section distributions are produced in two sets of jet
kinematic variables and unfolded to correct for reconstruction effects.

\subsection{Efficiency corrections}
\label{sec:corrections}
\label{sec:corrs}
Multiple corrections are applied to the data to account for
inefficiencies arising from triggers, gap requirements, and
EM pile-up. Each of these sources is described below. For
other selections described in Section \ref{sec:eventsel} but not listed
in this section, including the $\sumGap < 9$ requirement and event
cleaning selections, the associated efficiency corrections are considered
to be negligible since they remove no events from the MC sample.

Potential trigger inefficiencies arise at L1 from the ZDC \znxn\ and
minimum \sumET\ requirements. Studies of the ZDC trigger
indicate it has an inefficiency of less than 0.1\%, so no correction
is required. The efficiency due to the L1 $\sumET > 5$~\GeV requirement, \effET,
depends on \HT because the jets produced in the \gamA processes
contribute significantly to \sumET. The inefficiency arises
because the calculation of \sumET used in the trigger can
differ substantially from the offline values at low total event energy. These
differences yield substantially smaller \sumET values in the trigger, which results in
inefficiency due to the $\sumET > 5$~\GeV requirement. %
This inefficiency is a few percent for \HT\ values less than $40$~\GeV, less than
one percent for \HT values above $40$ \GeV, and negligible for $\HT > 50$~\GeV.

Inefficiencies in the HLT jet trigger arise from an initial
filtering step performed in the HLT jet reconstruction. These inefficiencies are only
significant near mid-rapidity due to differences in the energy scale
between the particle-flow jets used in the analysis and the calorimeter jets
used in the trigger. Single-jet efficiencies, \effjet, are evaluated
separately for the 10 and 20~\GeV thresholds as a function of \etajet
and \pTjet, and a per-event jet trigger efficiency correction is obtained using
\begin{equation}
C_{\text{trig}}^{\text{jet}} = \frac{1}{1 - \prod_{i}\left(1-{\effjet}_i\right)},
\end{equation}
where $i$ runs over the jets -- usually two -- in the final state. Since the jet
trigger requires only a single jet, the event-level efficiency is much higher than
the single-jet efficiency, due to the possibility of any jet satisfying the
trigger requirement. The inefficiency is about $10\%$ for \HT\ values less than
$35$~\GeV, less than $2\%$ for \HT\ values above $35$~\GeV, and negligible
for $\HT > 40$~\GeV. No trigger efficiency correction is applied to the MC simulation
sample since a trigger is not required to select its events. %
The event selection efficiency, \effevt, is computed as a function of
the complete set of particle-level kinematic variables used to
characterize the final state, either $(\yjets,\mjets,\HT)$ or
$(\xA,\zg,\HT)$. The efficiency is evaluated using the portion of the MC \gamA sample
that falls within the UPC \gamAjets
acceptance defined in the previous section, and is separately tabulated
for each of the kinematic intervals defined in Table~\ref{tbl:kinranges}. The event
selection efficiencies, \effevt, account for losses
of events due to the application of both non-physics background requirements and
gap requirements. The migration from truth to reconstructed values in the measured kinematic variables is
corrected as part of the unfolding procedure and does not contribute
to \effevt. It is important to note that, although events are selected using
gap requirements, they are corrected to a fiducial cross-section definition without any
gap selection. Thus, when comparing the results of this measurement to theoretical
calculations, the theoretical comparison should not have any gap selections applied.

Because the non-physics background requirements
cause negligible loss of MC \gamA\ events, \effevt
mainly accounts for losses due to the gap selections, and these are
dominated by the $\sumGapGam > 2.5$ requirement. For events lying well
within the kinematic acceptance, the losses due to the gap selections
arise primarily from detector response and imperfect correspondence between
particle-level and reconstructed \sumGapGam or \edgeGapA
values. However, near the edge of the fiducial acceptance, the
event selection efficiency also accounts for the loss of some
particle-level events within the fiducial acceptance. As a
result, the efficiency correction has the largest impact near the
acceptance edge or for large values of \zg and \yjets where the
jets restrict the range of \sumGapGam. The efficiency correction is
typically $10$--$20\%$ in the highest \zg interval measured, $0.015 < \zg < 0.027$.
In the next-highest interval, the correction is typically $5$--$10\%$ and is less than $1\%$
in other \zg intervals.

The \znxn\ requirement introduces an inefficiency due to the presence of
EM pile-up where independent dissociative
\PbPb\ collisions, occurring in the same bunch crossing as an event of
interest, produce neutrons in the photon-going direction of the \gamAjets event.
The rate at which these processes occur is estimated using the single EM dissociation
cross-section, which is used to calculate the per-event probability that
an event is lost due to EM pile-up following the procedure used in
Refs.~\cite{HION-2016-02,HION-2021-16}. The correction, \CEM, averaged over the full
data set, is $1.070 \pm 0.003$ and is independent of the \gamAjets
kinematics. It is evaluated and applied event-by-event using
the luminosity for the associated bunch crossing measured by ATLAS \cite{DAPR-2021-01}.

The corrections described above are combined into a single
correction per event,
\begin{equation}
C = \frac{1}{\effET \effevt}\times C_{\text{trig}}^{\text{jet}}\times \CEM,
\end{equation}
which is applied to every event included in the analysis, resulting in a
corrected yield, \Ncorr, which is used to populate reconstructed kinematic
distributions prior to unfolding.

\subsection{Backgrounds}
\label{sec:BkgdEst}

\begin{figure}[!tb]
\centerline{
\includegraphics[width=0.99\textwidth]{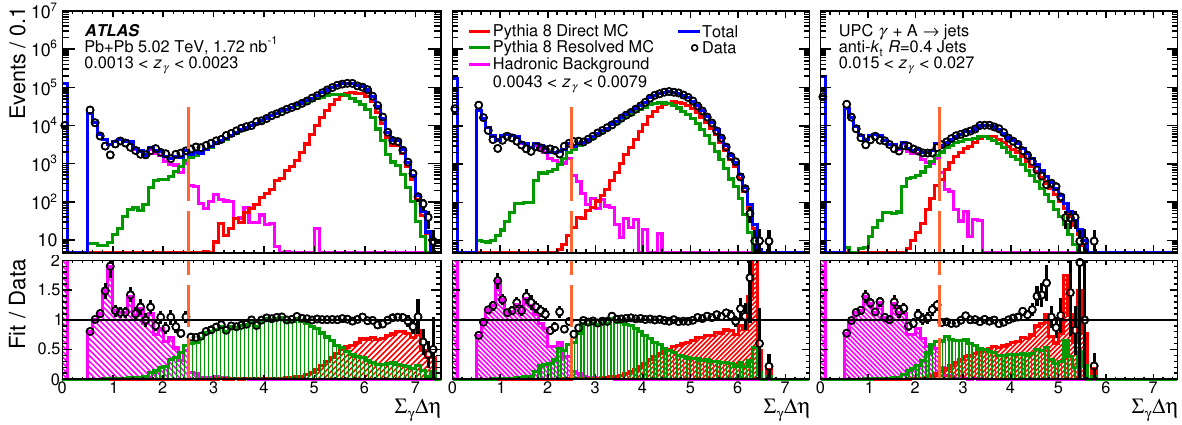}
}
\caption{
Template fits of the \sumGapGam\ distributions for several bins in \zg, with nominal analysis selections applied. The \pythia Direct and Resolved MC samples provide the two contributions to the signal, and the hadronic background template %
is derived from a fitted combination of MC simulation and \pp data. The bottom panel shows the ratio of the template fit results to the data in open markers and the ratio to each template component as the hatched bands. The orange dashed line denotes the nominal gap selection.
}
\label{fig:PhotonGapTempFits}
\end{figure}

\begin{figure}[!tb]
\centerline{
\includegraphics[width=0.99\textwidth]{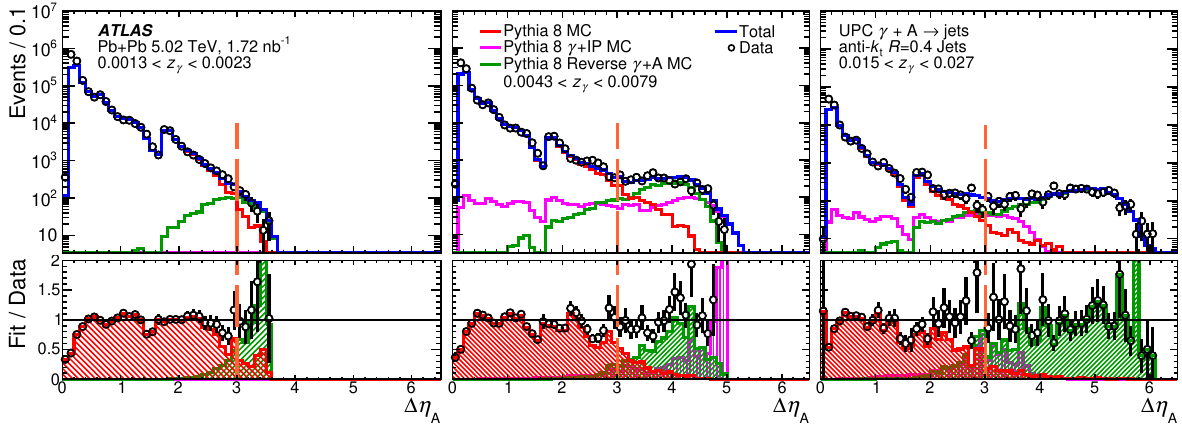}
}
\caption{
Template fits of the \edgeGapA\ distributions for several bins in \zg, with nominal analysis selections applied. The \pythia MC sample corresponds to photonuclear MC, the $\gamma + IP$ sample corresponds to the coherent photo-diffractive MC, and the Reverse \gamA\ MC sample corresponds to \gamA\ MC with its coordinate system inverted. The bottom panel shows the ratio of the template fit results to the data in open markers and the ratio to each template component as the hatched bands. The orange dashed line denotes the nominal gap selection.
}
\label{fig:NucGapTempFits}
\end{figure}

\begin{figure}[!tb]
\centerline{
\includegraphics[width=0.99\textwidth]{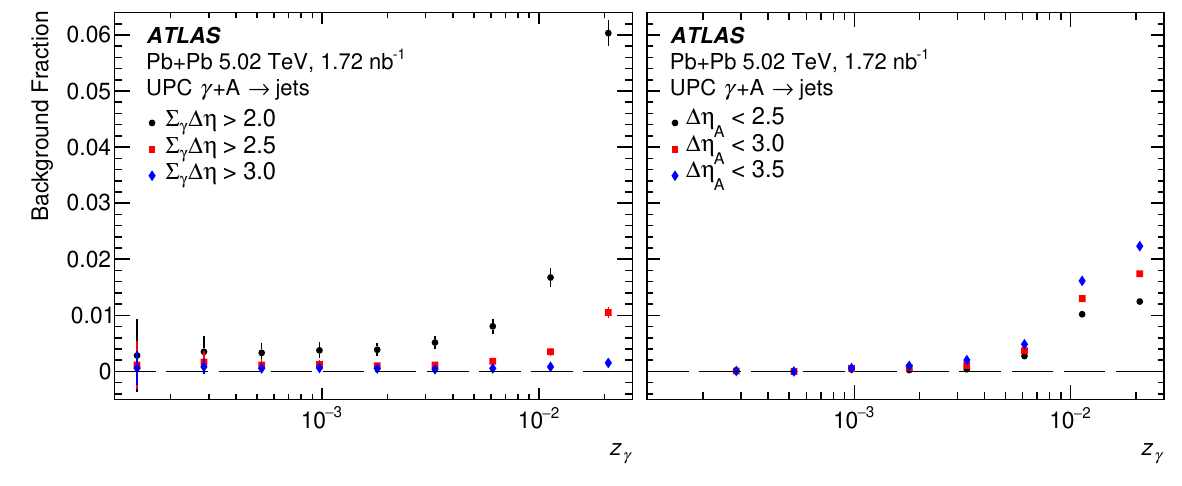}
}
\caption{
Residual background fractions for several different possible gap selections, extracted from template fits of the \sumGapGam and \edgeGapA distributions. The red markers indicate the nominal selections which are applied in this measurement, and the other markers show variations demonstrating the sensitivity of the background rate to the choice of selection. Error bars represent statistical uncertainties only.
}
\label{fig:BkgFracs}
\end{figure}

Backgrounds to this measurement arise primarily from hadronic or
photon-induced processes that mimic some part of the signature of
inclusive \gamA\ hard-scattering events. A key element of backgrounds arising
from photon-induced processes is independent nuclear
break-up via Coulomb excitation (\BU), which can cause processes that would
otherwise have a \znzn\ topology to satisfy the \znxn\ selection.
Additionally, the possibility is considered that the struck nucleus may
not always emit neutrons (\NBU). Either of these two possibilities, \BU or \NBU,
allow for contamination to occur between neutron topologies, causing processes
that do not typically have a \znxn\ topology to contribute backgrounds to
this measurement. Three such physical processes provide backgrounds sufficiently
large to require a dedicated veto: hadronic \PbPb\ collisions, \gamPomJets (diffractive photoproduction of jets), and
\znzn\ \gamAjets. Backgrounds resulting from these processes cannot be fully removed
via ZDC selections, so rapidity gap selections are employed to reject them.
Figures \ref{fig:PhotonGapTempFits}, \ref{fig:NucGapTempFits}, and \ref{fig:BkgFracs}
illustrate the efficacy of these selections.\par

The gap requirement in the photon-going direction ($\sumGapGam > 2.5$) mitigates
the most prominent background, hadronic \PbPb\ collisions with an \NBU
condition. This background yields the steeply falling distribution at small
\sumGapGam visible in Figure \ref{fig:PhotonGapTempFits}, and it is
modeled phenomenologically using a template distribution derived from the \pp
data and \pythia MC simulation. This description
is expected to be imperfect due to the different ratios of hadronic and diffractive
processes in \pp and \PbPb\ collisons. However,
the measurement itself is insensitive to the modeling of this background given
the low rate of contamination shown in Figure \ref{fig:BkgFracs}. Since the background
contamination rates are so small, no subtraction is applied and no additional uncertainty is considered on the
measurement resulting from imperfections in these template fits.

The gap requirement in the nucleus-going direction ($\edgeGapA < 3$) is employed to remove
backgrounds resulting from two sources: \gamPomJets and \NBU\ \gamAjets, both
of which may contaminate the \znxn\ sample in the \BU case. The \gamPomJets
background is simulated using a \pythia MC sample where the pomerons are emitted
coherently by the entire nucleus, softening the pomeron energy spectrum.
For the \NBU \gamAjets background, the contamination is only important when uncorrelated
nuclear breakup causes the photon-going direction to be mis-identified by
the ZDCs, manifesting as ``reverse'' \gamAjets events. These backgrounds are
modeled in template fits using a \pythia \gamAjets sample, with its coordinate system inverted
such that the photon-emitting nucleus moves in the negative direction. A template
fit to the \edgeGapA distribution is shown in Figure~\ref{fig:NucGapTempFits}, demonstrating
the good description these background models provide for the observed distributions.
Figure~\ref{fig:BkgFracs} shows the background contamination at different \zg\ values
for several edge gap requirements, including the
region used for the event selection ($\edgeGapA < 3$ and $\sumGapGam > 2.5$).  This figure demonstrates that the
rate of background contamination is small enough to be neglected, so no subtraction
is applied to the data. Studies of \NBU \gamAjets processes in a separate \znzn\ sample
indicate that they occur at a rate of about $4\%$ of the \znxn\ case. Since this rate
is small compared to the total uncertainties on the measurement described in Section
\ref{sec:syst}, no attempt is made to correct for it.

\begin{figure}[!tb]
\centerline{
\includegraphics[width=0.8\textwidth]{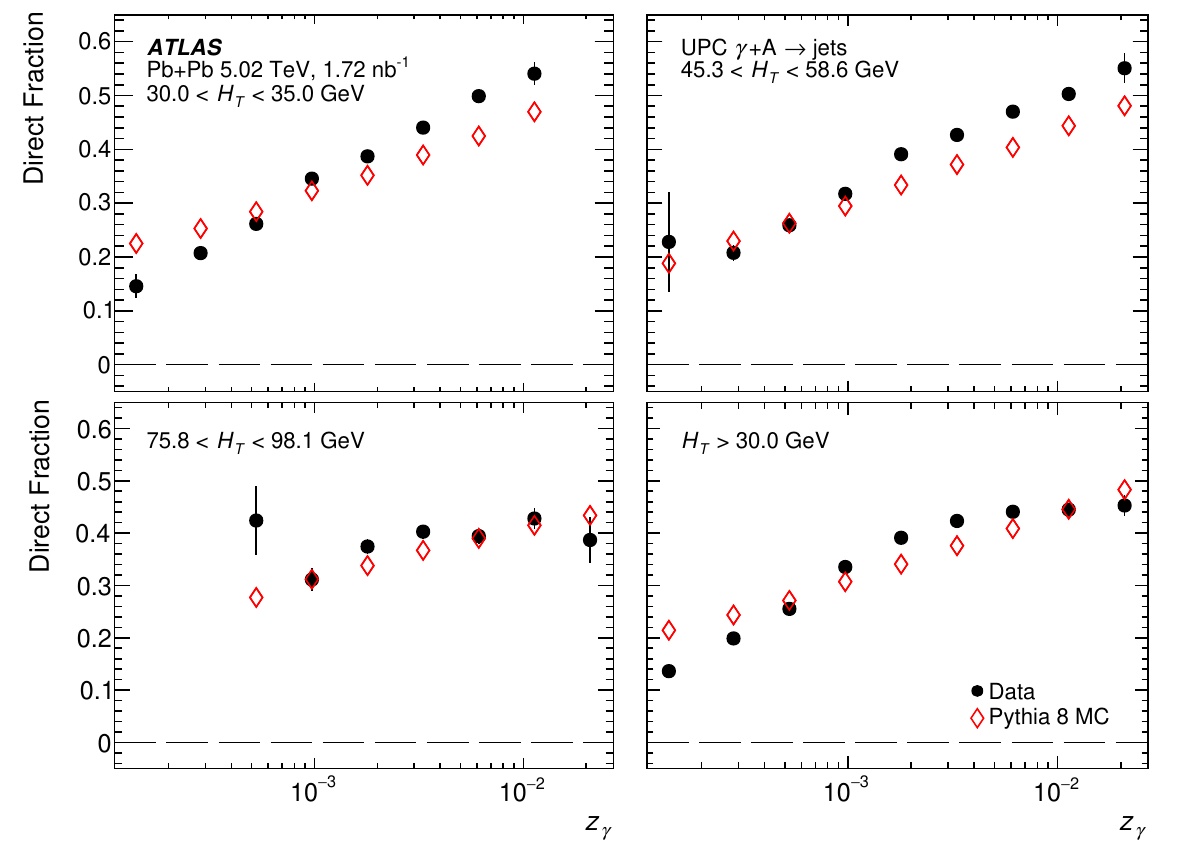}
}
\caption{
The fraction of events resulting from direct photon processes, extracted from template fits of the \sumGapGam distributions, compared to the results from \pythia using CJKL photon PDFs. Several different intervals in $H_\text{T}$ are shown. Error bars represent statistical uncertainties only.
}
\label{fig:DirFracs}
\end{figure}

The different shapes of the direct and resolved photon processes seen in Figure~\ref{fig:PhotonGapTempFits} can be used in a template fit to extract the rates of these processes in data as a function of \zg, thereby allowing a comparison with the predictions from the \pythia MC.
The shapes of the \sumGapGam templates derived from
\pythia are from LO modeling and could differ from data or from next-to-leading-order (NLO) calculations.
However, the difference between the LO and NLO cross-sections may be primarily a
normalization effect.
The good agreement between data and the template fits suggests that this may be the case.
The template fit then provides useful
input on the accuracy of \pythia in predicting the relative contributions from direct and resolved processes.
Figure~\ref{fig:DirFracs} shows the fraction of direct
\gamA processes obtained from the \sumGapGam template
fits. In general, the data show a slightly smaller fraction of
direct processes than \pythia at small \zg and a slightly higher
fraction at larger \zg, with better agreement at large \HT.

\FloatBarrier
\subsection{Data-MC comparison}
With the application of the corrections described above and a
data-driven correction for breakup of the photon-emitting nucleus (see Section \ref{sec:breakup}),
distributions of \gamA events from the data can be compared
to the same distributions obtained from the \pythia
MC sample. This is done by scaling differential cross-sections
obtained from \pythia by the integrated luminosity used in the
measurement and comparing the result to the differential distribution
of \Ncorr, the corrected yields, in the data. The
gap selections used in the analysis are
not applied to the \pythia sample since the data are already corrected for effects described in Section \ref{sec:corrs}, but the
same requirements on reconstructed jets are applied  and the kinematic
quantities for \pythia are obtained from reconstructed jets.

\begin{figure}[!tbp]
\centerline{
\subfloat[]{\includegraphics[width=0.49\textwidth]{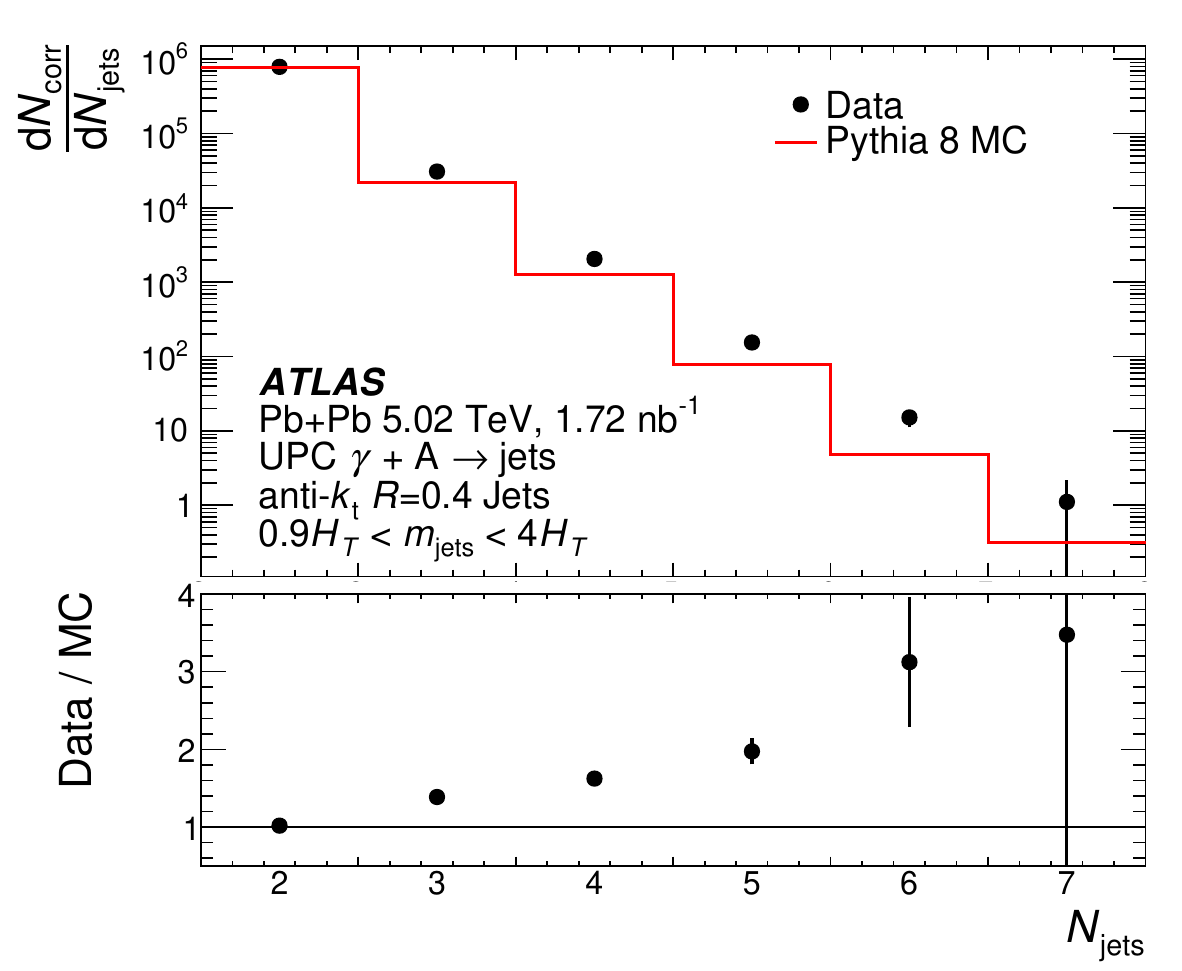}}
\subfloat[]{\includegraphics[width=0.49\textwidth]{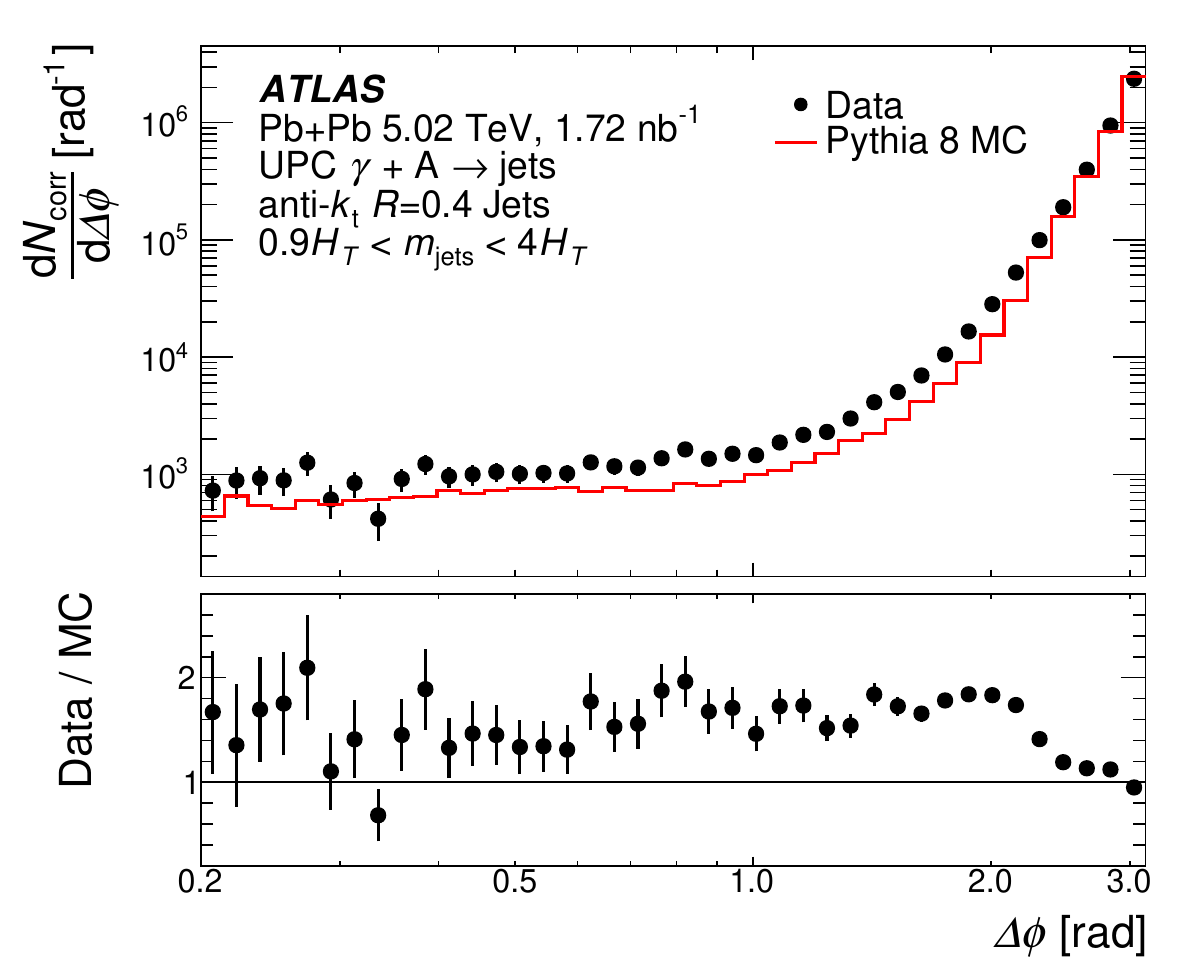}}
}
\caption{
Comparison of data and MC distributions for UPC \gamAjets production, within the fiducial acceptance,
for (a) the jet multiplicity and (b) the $\Delta\phi$ of the
leading dijet pair. The distributions are shown for reconstructed jet
kinematics after applying all event selections and
efficiency corrections. The bottom panels show the
ratios of the data and MC distributions. The error bars in
both panels show statistical uncertainties only.
}
\label{fig:CompsDPhi}
\end{figure}

Figure~\ref{fig:CompsDPhi} shows distributions
of the jet multiplicity and the azimuthal angle separation, \dphi, between the
two jets having the highest \pTjet values in the event. The \pythia
prediction systematically underestimates the rate for events with more
than two jets, likely because it only includes LO hard-scattering
matrix elements. This aspect of \pythia is most likely responsible for the
disagreement in the \dphi distribution for $\dphi \lesssim 2.5$ rad.

\begin{figure}[!tbp]
\centerline{
\subfloat[]{\includegraphics[width=0.49\textwidth]{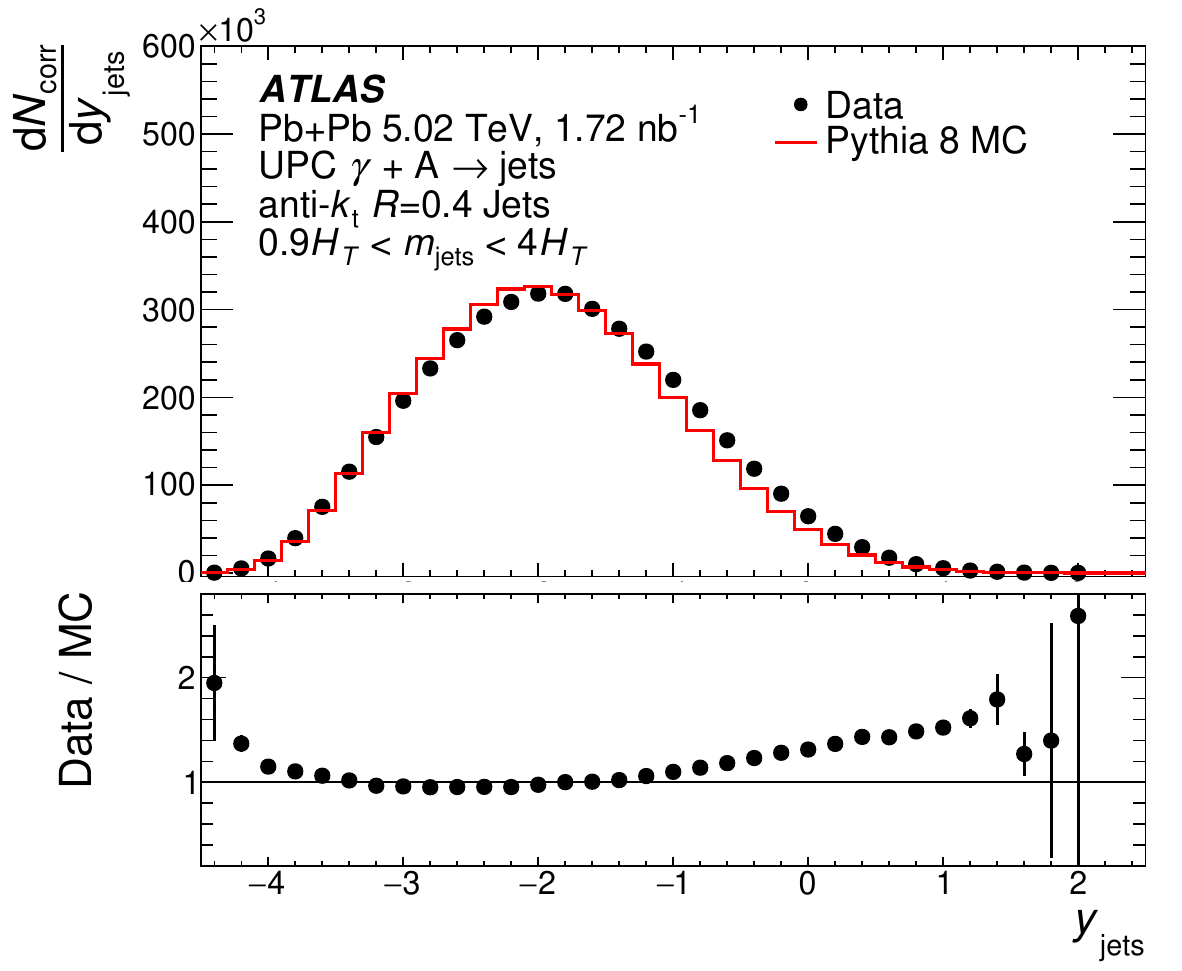}}
\subfloat[]{\includegraphics[width=0.49\textwidth]{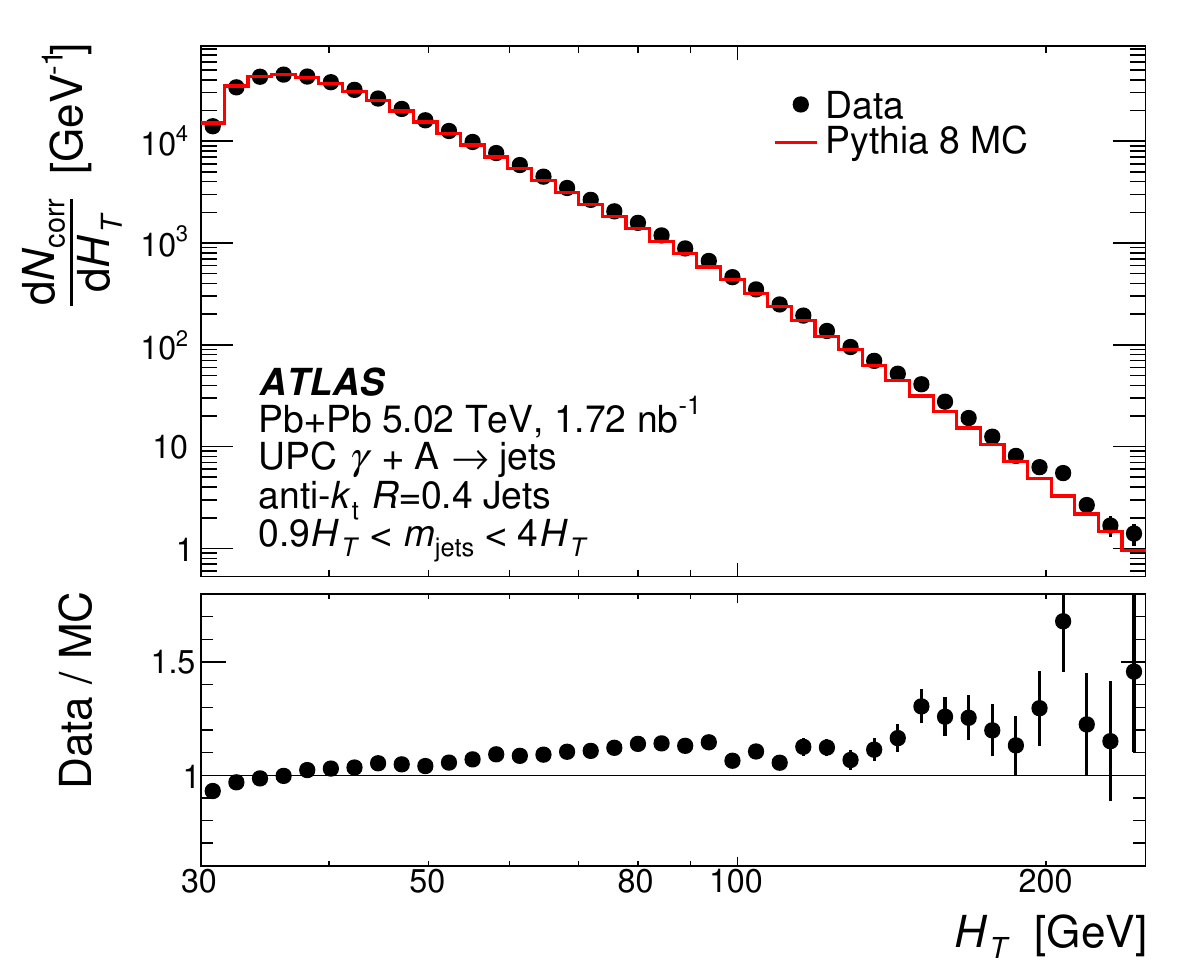}}
}
\caption{
Comparison of data and MC distributions for UPC \gamAjets production, within the fiducial acceptance,
for (a) \yjets and (b) \HT. The distributions are shown for reconstructed jet
kinematics after applying all event selections and
efficiency corrections.
The bottom panels show the ratios of the data and MC distributions. The error bars in
both panels show statistical uncertainties only.
}
\label{fig:CompsYjetsHT}
\end{figure}

Figure~\ref{fig:CompsYjetsHT} shows a comparison of \yjets and \HT
distributions between \pythia and data with corrections as described in
Section~\ref{sec:corrections}. In both MC and data, the \yjets
distributions are shifted toward the nucleus-going direction because the typical
photon energy is much smaller than that of the parton from which it
scatters. The \yjets\ distribution in data appears to be enhanced
compared to the MC at both backward ($\yjets < -4$) and forward ($\yjets > -1$)
\yjets values, while the data and MC agree well for intermediate values of \yjets. Nonetheless,
the level of agreement is sufficiently good that \pythia
can be used to evaluate the migration in \kinsetjet and \kinsetpart
due to jet response, which is needed to unfold the measured cross-sections.

\FloatBarrier
\subsection{Fiducial cross-sections}
\label{sec:crosssecdef}
The measured photonuclear jet cross-sections, corrected for efficiency
before unfolding for detector response, are calculated as
\begin{align}
\sigtridiffJetKin[\text{meas}] &\equiv \frac{1}{\mathcal{L}}\frac{\Delta\Ncorr}{\Delta V\kinsetjet},\label{eq:diffcsJetKin}\\
\sigtridiffPartKin[\text{meas}] &\equiv \frac{1}{\mathcal{L}}\frac{\Delta\Ncorr}{\Delta V\kinsetpart},\label{eq:diffcsPartKin}
\end{align}
where $\Delta \Ncorr$ represents the corrected number of \gamAjets events
measured within the acceptance volume, $\Delta V$ is the geometric acceptance volume as a function of its three-dimensional
bin defined in Table~\ref{tbl:kinranges}, and $\mathcal{L}$ is the
integrated luminosity. The calculation of $\Delta V$ is described in
greater detail in Appendix \ref{sec:fiducialaccept}. To account for the different \pTjet and \etajet
acceptances and the different sampled luminosities of the jet
triggers, the cross-sections are separately evaluated  in different
intervals of leading and sub-leading jet \pTjet and \etajet and then
summed over the fiducial acceptance of those variables.

\begin{figure}[!tbp]
\centerline{
\includegraphics[width=0.49\textwidth]{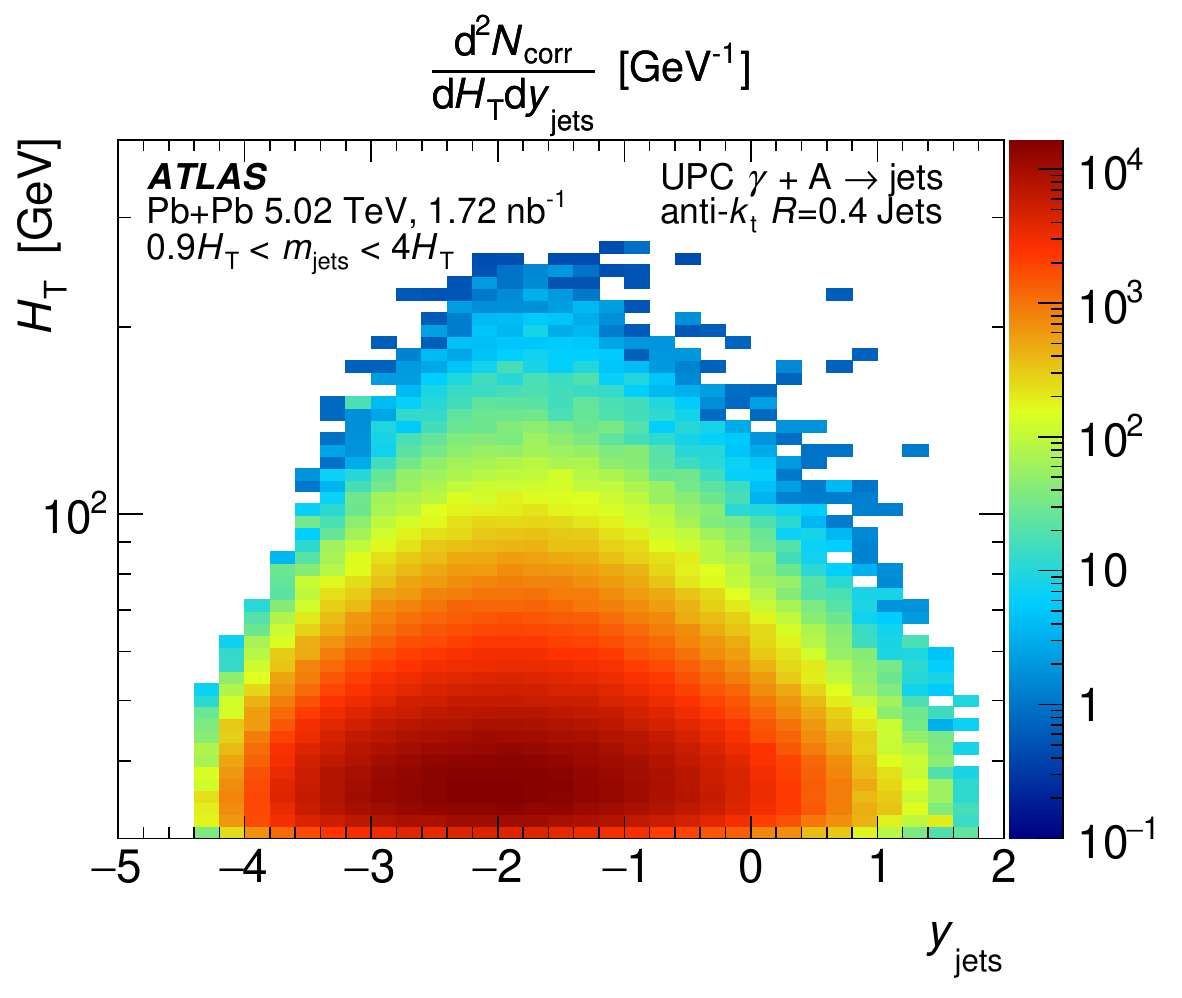}
\includegraphics[width=0.49\textwidth]{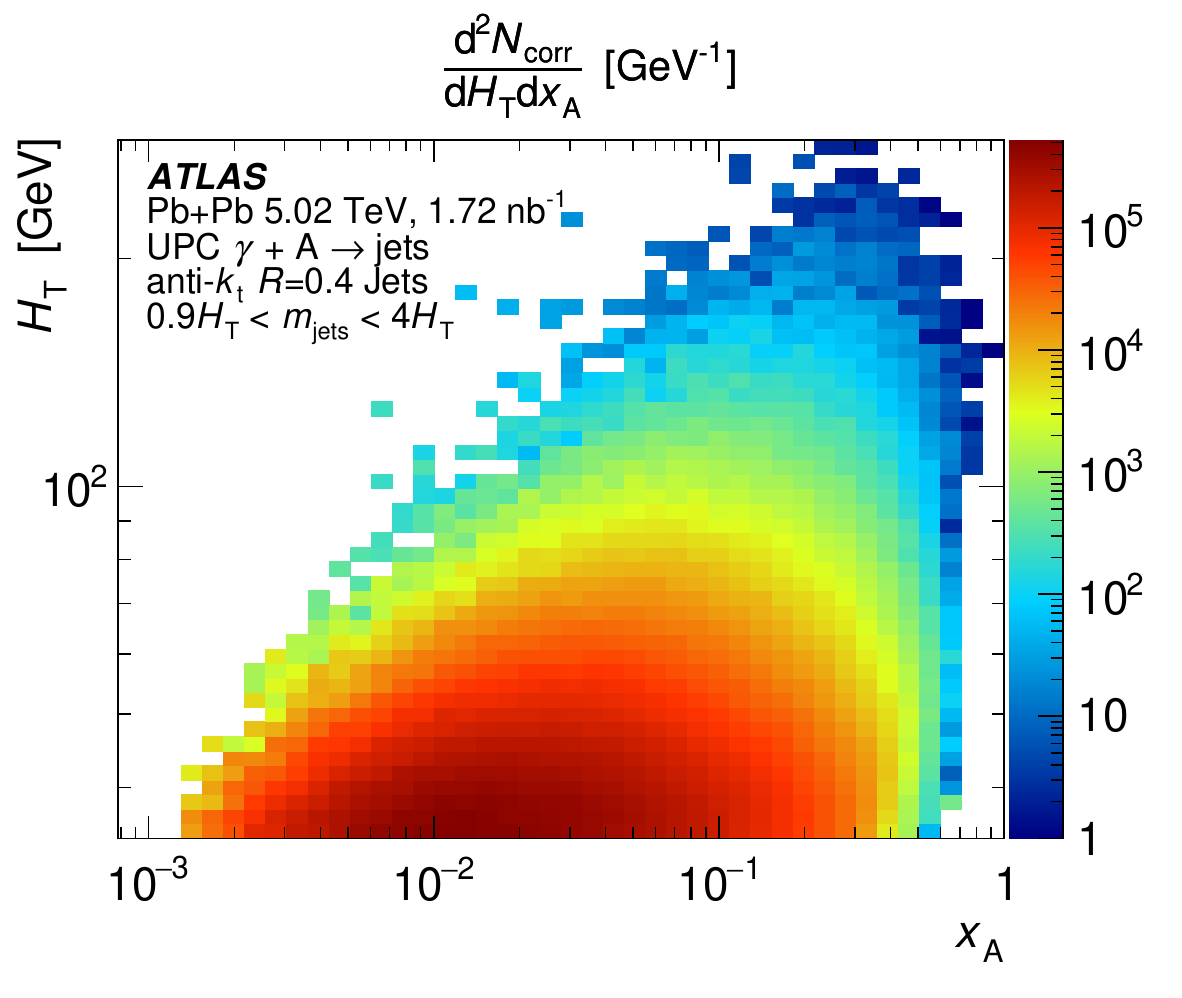}
}
\centerline{
\includegraphics[width=0.49\textwidth]{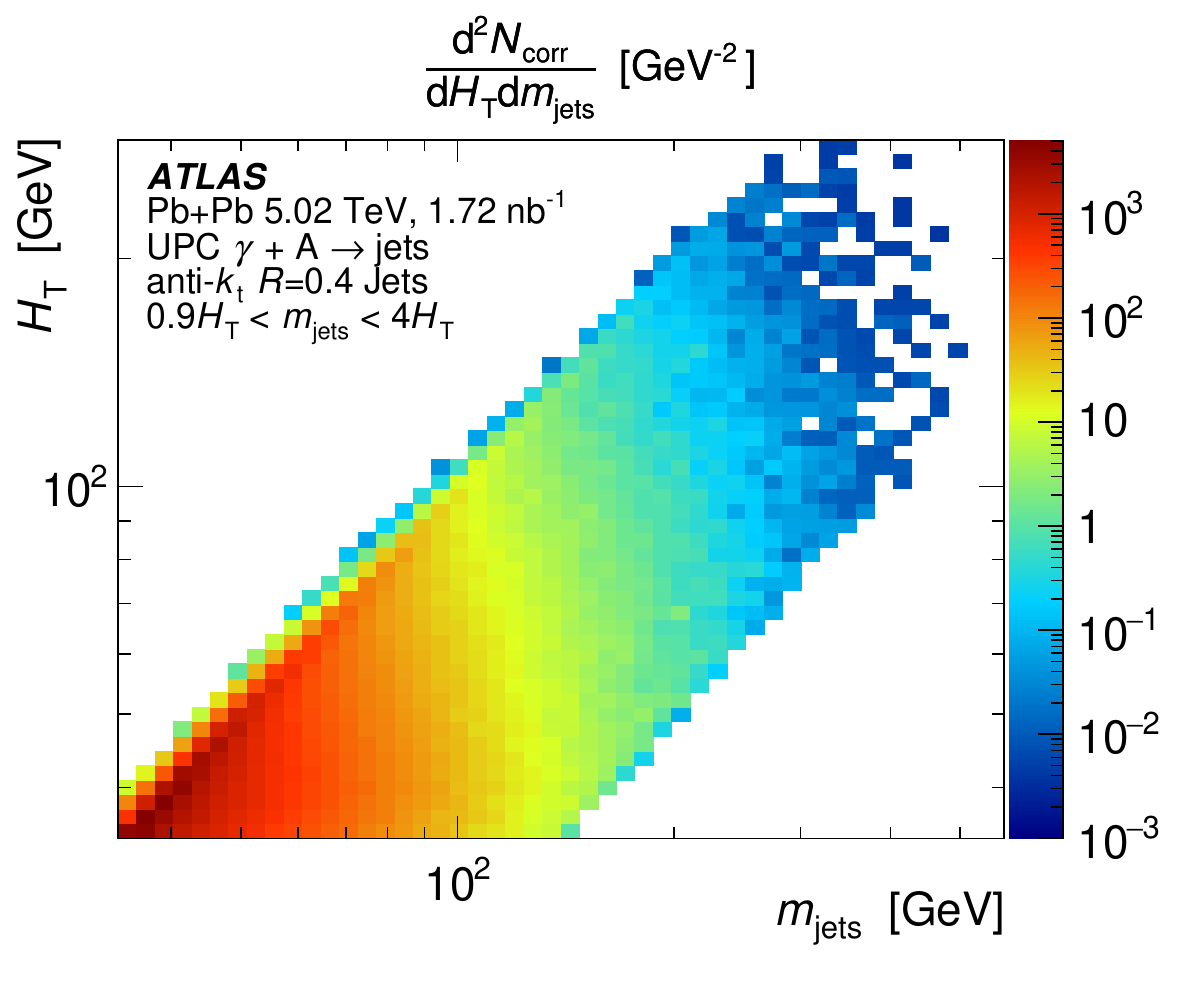}
\includegraphics[width=0.49\textwidth]{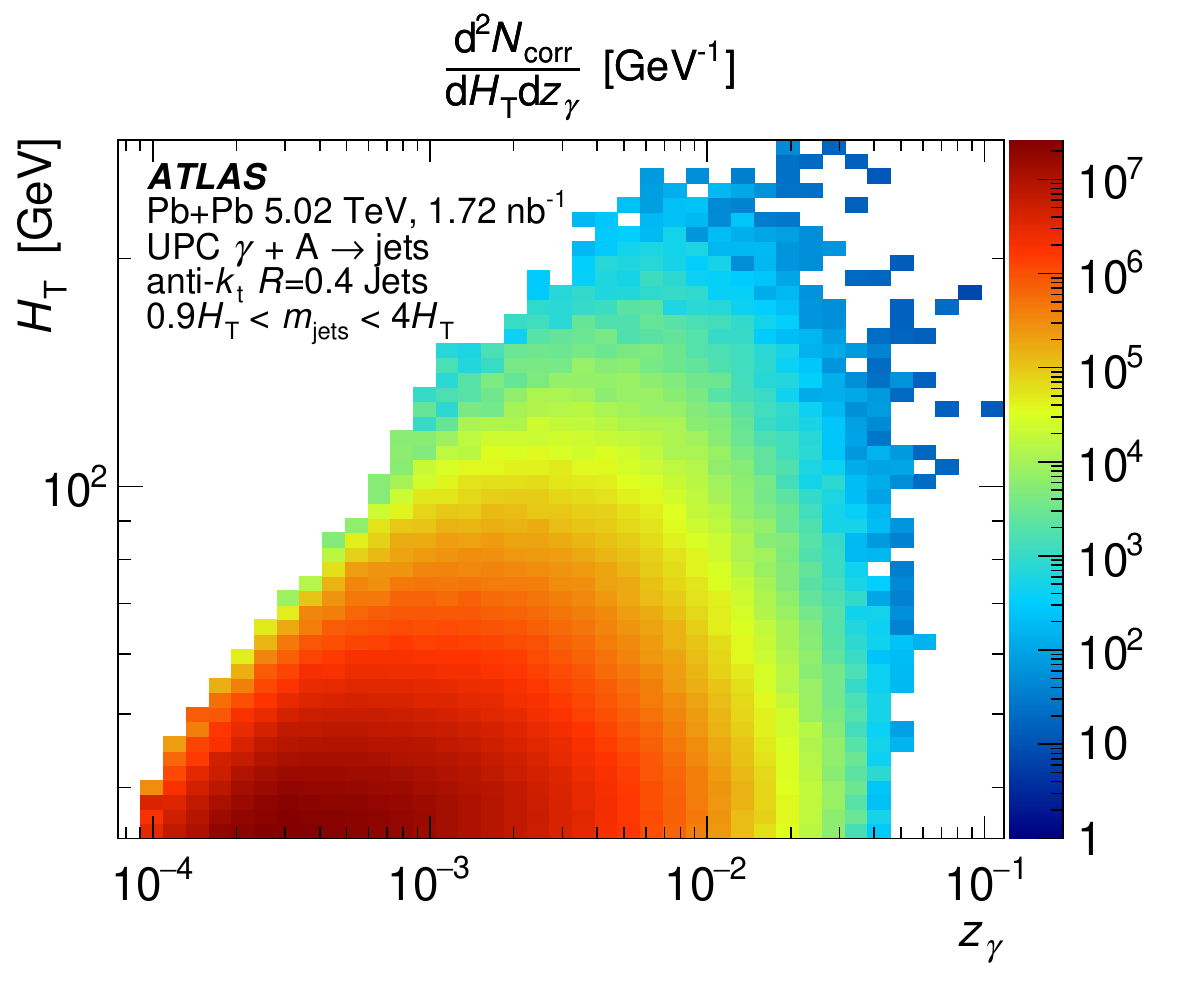}
}
\centerline{
\includegraphics[width=0.49\textwidth]{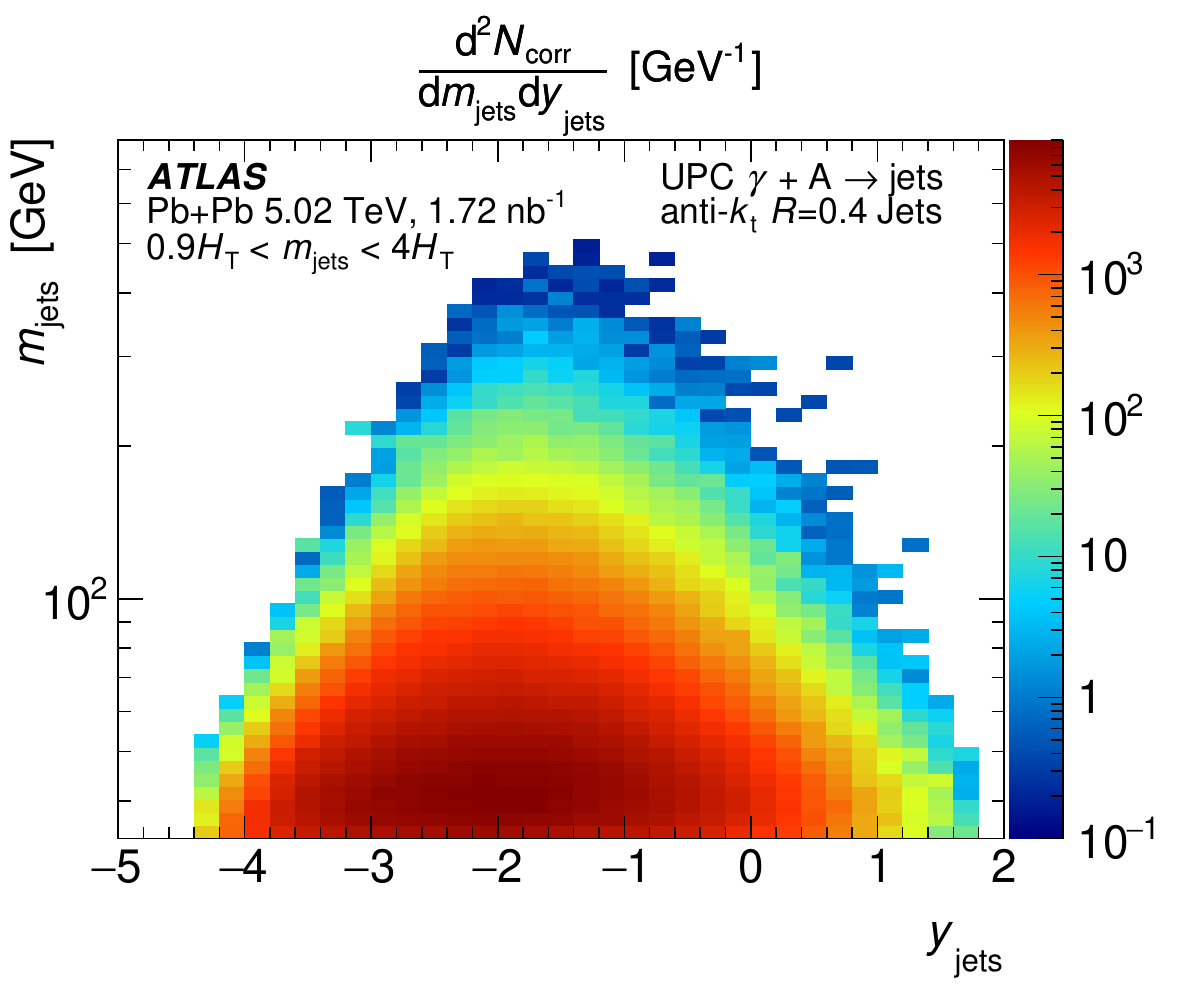}
\includegraphics[width=0.49\textwidth]{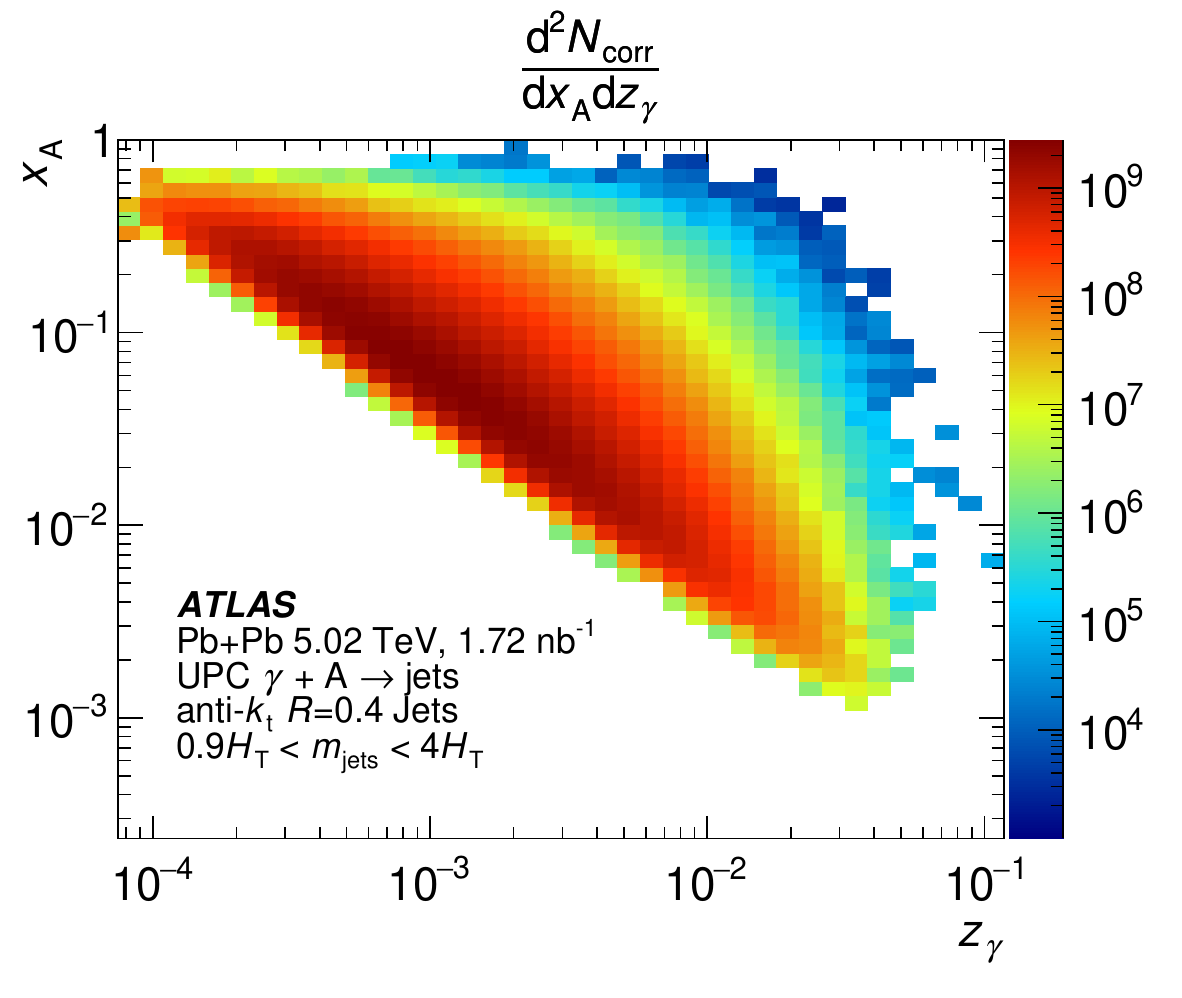}
}
\caption{
Two-dimensional measured differential distributions of \Ncorr for
\gamAjets production in \PbPb\ UPC interactions obtained by
integrating the three-dimensional cross-sections obtained from
Eqs.~(\ref{eq:diffcsJetKin}) and (\ref{eq:diffcsPartKin}) over the
acceptance of the third variable. The left column shows results using
\kinsetjet and the right column results using \kinsetpart. These results
are not unfolded for detector response.
}
\label{fig:CrossSecMeas2D}
\end{figure}

To visualize the kinematic coverage of the measurement,
two-dimensional (2-D) cross-sections are obtained from
Eqs. (\ref{eq:diffcsJetKin}) and (\ref{eq:diffcsPartKin}) by integrating
the third variable over the fiducial acceptance of the
measurement. The 2-D cross-sections are evaluated using finer
binning than that given in Table~\ref{tbl:kinranges}, in order to
better illustrate the kinematic coverage of the measurement. The
results are presented in Figure~\ref{fig:CrossSecMeas2D} for both sets
of kinematic variables, scaled by the integrated luminosity to provide
the corrected number of events. The left column shows the three
possible sets of 2-D cross-sections using \kinsetjet and the right
column shows those using \kinsetpart.

\subsection{Unfolding procedure}
The effects of detector response on the measured cross-sections are
corrected using an iterative Bayesian unfolding
procedure~\cite{DAgostini:1994fjx,Lucy:1974,Richardson:1972}, as implemented using
the \verb=RooUnfold= software package~\cite{Adye:2011gm}. The procedure accounts for both
bin migration within the fiducial region and migration into and
out of the fiducial region via a response matrix populated from the MC sample, which maps
between the reconstructed and truth-level kinematic variables. This procedure works by unfolding the
measured data using an assumed prior distribution as the true
distribution. The output of the unfolding procedure is a set of posterior weights,
which are used to construct an updated estimate
for the true distribution.  The unfolding is performed in three dimensions to properly account for bin
migration effects that are correlated across one or more
dimensions. These effects arise since the detector response in one unfolding variable is correlated with the
value of other variables. The \kinsetjet\  and \kinsetpart\ distributions are unfolded separately.
The performance of the unfolding procedure
was validated through studies of its efficiency, purity, and closure in simulated events.

The number of iterations used in the iterative Bayesian unfolding procedure must be specified.
As the number of iterations is increased,
the sensitivity of the unfolded results to the prior is reduced
but the statistical uncertainties of the results typically increase.
The residual sensitivity is evaluated by
unfolding with an alternative prior, and is used to evaluate the systematic
uncertainty discussed in Section~\ref{sec:systematics}.
The number of iterations is chosen by first considering
the general stability of the unfolded distributions under additional
iterations and then by quantitatively comparing the systematic uncertainty from the residual
sensitivity to the prior with the statistical uncertainty.
The unfolded results for both sets of kinematic variables are based on five iterations.

Statistical uncertainties on the unfolded results are evaluated in the data
by studying the impact of 1000 possible stochastic
variations generated from the input distributions. The variations are
each unfolded in the same manner as the data, and the statistical
covariance is computed for the unfolded cross-section for each pair of
bins. Statistical uncertainties due to possible statistical
fluctuations in the response matrix are obtained in an analogous
fashion, by generating alternative samplings of the response distributions.
\FloatBarrier

% End of text imported from the .//tex/analysis.tex input file

%

%
\section{Systematic uncertainties}
\label{sec:systematics}
\label{sec:syst}

% The next lines are included from the .//tex/systematics.tex input file
Point-to-point systematic uncertainties are assigned to the data from four different sources: event selection efficiency, JES uncertainty, JER uncertainty, and residual sensitivity to the unfolding prior. The event selection efficiency accounts for improper modelling of the impact of the gap selection when event selection efficiencies are evaluated in \pythia. The JES systematic uncertainties account for the impact of imperfect knowledge of the \textit{in-situ} JES calibration on the measured cross-sections. The JER uncertainties similarly describe the impact of imperfect knowledge of differences in the JER between data and MC simulation. The systematic uncertainty for residual sensitivity to the unfolding prior accounts for the residual dependence on the prior used in constructing the response matrices due to the finite number of iterations used in the Bayesian unfolding procedure. Additional uncertainties due to the luminosity measurement and correction for EM pile-up, which are fully correlated across all bins, are propagated to the result. Uncertainties due to the trigger efficiency corrections were found to be negligible, due to the already small magnitude of these corrections within the fiducial acceptance of the measurement.\par

The event selection uncertainties are assessed for both gap requirements ($\edgeGapA < 3$ and $\sumGapGam > 2.5$), following the same procedure in each case, separately for each selection. Each gap requirement is individually tightened by $0.5$ units and the cross-sections are measured with this new event selection. The event selection efficiency estimated in the simulation is also modified for this new selection, and the corrected cross-sections are unfolded using the nominal procedure. The resulting cross-sections are then compared to the nominal result to extract the uncertainty. By varying the efficiency correction as well, this method assigns uncertainties for the modified event selection efficiency incorrectly accounting for the effect of varying the selection in data, which provides a metric for mis-modeling of the efficiency. Uncertainties due to residual backgrounds are neglected, since the rates of these backgrounds are found to be negligible in the studies shown in Section \ref{sec:BkgdEst}.\par

The JES and JER systematic uncertainties are propagated from the uncertainties obtained using the procedure described in Appendix \ref{sec:JetResp}. For each source of uncertainty on the JES, a nuisance parameter variation is introduced, where the reconstructed \pTjet\ of each jet in simulation is multiplied by a scale factor as a function of its \etajet and \pTjet.  These nuisance parameter variations are used to construct modified response matrices, where the truth-level jet information is the same, but the varied jets are used to determine the reconstructed kinematics. The cross-sections are unfolded with the modified response matrices and compared to the nominal result in order to determine the JES uncertainty for that nuisance parameter.\par

In the case of the JER, the procedure is similar, except that the jet transverse momenta are smeared as a function of \etajet and \pTjet according to each nuisance parameter. Additionally, in cases where the \textit{in-situ} measurement of the resolution in data is smaller than in the MC sample, it is necessary to smear data distributions instead of the MC response matrices. A pseudo-data smearing procedure, where cross-section distributions are populated from reconstructed MC events and smeared, rather than directly smearing the data, is applied in order to reduce the impact of statistical fluctuations on the uncerainty. These smeared distributions are unfolded with similarly varied response matrices, and the results are then compared to the nominal unfolded MC distributions, in order to translate fractional uncertainties bin-by-bin to the nominal cross-sections.\par

The systematic uncertainty for residual sensitivity to the unfolding prior is determined by re-weighting the prior distribution in the response matrices in order to match the reconstructed data distribution. The ratios of the reconstructed cross-sections between data and MC simulation are evaluated as a function of the three variables used in the unfolding, and the results are smoothed using a Gaussian kernel. The ratios are then used to re-weight the response matrix as a function of the truth-level jet kinematics. The cross-sections are unfolded with the same number of iterations using these modified response matrices, and the results of this procedure are compared to the nominal results in order to derive the systematic uncertainty. %
\par

\begin{figure}[t]
\centerline{
\includegraphics[width=0.99\textwidth]{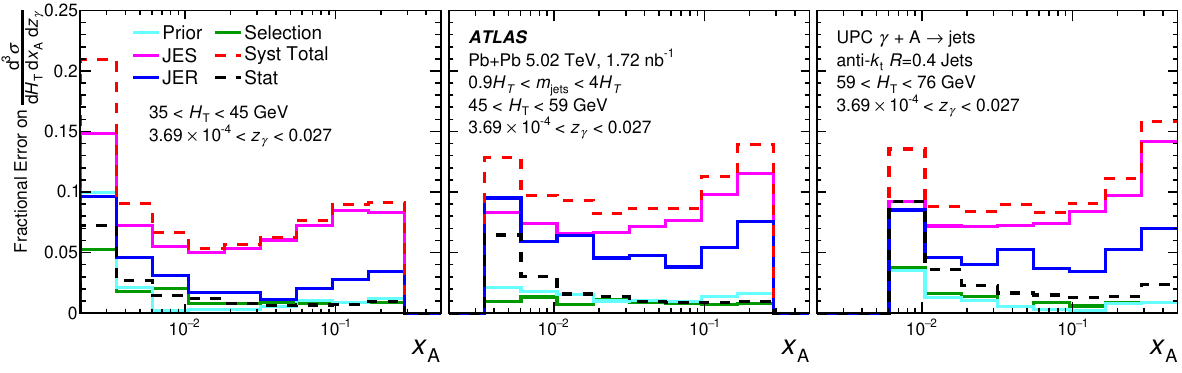}
}
\caption{ A breakdown of the different systematic uncertainty contributions as a function of \xA, for a selection of bins in \HT, summed over \zg. }
\label{fig:SystSummary}
\end{figure}

The uncertainty on the integrated luminosity of the data sample is
$2.0\%$. It is derived from the calibration of the luminosity
scale using $x$-$y$ beam-separation scans, following a methodology
similar to that detailed in Ref.~\cite{DAPR-2021-01}, using the
LUCID-2 detector for the baseline luminosity measurements
\cite{LUCID2}. An additional uncertainty is applied to the measured total cross-section
due to uncertainty on the rate of EM pile-up, which is not correlated
with event kinematics. This uncertainty is obtained by varying the \PbPb\ dissociative
cross-section within its uncertainties, yielding a fractional uncertainty of $0.3\%$.\par

A summary of the different sources of systematic uncertainty for a selection of bins in \HT and summed over the full range in \zg are shown as function of \xA in Figure \ref{fig:SystSummary}, demonstrating the contributions from each source of systematic uncertainty, compared to the statistical uncertainty. The statistical uncertainty due to the unfolding response matrix is included in the total statistical uncertainty, but is negligible. The \textit{in-situ} measurement of the JES typically provides the largest source of systematic uncertainty, but in some regions near the edges of the phase-space, the JER uncertainties grow larger. The event selection uncertainties are never dominant, and they are only substantial in the kinematic regions where the jets are near the edge of the photon-going gap (large \zg, large positive \yjets). The sensitivity to the unfolding prior is smaller than the statistical uncertainty in all bins except for at the very edge of the phase-space, where the unfolding becomes more difficult to control at the percent level. Additionally, correlations between the bins of the systematic uncertainties are assessed, where each JES or JER nuisance parameter is taken as $100\%$ correlated between bins but uncorrelated with other nuisance parameters. The systematic uncertainty due to the event selection efficiency and sensitivity to the unfolding prior may have some unknown bin-to-bin correlation, so they are conservatively taken as completely uncorrelated bin-to-bin.\par

\FloatBarrier

% End of text imported from the .//tex/systematics.tex input file

%

%
\section{Nuclear breakup}
\label{sec:breakup}

% The next lines are included from the .//tex/breakup.tex input file
\label{sec:ResultsBU}

In order to provide a useful evaluation of the measured
cross-sections, the data are compared to particle-level
triple-differential cross-sections from \pythia. However, because the
measurement nominally uses the \znxn\ condition to identify \gamA
events, the \pythia cross-sections must first be corrected for the effects of
breakup of the photon-emitting nucleus, which can depend on the \gamA kinematics,
especially \zg. This section describes how the methods described in
Sections \ref{sec:analysis} and \ref{sec:syst}
are applied to the \xnxn\ sample in order to obtain an experimental
measurement of the photon-going breakup likelihood. This result is then applied to
the \pythia cross-sections in order to allow a comparison with unfolded data.\par

The fraction of photonuclear jet events without breakup of the photon-emitting nucleus
is measured as a function of \zg using two samples: the nominal
analysis \znxn sample and an additional \xnxn sample. In the latter
sample, the photon-going direction is selected using the side of the
detector with the largest \sumGap. In order to reduce potential background
contamination from hadronic processes, a more strict gap requirement,
$\edgeGapGam > 2.5$, is imposed in the photon-going direction for both samples.
Studies of the gap distributions in \xnxn events indicated that this
selection is necessary to achieve sufficient purity to measure \gamAjets
events in \xnxn collisions. This requirement preferentially removes
resolved-photon events, but the probability that the photon-emitting nucleus breaks up is
not sensitive to the nature of the hard-scattering process, due to
the large separation in energy scales between the hard-scattering and
soft photon exchange \cite{Baltz:2002pp}.
An assessment of the acceptance and \effevt, separate from the one used in
the nominal analysis, is performed for both the \xnxn and the
corresponding \znxn\ sample. Corrected cross-sections from
each sample are then unfolded
in a modified three-dimensional binning, providing finer intervals
in \zg, and the fraction of events which do not break up is computed
from the unfolded cross-sections as
\begin{equation}
f_{\text{no BU}} \equiv
\frac{\left.\text{d}\sigma/\text{d}\zg\right|_{\znxn}}{\left.\text{d}\sigma/\text{d}\zg\right|_{\xnxn}
+ \left.\text{d}\sigma/\text{d}\zg\right|_{\znxn}} .
\end{equation}
\begin{figure}[!htb]
\centerline{
\includegraphics[width=0.85\textwidth]{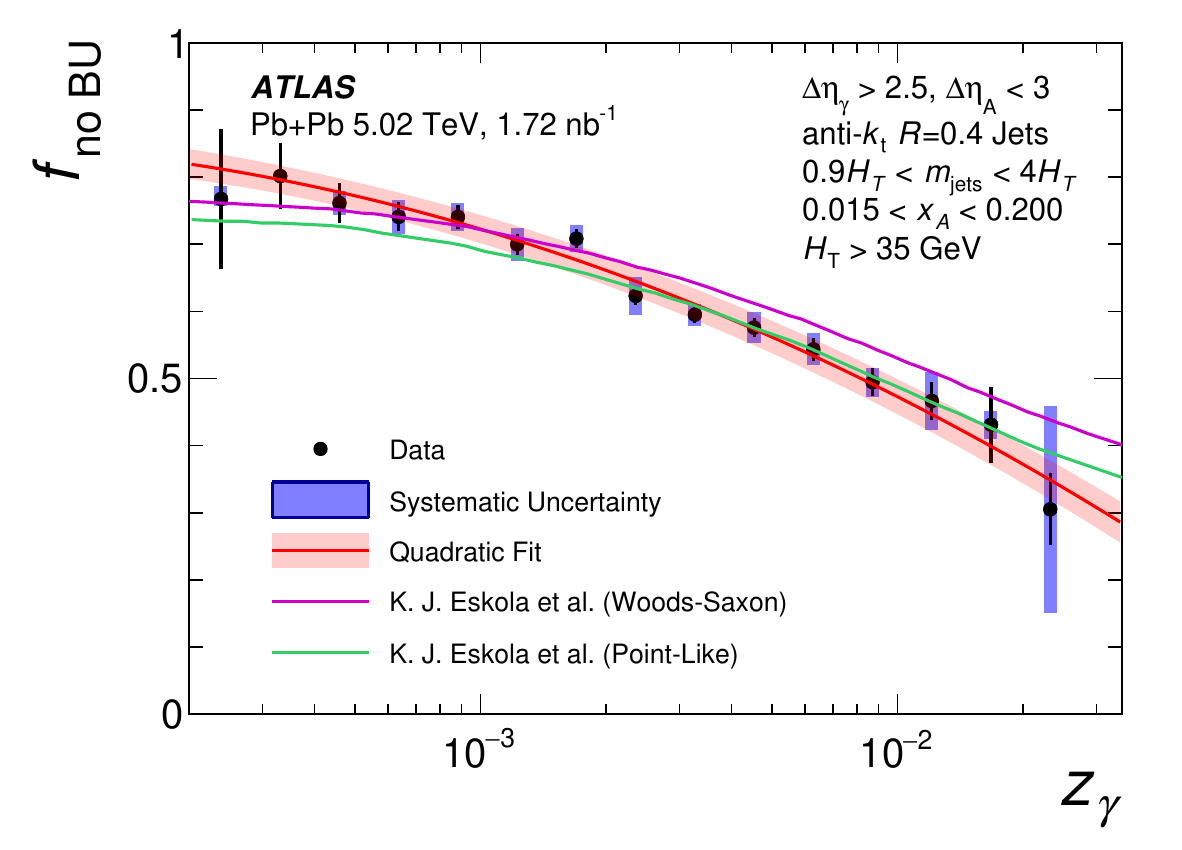}
}
\caption{ The fraction of photonuclear jet events passing the fiducial requirements in which the photon-emitting nucleus does not break up as a function of \zg. The systematic uncertainties are shown as shaded blue bands, and the error bars are statistical uncertainties. A quadratic fit in $\ln(\zg)$ is shown in red. These results indicate a strong dependence of the breakup rate on \zg and an overall high rate of breakup due to additional Coulomb interactions, reaching $70\%$ at high \zg. Results are compared to theoretical calculations from Ref. \cite{Eskola:2024fhf}.}
\label{fig:BreakupFracResults}
\end{figure}

Full systematic uncertainties are then propagated through the ratio by applying systematic variations to each sample; most of the uncertainties largely cancel in the ratio. Figure \ref{fig:BreakupFracResults} shows the measured no-breakup fraction as a function of \zg.
A quadratic fit in $\ln(\zg)$ is used to smooth the variations. The quadratic fit is applied as a correction to theoretical predictions shown in Section \ref{sec:Xsec3D} as a function of \zg. The breakup fraction shows a substantial variation with \zg, which is expected to arise because higher photon energies are more likely at smaller impact parameters, and the probability of additional Coulomb excitations is also correlated with the impact parameter of the collision. Theoretical calculations from Ref.~\cite{Eskola:2024fhf} describe the data by modelling the EM breakup probability as a function of impact parameter in a model of NLO \gamAjets production, and they are compared to the data in Figure \ref{fig:BreakupFracResults}. These calculations describe the measured fraction quite well but no single prediction captures the shape over the full range in \zg. Since \HT and \xA are much more weakly correlated with impact parameter, no additional dependence on these variables is considered.\par

\FloatBarrier

% End of text imported from the .//tex/breakup.tex input file

%

%
\section{Results}
\label{sec:results}

% The next lines are included from the .//tex/results.tex input file
This section provides fully corrected triple-differential cross-sections for photon-nucleus
collisions with no breakup in the photon-going direction, \sigtridiffJetKin and
\sigtridiffPartKin. Namely, the measured cross-sections obtained from
Eqs. (\ref{eq:diffcsJetKin}) and (\ref{eq:diffcsPartKin}) in the kinematic
intervals described in Table~\ref{tbl:kinranges} are unfolded using
the detector response obtained from \pythia events with full detector simulation,
and systematic uncertainties
described in Section~\ref{sec:systematics} are evaluated for each
three-dimensional bin. More information about which bins are reported, the
calculated mean value of each bin, and the phase-space volume used to define
the triple-differential cross-section are provided in Appendix~\ref{sec:fiducialaccept}.

In order to separate uncertainties correlated across bins from those that are
uncorrelated, the uncertainties of the results shown in this section are represented by three components: statistical, scale, and residual systematic uncertainties. The scale
uncertainty is computed by separating out the part of the correlated uncertainty that corresponds
to a single re-scaling of the entire distribution. This component is taken to be the smallest fractional
uncertainty on any single bin in that selection, for each uncertainty that is $100\%$ correlated bin-to-bin.
These scale uncertainties are added in quadrature and represented as a light red band, while the remaining
systematic uncertainty is represented as a gray band and the statistical uncertainty is shown in
yellow. Although these residual systematic uncertainties do have some correlation, the conservative assumption
should be made that they are uncorrelated when interpreting the uncertainties.

\subsection{Sensitivity to nuclear PDF effects}

\begin{figure}[!tb]
\centerline{
\includegraphics[width=0.95\textwidth]{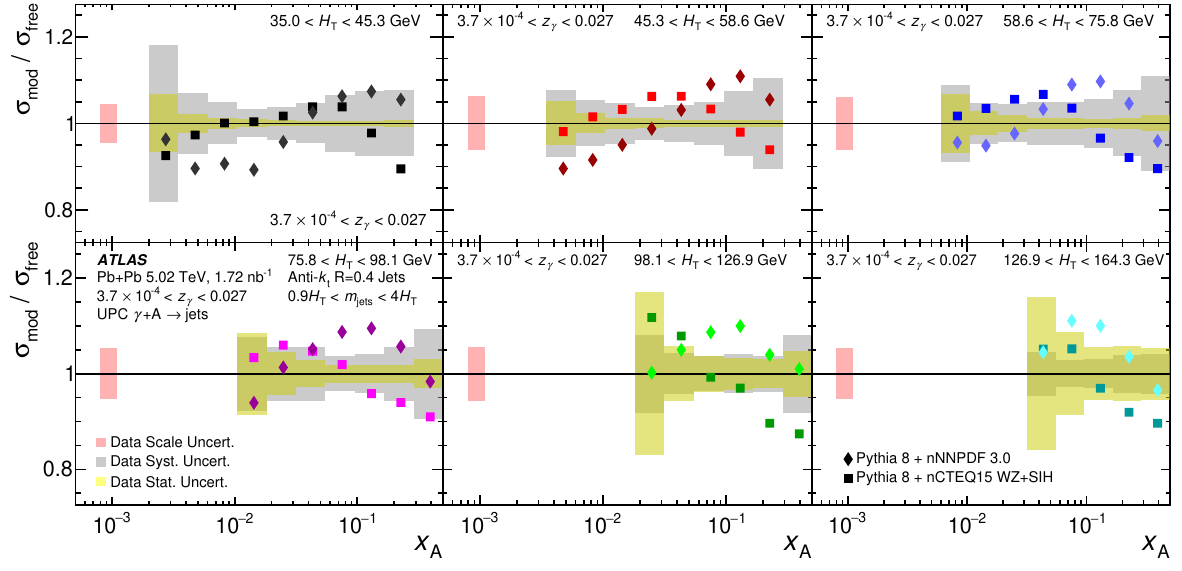}
}
\caption{
The impact of nPDF modifications on the photo-nuclear jet cross-section, represented by the ratio of the cross-section calculated with modified nPDFs ($\sigma_\text{mod}$) to the cross-section for un-modified free nucleon PDFs ($\sigma_\text{free}$). These effects are shown for nCTEQ15 WZ+SIH (squares) and nNNPDF 3.0 (diamonds) fits, where ratios are taken with respect to CT18 and NNPDF 3.1, respectively. Results are integrated over the \zg range $3.7\times10^{-4} < \zg < 0.027$, and each panel shows a separate range in \HT. The total uncertainty on this measurement in each bin is compared to the size of these modifications, where the light red bands show the total scale uncertainty, the grey bands are the quadrature sum of the residual systematic uncertainty, and the yellow bands show the statistical uncertainty.
}
\label{fig:SystCompnPDF}
\end{figure}

The results presented in this section for UPC \gamAjets cross-sections are compared to a
theoretical calculation from LO \pythia which was produced using nCTEQ15 WZ+SIH \cite{Duwentaster:2021ioo} PDFs.
While discrepancies may arise between the data and this theory calculation due to
other sources, such as higher-order QCD effects or uncertainty on the photon flux, the primary goal of this
comparison is to evaluate any inconsistencies that can be attributed to
differences in the measured nPDFs. An understanding of the size of these effects and differences
between the modifications in competing fits is important in assessing
the impact of the data on constraining these effects. Figure \ref{fig:SystCompnPDF}
shows the ratio of photonuclear jet cross-section calculations between a PDF set with
nuclear modifications (nCTEQ15 WZ+SIH, nNNPDF 3.0 \cite{AbdulKhalek:2022fyi}) and without modifications (CT18 \cite{Hou:2019efy}, NNPDF 3.1 \cite{PDF4LHCWorkingGroup:2022cjn}). Both
the numerator and denominator in these ratios are corrected for small isospin effects arising
from differences in the parton content of protons and neutrons. The
total uncertainty on the data measurement in these bins is shown for comparison, demonstrating
the sensitivity of these data both to nPDF effects of this size and the existing differences
in current nPDF models. Studies of the potential impact of this measurement on nPDF uncertainties
\cite{Guzey:2019kik} indicate that the nPDF uncertainties are typically $10\%$ or larger in this kinematic
region, substantially smaller than the point-to-point uncertainty achieved in this measurement.\par

\subsection{Photonuclear jet cross-sections}
\label{sec:Xsec3D}
This section presents the primary results of the analysis described in
this paper, the triple-differential cross-sections for jet
production in \PbPb\ UPCs satisfying
the \znxn\ condition and the fiducial requirements presented in
Section \ref{sec:eventsel} and Appendix~\ref{sec:fiducialaccept}.
The measured cross-sections are compared to LO \pythia cross-sections
with nCTEQ15 WZ+SIH nPDFs, obtained from the sample described in
Section~\ref{sec:samples}, multiplied by $f_\text{no BU}$, the \zg-dependent
probability of no breakup determined empirically from data, as described
in Section \ref{sec:ResultsBU}. The \pythia results are represented by the
dashed lines. The ratios between the data and the \pythia
calculations for selected intervals in \HT\ or \xA\ are shown in the
panels below each figure. As these results present the first measurement of photonuclear jet cross-sections, no
attempt is made to incorporate theoretical uncertainties corresponding to effects such as
renormalization scale, nPDFs, or the photon flux, since these uncertainties may be
improved in subsequent theoretical comparisons.\par

\begin{figure}[!tb]
\centerline{
\includegraphics[width=0.49\textwidth]{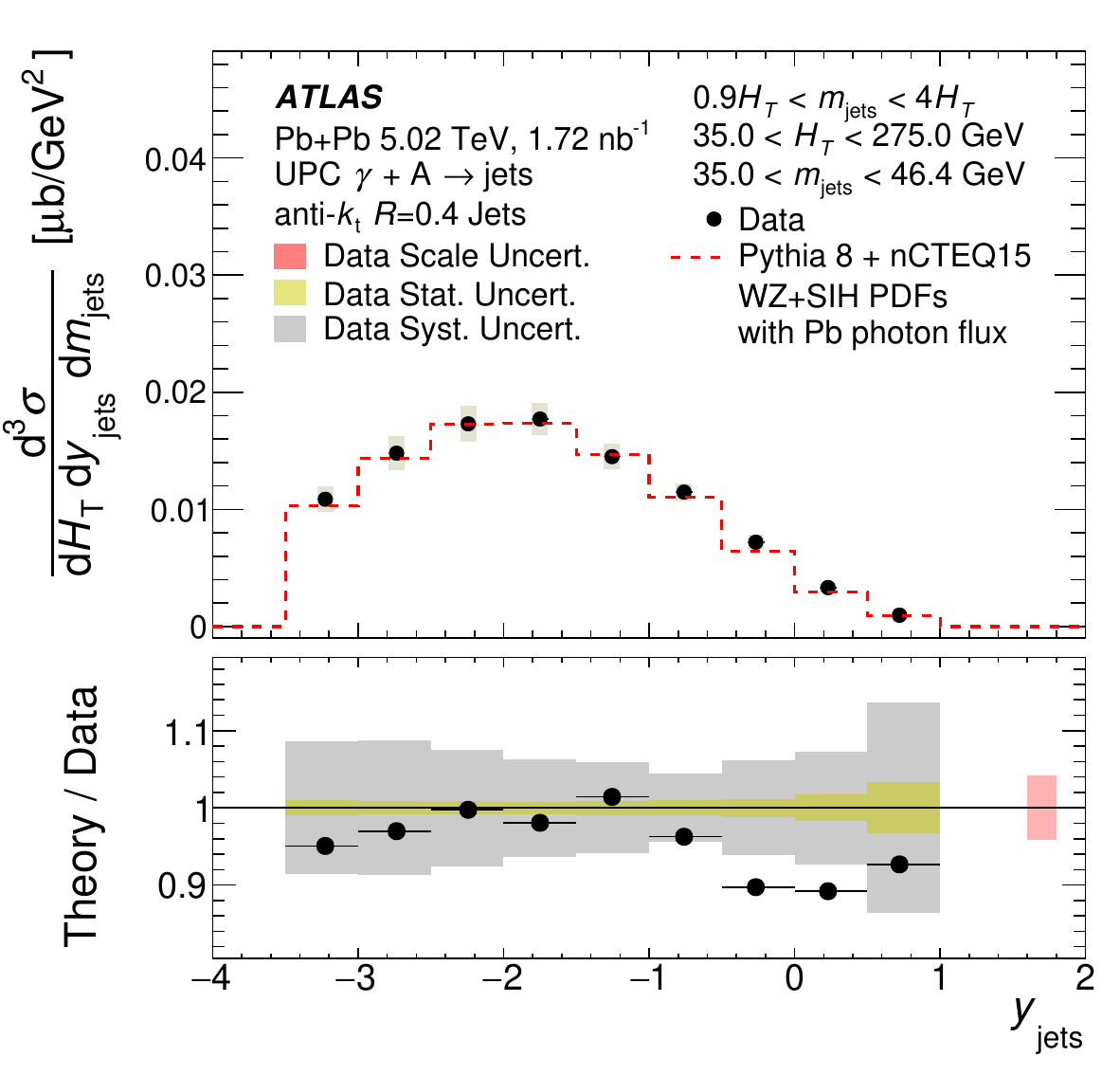}
\includegraphics[width=0.49\textwidth]{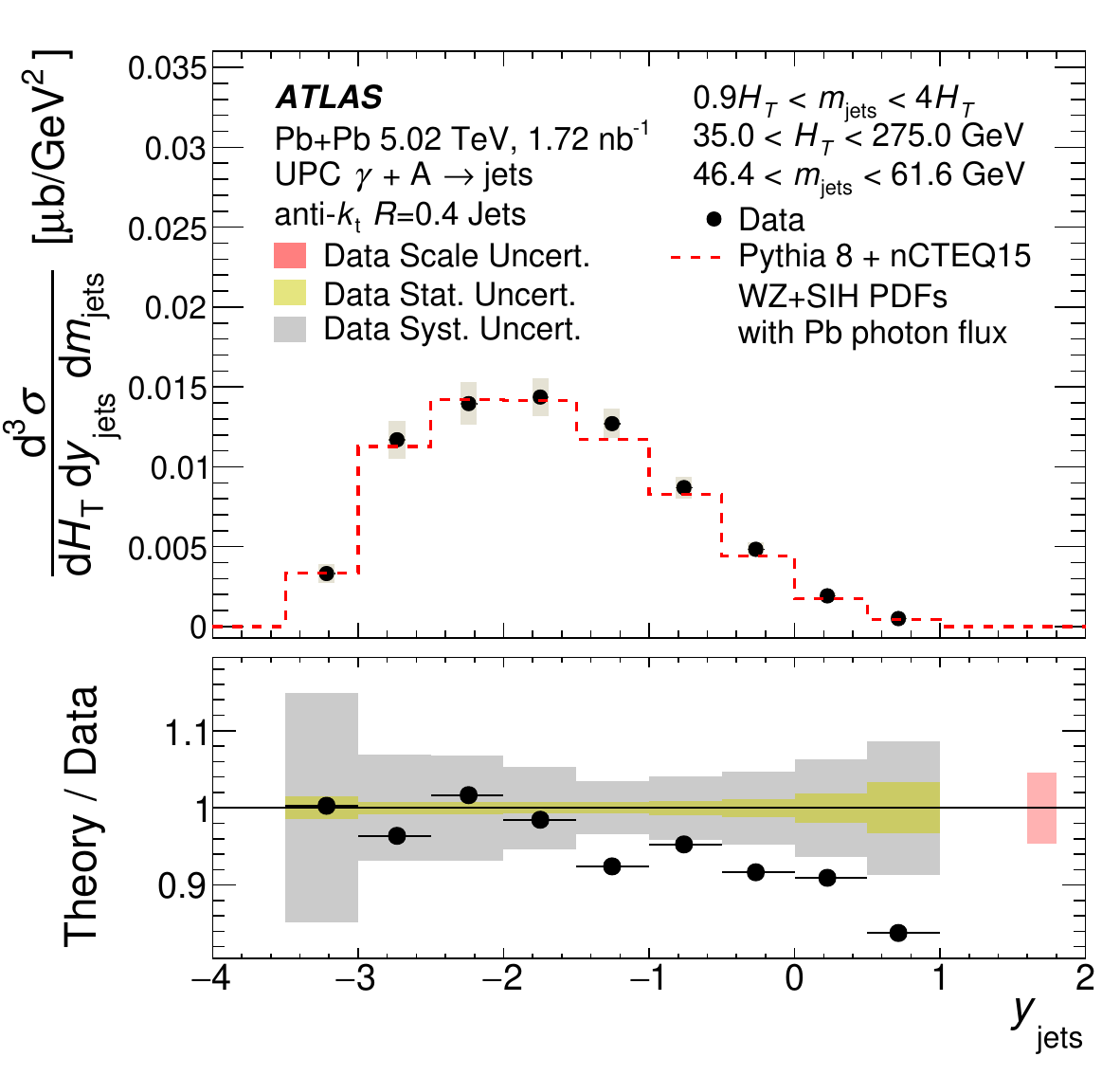}
}
\centerline{
\includegraphics[width=0.49\textwidth]{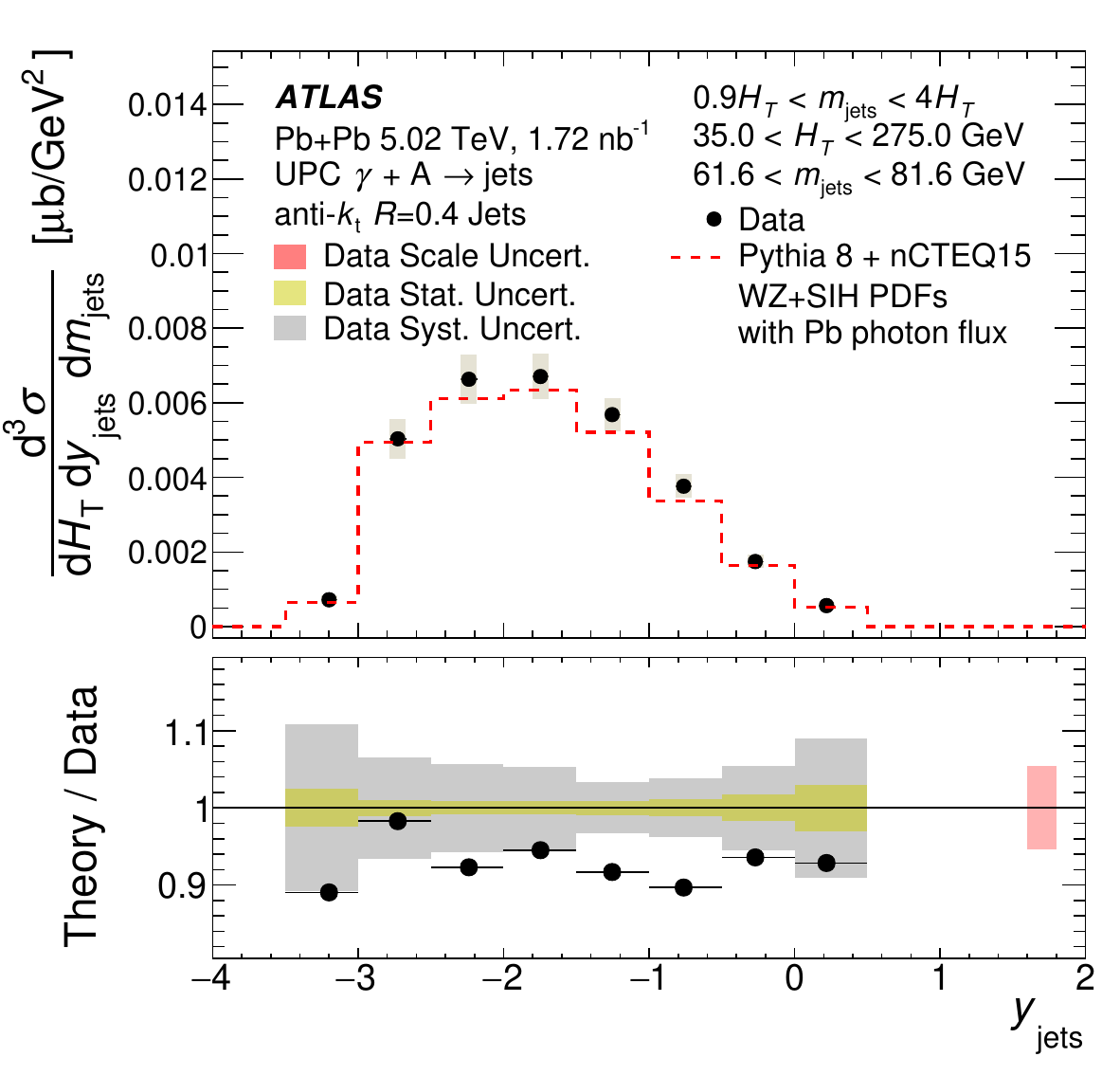}
\includegraphics[width=0.49\textwidth]{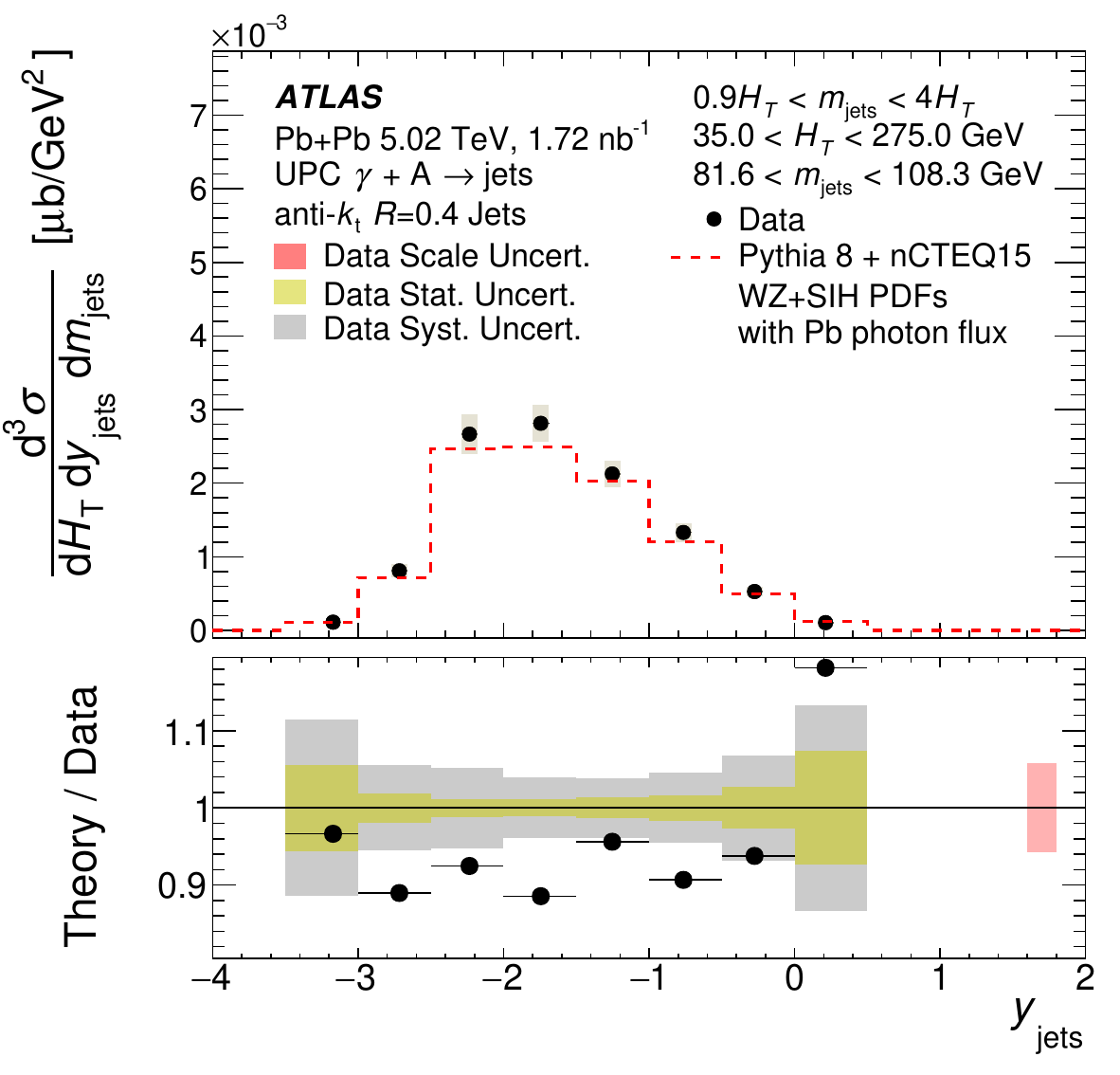}
}
\caption{
Triple-differential cross-sections, \sigtridiffJetKin, as a
function of \yjets, in four \mjets intervals with \mjets  increasing
from the top left to the bottom right. The cross-sections are shown
for the selected intervals in \mjets and integrated over
the \HT acceptance. For each plot, systematic uncertainties are shown in
the upper panel as shaded boxes, while statistical uncertainties shown as vertical lines
are usually smaller than the size of the markers. A theoretical
comparison is shown to cross-sections computed using \pythia with
nCTEQ15 WZ+SIH PDFs, a photon flux from \textsc{Starlight}, and a \zg-dependent data-driven
breakup fraction. The bottom
panels show the ratio between the theory prediction and the data.
The light red bands in the ratio panels are the quadrature sum of scale uncertainties on the cross-section, while the gray band shows the residual systematic uncertainty. The yellow band  shows the point-to-point statistical uncertainty.
}
\label{fig:ResultsYjetsComp1}
\end{figure}

\begin{figure}[!tbp]
\centerline{
\includegraphics[width=0.85\textwidth]{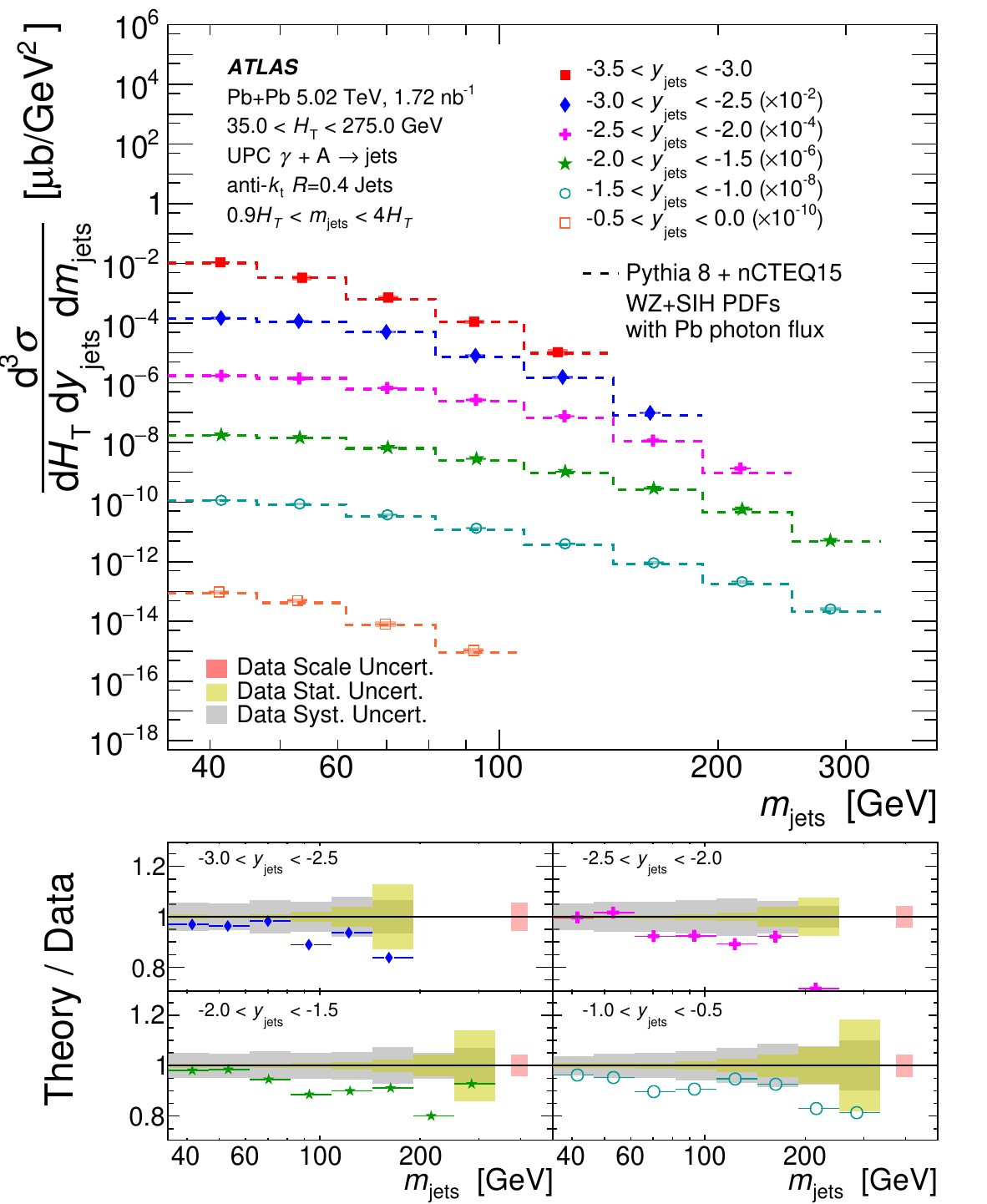}
}
\caption{
Triple-differential cross-sections, \sigtridiffJetKin, as a
function of \mjets, in several \yjets intervals. The cross-sections are integrated over
the \HT acceptance. Systematic uncertainties are shown in
the upper panel as shaded boxes, while statistical uncertainties shown as vertical lines
are usually smaller than the size of the markers. A theoretical
comparison is shown to cross-sections computed using \pythia with
nCTEQ15 WZ+SIH PDFs, a photon flux from \textsc{Starlight}, and a \zg-dependent data-driven
breakup fraction. The bottom
panels show the ratio between the theory prediction and the data for a selection of the \yjets intervals.
The light red bands in the ratio panels are the quadrature sum of scale uncertainties on the cross-section, while the gray band shows the residual systematic uncertainty. The yellow band  shows the point-to-point statistical uncertainty.
}
\label{fig:ResultsMjetsComp}
\end{figure}

The presentation of results begins with cross-sections evaluated
using \kinsetjet which are closely related to the measured jet kinematics.
Figure~\ref{fig:ResultsYjetsComp1} shows cross-sections as a function
of \yjets in different intervals of \mjets that have been integrated
over the \HT acceptance in each \yjets and \mjets interval. The
comparison with \pythia + nCTEQ15 WZ+SIH shows a systematic difference as a function
of \yjets. For small \mjets, the cross-section in the data is larger than that in \pythia
at forward rapidities and matches the results from \pythia at backward
rapidities, while at larger \mjets, the ratio is largely independent of \yjets
but the theory under-predicts the total cross-section. The difference between data and MC varies with \HT,
becoming smaller near the maximum in the \yjets distribution with
increasing \HT. These effects are
also visible in Figure~\ref{fig:ResultsMjetsComp}, where the theory under-predicts
the data at larger \mjets.

\begin{figure}[!tbp]
\centerline{
\includegraphics[width=0.85\textwidth]{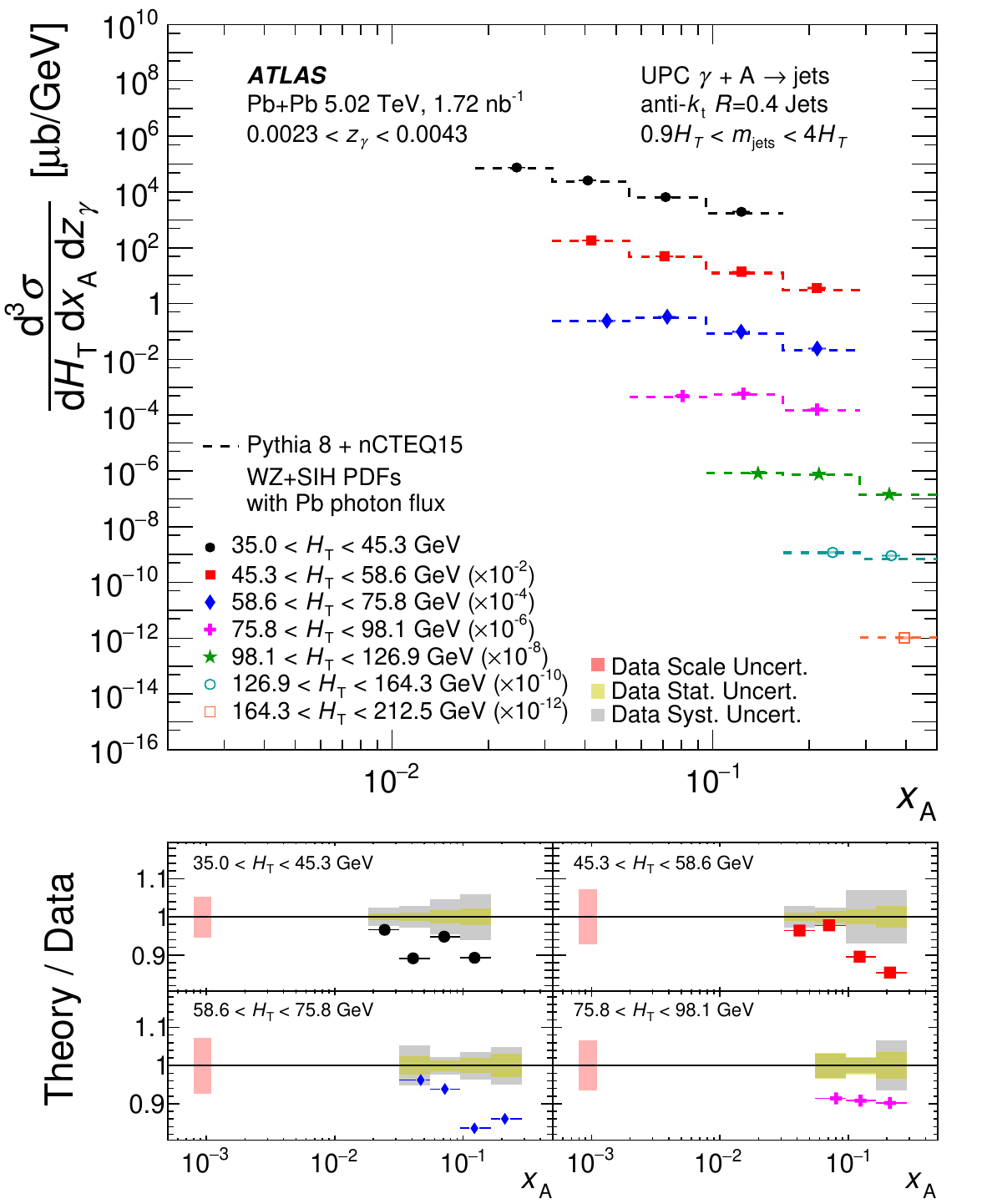}
}
\caption{ Triple-differential cross-sections, \sigtridiff, as a
function of $\xA$ for different bins of $\HT$ for events with emitted
photon energies in the kinematic range $0.0023 < \zg <
0.0043$.  In the upper panel, systematic uncertainties are shown as
shaded boxes, while statistical uncertainties shown as vertical lines
are usually smaller than the size of the markers. A theoretical
comparison is shown to cross-sections computed using \pythia with
nCTEQ15 WZ+SIH PDFs, a photon flux from \textsc{Starlight}, and a \zg-dependent
breakup fraction. The bottom
panels show the ratio between the theory prediction and data for a representative subset of the
bins of \HT. The lorentz-invariant variables \xA, \zg, and \HT are un-ambiguously defined in terms of the particle-level jet kinematics and correspond at leading order to the Bjorken-$x$, emitted photon energy fraction,
and collision $Q^2$, respectively.
The light red bands in the ratio panels are the quadrature sum of scale uncertainties on the cross-section, while the gray band shows the remaining systematic uncertainty. The yellow band  shows the point-to-point statistical uncertainty.  }
\label{fig:Results_xA_TheoryComp_SlicesHT_zGamma7}
\end{figure}

\begin{figure}[!tbp]
\centerline{
\includegraphics[width=0.85\textwidth]{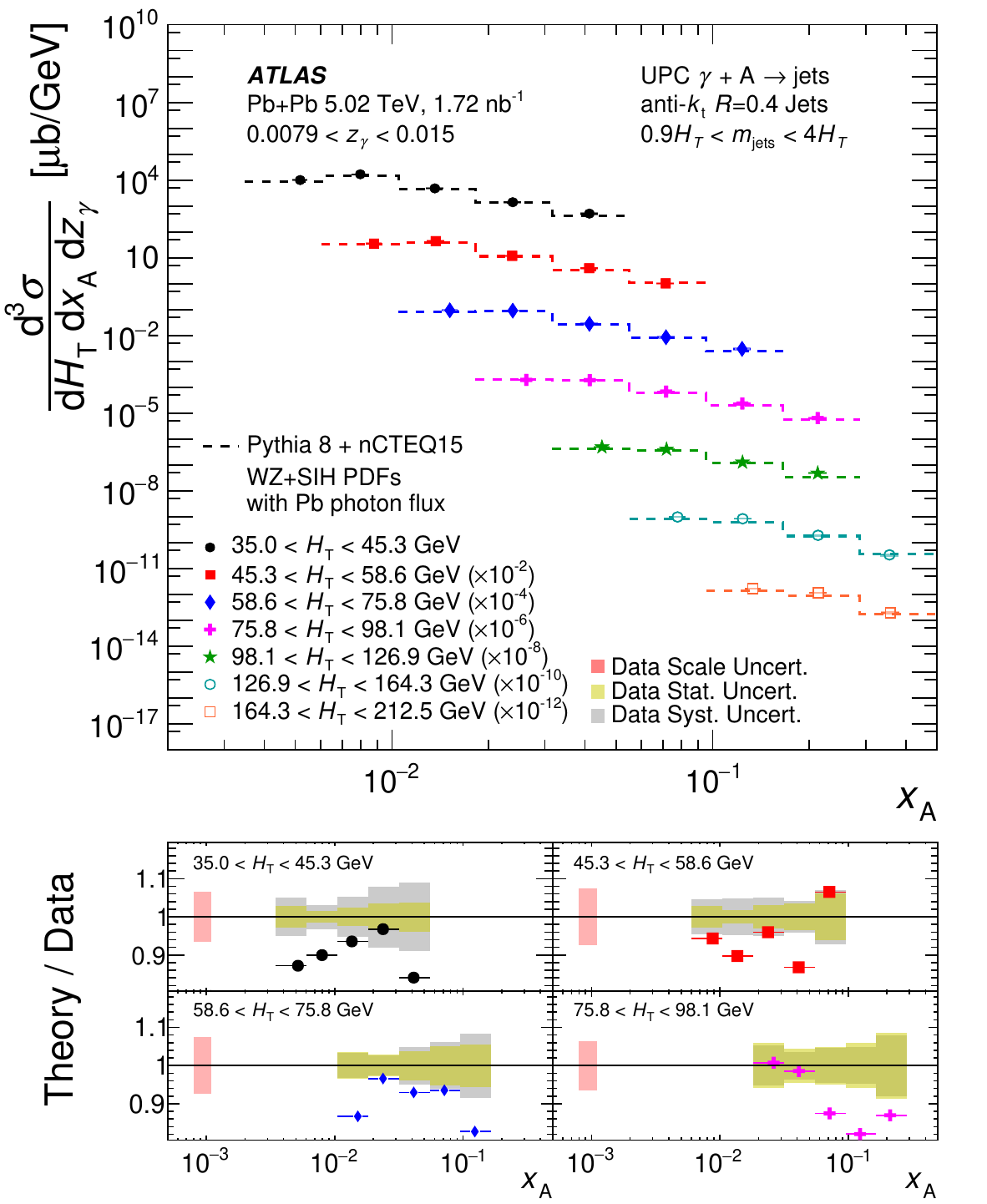}
}
\caption{ Triple-differential cross-sections, \sigtridiff, as a
function of $\xA$ for different bins of $\HT$ for events with emitted
photon energies in the kinematic range $0.0079 < \zg <
0.015$.  In the upper panel, systematic uncertainties are shown as
shaded boxes, while statistical uncertainties shown as vertical lines
are usually smaller than the size of the markers. A theoretical
comparison is shown to cross-sections computed using \pythia with
nCTEQ15 WZ+SIH PDFs, a photon flux from \textsc{Starlight}, and a \zg-dependent
breakup fraction. The bottom
panels show the ratio between the theory prediction and data for a representative subset of the
bins of \HT. The lorentz-invariant variables \xA, \zg, and \HT are un-ambiguously defined in terms of the particle-level jet kinematics and correspond at leading order to the Bjorken-$x$, emitted photon energy fraction,
and collision $Q^2$, respectively.
The light red bands in the ratio panels are the quadrature sum of scale uncertainties on the cross-section, while the gray band shows the remaining systematic uncertainty. The yellow band  shows the point-to-point statistical uncertainty.  }
\label{fig:Results_xA_TheoryComp_SlicesHT_zGamma9}
\end{figure}

In the following figures, the measured cross-sections are presented in terms
of \kinsetpart that more closely reflect the kinematics of the partons
participating in the hard-scattering process. The results are presented as a function of \xA
for different \HT intervals in four \zg\ ranges: $0.0023 < \zg
< 0.0043$, $0.0079 < \zg < 0.015$, $0.0043 < \zg
< 0.0079$, and $0.015 < \zg < 0.027$ in
Figures~\ref{fig:Results_xA_TheoryComp_SlicesHT_zGamma7},
\ref{fig:Results_xA_TheoryComp_SlicesHT_zGamma8},
\ref{fig:Results_xA_TheoryComp_SlicesHT_zGamma9}, and
\ref{fig:Results_xA_TheoryComp_SlicesHT_zGamma10}, respectively, where
the latter two figures are shown in Appendix \ref{sec:morePlots}. Figure
\ref{fig:Results_xA_TheoryComp_SlicesHT_zGammaAll} demonstrates the impact of
integrating over the full range in \zg.
The cross-sections in the different \HT intervals are shown scaled by
different powers of ten to allow them to be presented in the same figure.
Due to the correlation of the acceptance in \xA with \zg, measuring individual
\zg\ intervals illustrates the role of the photon energy in probing different features of
the nPDFs. These individual intervals also allow for a more
robust separation of correlated systematic uncertainties.

The breakup-adjusted cross-sections provide substantial information about differences
between the data and \pythia, where the uncertainties on the measurement are substantially smaller
than the existing nPDF uncertainties. Broadly, the data suggest that the nPDFs
in nCTEQ are too small at lower \HT and lower \xA, but that the agreement improves at higher \xA. From
the lower \zg intervals, it can be observed that the cross-section in the
anti-shadowing region from \pythia+ nCTEQ agrees with the data, whereas at high \zg,
the theoretical calculation in the nuclear shadowing region under-predicts the data. These features
are most prominent at low \HT, while the theory typically under-predicts the data for the higher \HT intervals
without substantial dependence on \xA. Only in the highest
\xA interval, $0.288$--$0.5$, is any substantial modification from the EMC effect expected, and it is
typically in good agreement with the data.\par

An alternative presentation of the results shown above
plotted as a function of \HT instead of \xA is provided in Figures~
\ref{fig:Results_HT_TheoryComp_SlicesxA_zGamma7},
\ref{fig:Results_HT_TheoryComp_SlicesxA_zGamma8},
\ref{fig:Results_HT_TheoryComp_SlicesxA_zGamma9}, and
\ref{fig:Results_HT_TheoryComp_SlicesxA_zGamma10} in Appendix \ref{sec:morePlots}. Figure
\ref{fig:Results_HT_TheoryComp_SlicesxA_zGammaAll} shows the
analogous results integrated over the full range in \zg.
The cross-sections in the restricted intervals of \zg and \xA
vary slowly with \HT except at the upper limit of the \HT range.\par

\begin{figure}[!tbp]
\centerline{
\includegraphics[width=0.85\textwidth]{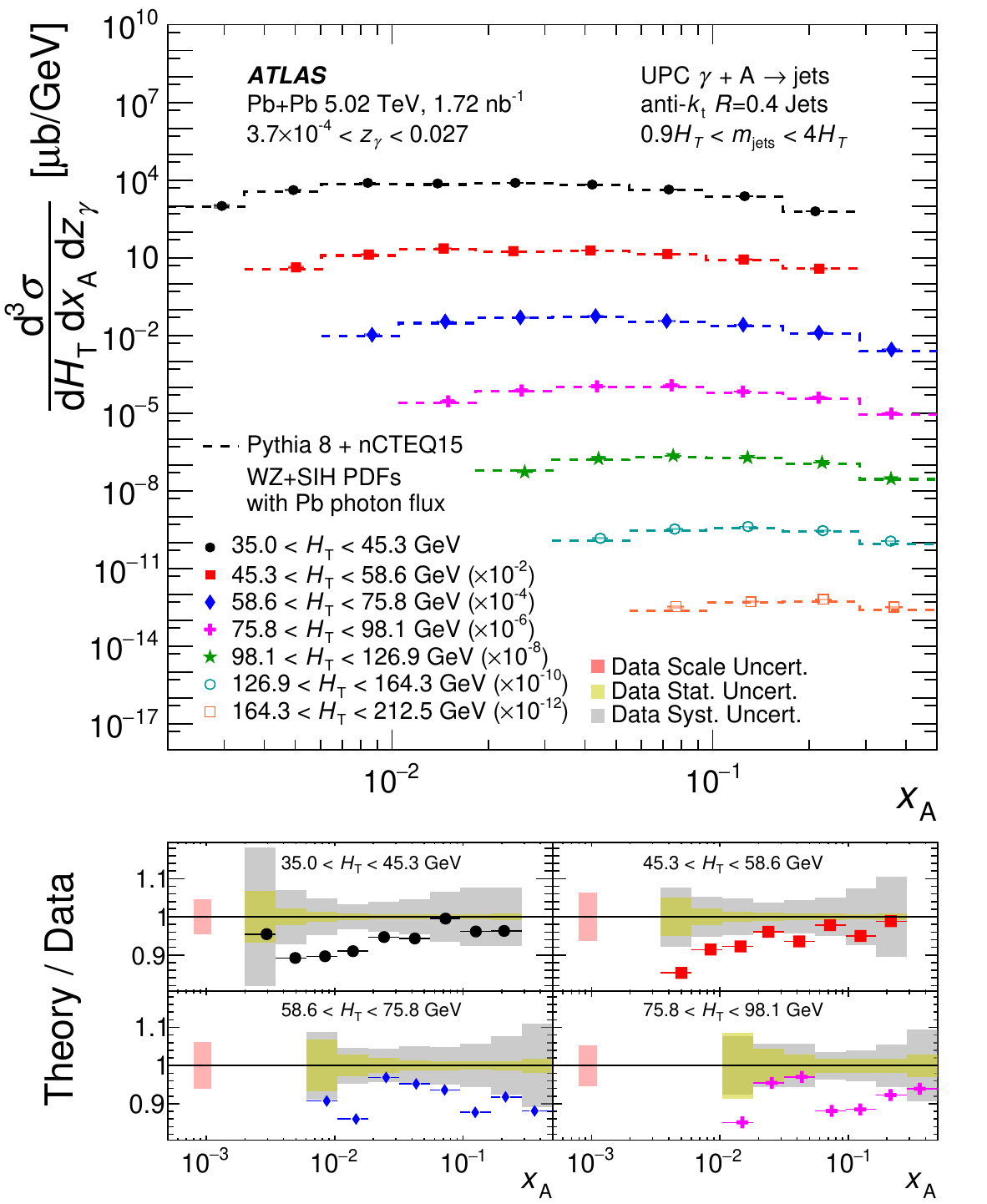}
}
\caption{ Triple-differential cross-sections, \sigtridiff, as a
function of $\xA$ for different bins of $\HT$ for events with emitted
photon energies in the kinematic range $3.7\times10^{-4} < \zg <
0.027$.  In the upper panel, systematic uncertainties are shown as
shaded boxes, while statistical uncertainties shown as vertical lines
are usually smaller than the size of the markers. A theoretical
comparison is shown to cross-sections computed using \pythia with
nCTEQ15 WZ+SIH PDFs, a photon flux from \textsc{Starlight}, and a \zg-dependent
breakup fraction. The bottom
panels show the ratio between the theory prediction and data for a representative subset of the
bins of \HT. The lorentz-invariant variables \xA, \zg, and \HT are un-ambiguously defined in terms of the particle-level jet kinematics and correspond at leading order to the Bjorken-$x$, emitted photon energy fraction,
and collision $Q^2$, respectively.
The light red bands in the ratio panels are the quadrature sum of scale uncertainties on the cross-section, while the gray band shows the remaining systematic uncertainty. The yellow band  shows the point-to-point statistical uncertainty.  }
\label{fig:Results_xA_TheoryComp_SlicesHT_zGammaAll}
\end{figure}

\begin{figure}[!tbp]
\centerline{
\includegraphics[width=0.85\textwidth]{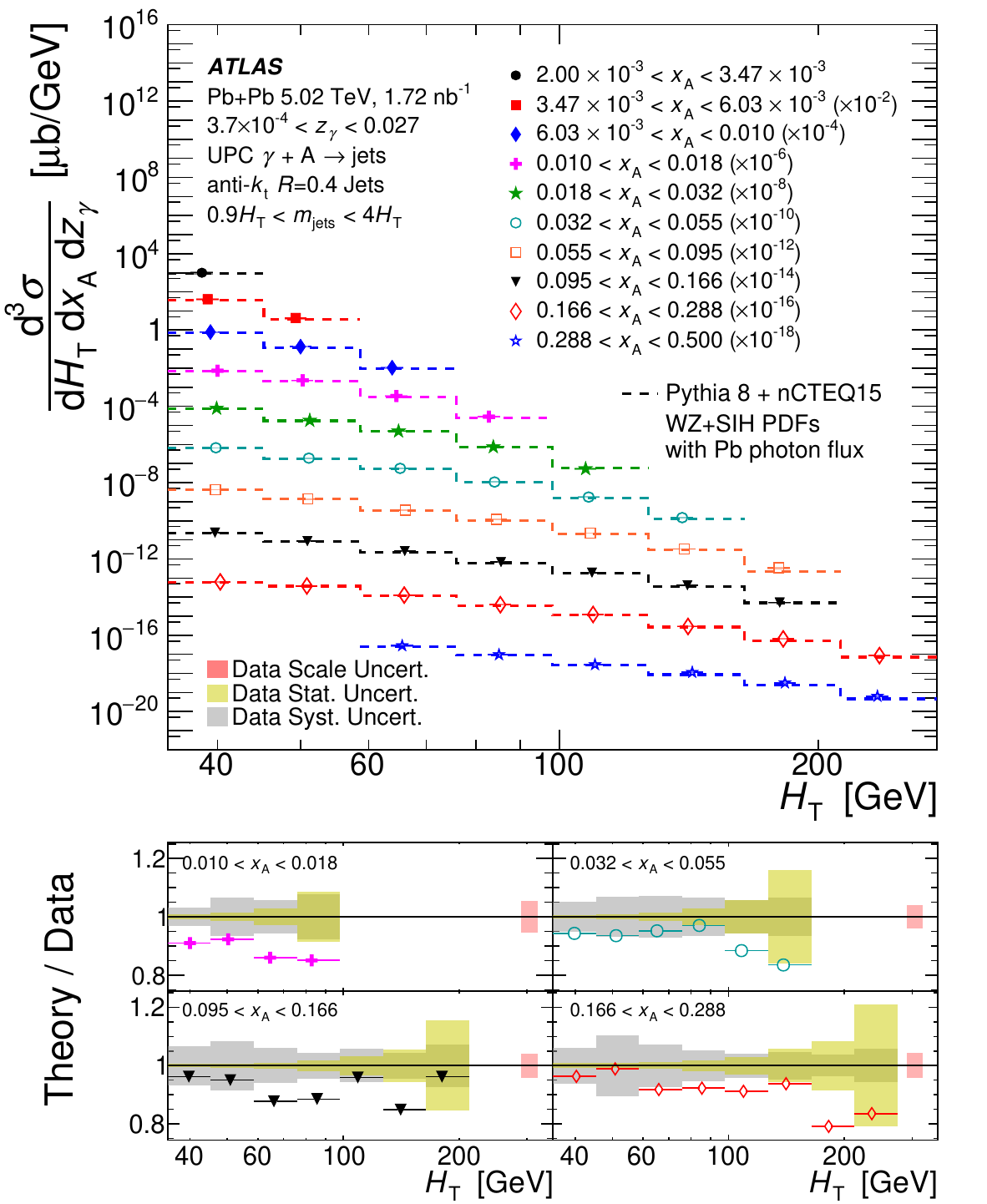}
}
\caption{ Triple-differential cross-sections, \sigtridiff, as a
function of $\HT$ for different bins of $\xA$ for events with emitted
photon energies in the kinematic range $3.7\times10^{-4} < \zg <
0.027$.  In the upper panel, systematic uncertainties are shown as
shaded boxes, while statistical uncertainties shown as vertical lines
are usually smaller than the size of the markers. A theoretical
comparison is shown to cross-sections computed using \pythia with
nCTEQ15 WZ+SIH PDFs, a photon flux from \textsc{Starlight}, and a \zg-dependent
breakup fraction. The bottom
panels show the ratio between the theory prediction and data for a representative subset of the
bins of \xA. The lorentz-invariant variables \xA, \zg, and \HT are un-ambiguously defined in terms of the particle-level jet kinematics and correspond at leading order to the Bjorken-$x$, emitted photon energy fraction,
and collision $Q^2$, respectively.
The light red bands in the ratio panels are the quadrature sum of scale uncertainties on the cross-section, while the gray band shows the remaining systematic uncertainty. The yellow band  shows the point-to-point statistical uncertainty.  }
\label{fig:Results_HT_TheoryComp_SlicesxA_zGammaAll}
\end{figure}

\begin{figure}[!tbp]
\centerline{
\includegraphics[width=0.85\textwidth]{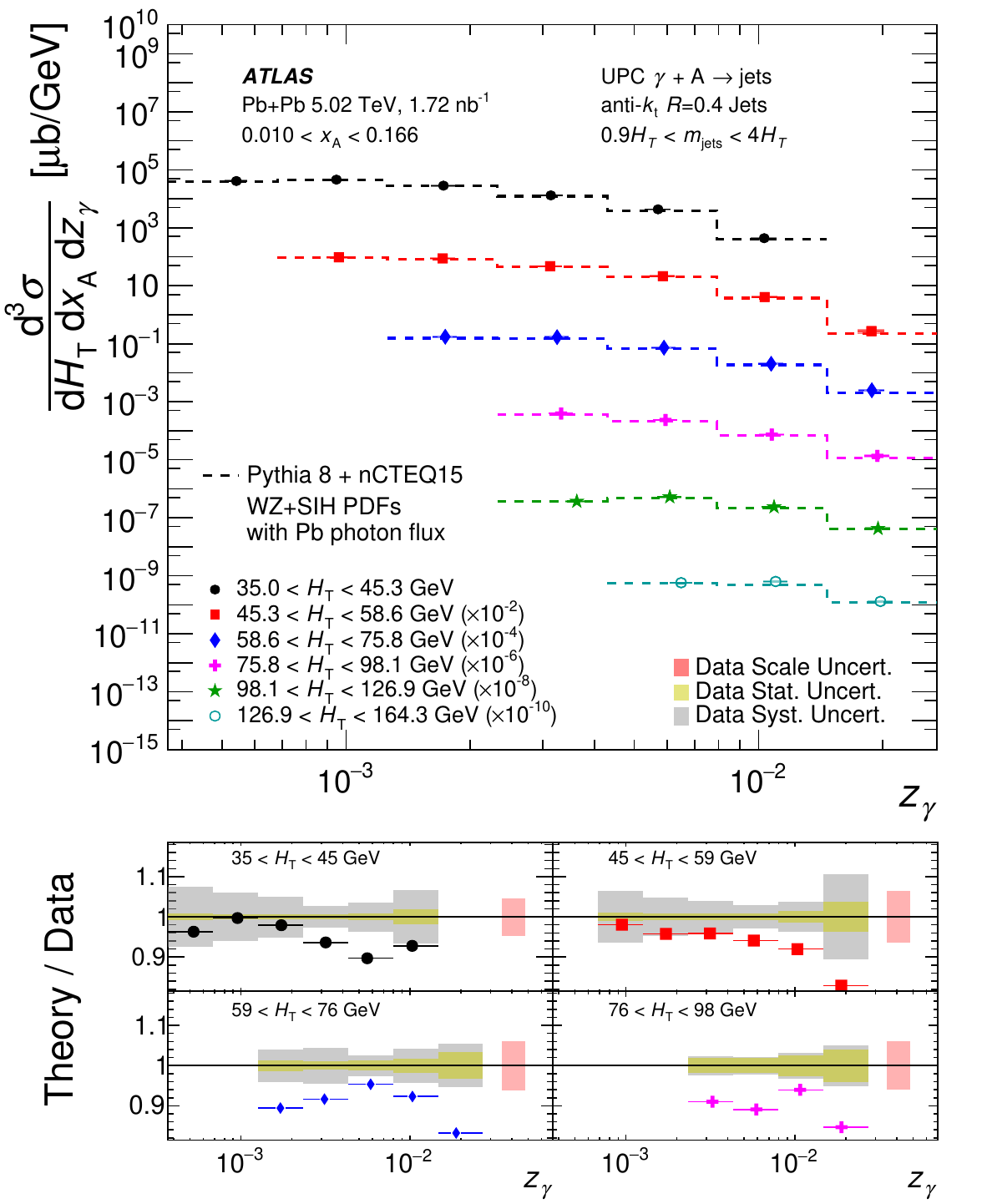}
}
\caption{ Triple-differential cross-sections, \sigtridiff, as a
function of \zg for different bins of \HT for events with struck
parton energies in the kinematic range $0.010 < \xA <
0.166$.  In the upper panel, systematic uncertainties are shown as
shaded boxes, while statistical uncertainties shown as vertical lines
are usually smaller than the size of the markers. A theoretical
comparison is shown to cross-sections computed using \pythia with
nCTEQ15 WZ+SIH PDFs, a photon flux from \textsc{Starlight}, and a \zg-dependent
breakup fraction. The bottom
panels show the ratio between the theory prediction and data for a representative subset
of intervals in \HT. The lorentz-invariant variables \xA, \zg, and \HT are un-ambiguously defined in terms of the particle-level jet kinematics and correspond at leading order to the Bjorken-$x$, emitted photon energy fraction,
and collision $Q^2$, respectively.
The light red bands in the ratio panels are the quadrature sum of scale uncertainties on the cross-section, while the gray band shows the remaining systematic uncertainty. The yellow band  shows the point-to-point statistical uncertainty.  }
\label{fig:Results_zGamma_TheoryComp_SlicesHT_AllxA}
\end{figure}

\begin{figure}[!tbp]
\centerline{
\includegraphics[width=0.85\textwidth]{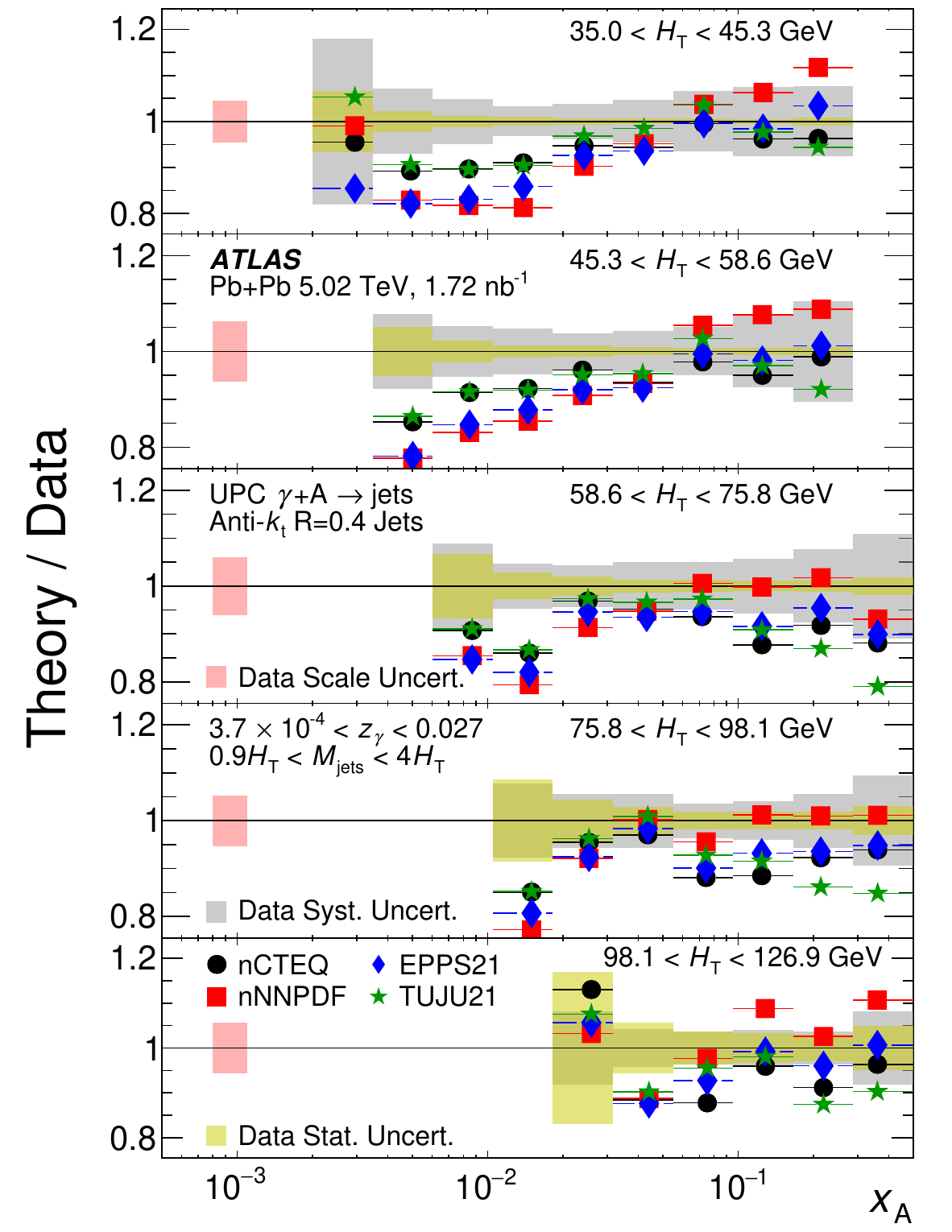}
}
\caption{ Ratios of the triple-differential cross-sections, \sigtridiff,
to theoretical preductions using several nPDF models as a
function of \xA for different bins of \HT for events with struck
parton energies in the kinematic range $3.7\times10^{-4} < \zg <
0.027$. The lorentz-invariant variables \xA, \zg, and \HT are un-ambiguously defined in terms of the particle-level jet kinematics and correspond at leading order to the Bjorken-$x$, emitted photon energy fraction,
and collision $Q^2$, respectively. Theoretical
comparisons for the cross-sections are computed using \pythia withh
a photon flux from \textsc{Starlight}, and a \zg-dependent data-driven breakup correction. Four different theoretical comparisons are shown corresponding to the nCTEQ 15 WZ+SIH, nNNPDF 3.0, EPPS21, and TUJU21 nPDF fits. The light red bands are the quadrature sum of scale uncertainties on the ratio, while the gray band shows the residual systematic uncertainty. The yellow band  shows the point-to-point statistical uncertainty.  }
\label{fig:Results_xA_CompPDFs_SlicesHT_AllzGamma}
\end{figure}

Figure~\ref{fig:Results_zGamma_TheoryComp_SlicesHT_AllxA} shows
the measured differential cross-section as a function of \zg in different
intervals of \HT for a fixed \xA range, $0.010 < \xA <
0.166$. The \zg dependence over a wide \xA interval should be
determined, primarily, by the photon flux and the precision of the cross-section
calculation for the hard-scattering process. Thus, this comparison should primarily indicate the
regions of phase space which are most impacted by the LO precision of \pythia or
uncertainty on the photon flux.
The results indicate that differences from \pythia arise without an obvious dependence on \zg,
and the discrepancies are slightly more significant at higher \HT.\par

Figure \ref{fig:Results_xA_CompPDFs_SlicesHT_AllzGamma} shows the ratio between measured
cross-sections and predictions using \pythia LO calculations with a simulated parton shower (LO+PS) and a variety of leading global nPDF fits. These fits include nCTEQ15 WZ+SIH, nNNPDF 3.0, EPPS21 \cite{Eskola:2021nhw}, and TUJU21 \cite{Helenius:2021tof}. The former three fits include different global datasets and employ different methodologies, while the TUJU21 fit is unique for restricting its inputs to only measurements with NNLO theory calculations available. These
comparisons demonstrate that no single fit describes the data best in all kinematic regions. At lower \HT, all models under-predict the
shadowing region, while nNNPDF slightly over-predicts the anti-shadowing region. For
intermediate \HT intervals, similar discrepancies persist at low \xA, while nNNPDF predicts
the cross-section in the anti-shadowing region better than other models. It shows signs of over-prediction at high \HT, though, as
observed for higher $Q^2$ in recent measurements of $t\bar{t}$ production in \PbPb\ collisions \cite{TOPQ-2023-32}.
These observations may be modified
when NLO corrections become available, or when theoretical uncertainties on the modelling of the
photon flux are included.\par

\FloatBarrier
\pagebreak

\FloatBarrier
\pagebreak

% End of text imported from the .//tex/results.tex input file

%

%
\section{Conclusion}
\label{sec:conclusion}

% The next lines are included from the .//tex/conclusions.tex input file
This paper presents measurements of photonuclear jet production in
ultra-peripheral \PbPb\ collisions at $\sqn = 5.02$~\TeV from a data
set collected with the ATLAS detector at the LHC corresponding to
an integrated luminosity of $1.72$~\inb. The measurement provides unique constraints on
nuclear parton distributions in a region of $x$ and $Q^2$ where
currently available data provide very little information.

In the analysis, candidate \gamA\ events are selected using a
combination of requirements on the neutron yields in the zero-degree calorimeters and
substantial rapidity gaps
in the photon-going direction.  The selection also has good acceptance for
resolved-photon events that can
populate forward rapidities with particles. Triple-differential cross-sections
are measured using two sets of kinematic variables, \kinsetjet and
\kinsetpart, the latter of which more directly reflect the kinematics
of the partons in the hard-scattering process, allowing for a clear
understanding of the kinematic ranges probed by this measurement. The measured
cross-sections are corrected for trigger and event selection
efficiencies and are unfolded to the particle level using an
evaluation of the detector response obtained from a \pythia \gamAjets
MC sample. These \pythia events were generated using nCTEQ15 nuclear
PDFs and a coherent nuclear photon flux. The resulting sample of events was re-weighted
to account for modifications to the photon flux resulting from the
ultra-peripheral (i.e. non-hadronic) event selection and the real
impact parameter distribution of \gamA processes.

The measured
cross-sections are compared to LO \pythia predictions, which are
corrected by the probability that the photon-emitting nucleus does not
break up due to additional soft photons exchanges that can excite, for example, the
giant dipole resonance in the nucleus. This no-breakup probability for
\gamA processes was measured in data using a sample of events with forward
neutron emission in both directions.
The comparisons between data and \pythia show systematic deviations
as a function of both \zg and \xA that may indicate limitations in the
theoretical description of the photon flux, but also, at fixed \zg,
deviations in the lead PDF relative to the nCTEQ15 WZ+SIH nPDF set used to
produce the \pythia sample. LO \pythia predictions using all included fits
under-predict the data in the shadowing region, while most fits
exhibit better agreement with the data in the anti-shadowing and EMC regions.
Extraction of information about nPDFs from this data, however, will rely on
more precise NLO pQCD calculations directly comparable to these results.
Such theoretical results should offer a direct path for including
the data presented here in global nuclear PDF fits, although the inclusion of parton shower
corrections is also important for a robust extraction of nPDF effects.
%
%

% End of text imported from the .//tex/conclusions.tex input file

%

%
\section*{Acknowledgements}
%

% The next lines are included from the .//acknowledgements/Acknowledgements.tex input file
%
%

%
%

We thank CERN for the very successful operation of the LHC and its injectors, as well as the support staff at
CERN and at our institutions worldwide without whom ATLAS could not be operated efficiently.

The crucial computing support from all WLCG partners is acknowledged gratefully, in particular from CERN, the ATLAS Tier-1 facilities at TRIUMF/SFU (Canada), NDGF (Denmark, Norway, Sweden), CC-IN2P3 (France), KIT/GridKA (Germany), INFN-CNAF (Italy), NL-T1 (Netherlands), PIC (Spain), RAL (UK) and BNL (USA), the Tier-2 facilities worldwide and large non-WLCG resource providers. Major contributors of computing resources are listed in Ref.~\cite{ATL-SOFT-PUB-2023-001}.

We gratefully acknowledge the support of ANPCyT, Argentina; YerPhI, Armenia; ARC, Australia; BMWFW and FWF, Austria; ANAS, Azerbaijan; CNPq and FAPESP, Brazil; NSERC, NRC and CFI, Canada; CERN; ANID, Chile; CAS, MOST and NSFC, China; Minciencias, Colombia; MEYS CR, Czech Republic; DNRF and DNSRC, Denmark; IN2P3-CNRS and CEA-DRF/IRFU, France; SRNSFG, Georgia; BMBF, HGF and MPG, Germany; GSRI, Greece; RGC and Hong Kong SAR, China; ISF and Benoziyo Center, Israel; INFN, Italy; MEXT and JSPS, Japan; CNRST, Morocco; NWO, Netherlands; RCN, Norway; MNiSW, Poland; FCT, Portugal; MNE/IFA, Romania; MSTDI, Serbia; MSSR, Slovakia; ARIS and MVZI, Slovenia; DSI/NRF, South Africa; MICIU/AEI, Spain; SRC and Wallenberg Foundation, Sweden; SERI, SNSF and Cantons of Bern and Geneva, Switzerland; NSTC, Taipei; TENMAK, T\"urkiye; STFC/UKRI, United Kingdom; DOE and NSF, United States of America.

Individual groups and members have received support from BCKDF, CANARIE, CRC and DRAC, Canada; CERN-CZ, FORTE and PRIMUS, Czech Republic; COST, ERC, ERDF, Horizon 2020, ICSC-NextGenerationEU and Marie Sk{\l}odowska-Curie Actions, European Union; Investissements d'Avenir Labex, Investissements d'Avenir Idex and ANR, France; DFG and AvH Foundation, Germany; Herakleitos, Thales and Aristeia programmes co-financed by EU-ESF and the Greek NSRF, Greece; BSF-NSF and MINERVA, Israel; NCN and NAWA, Poland; La Caixa Banking Foundation, CERCA Programme Generalitat de Catalunya and PROMETEO and GenT Programmes Generalitat Valenciana, Spain; G\"{o}ran Gustafssons Stiftelse, Sweden; The Royal Society and Leverhulme Trust, United Kingdom.

In addition, individual members wish to acknowledge support from Armenia: Yerevan Physics Institute (FAPERJ); CERN: European Organization for Nuclear Research (CERN PJAS); Chile: Agencia Nacional de Investigaci\'on y Desarrollo (FONDECYT 1230812, FONDECYT 1230987, FONDECYT 1240864); China: Chinese Ministry of Science and Technology (MOST-2023YFA1605700, MOST-2023YFA1609300), National Natural Science Foundation of China (NSFC - 12175119, NSFC 12275265, NSFC-12075060); Czech Republic: Czech Science Foundation (GACR - 24-11373S), Ministry of Education Youth and Sports (FORTE CZ.02.01.01/00/22\_008/0004632), PRIMUS Research Programme (PRIMUS/21/SCI/017); EU: H2020 European Research Council (ERC - 101002463); European Union: European Research Council (ERC - 948254, ERC 101089007), Horizon 2020 Framework Programme (MUCCA - CHIST-ERA-19-XAI-00), European Union, Future Artificial Intelligence Research (FAIR-NextGenerationEU PE00000013), Italian Center for High Performance Computing, Big Data and Quantum Computing (ICSC, NextGenerationEU); France: Agence Nationale de la Recherche (ANR-20-CE31-0013, ANR-21-CE31-0013, ANR-21-CE31-0022, ANR-22-EDIR-0002), Investissements d'Avenir Labex (ANR-11-LABX-0012); Germany: Baden-Württemberg Stiftung (BW Stiftung-Postdoc Eliteprogramme), Deutsche Forschungsgemeinschaft (DFG - 469666862, DFG - CR 312/5-2); Italy: Istituto Nazionale di Fisica Nucleare (ICSC, NextGenerationEU), Ministero dell'Università e della Ricerca (PRIN - 20223N7F8K - PNRR M4.C2.1.1); Japan: Japan Society for the Promotion of Science (JSPS KAKENHI JP22H01227, JSPS KAKENHI JP22H04944, JSPS KAKENHI JP22KK0227, JSPS KAKENHI JP23KK0245); Netherlands: Netherlands Organisation for Scientific Research (NWO Veni 2020 - VI.Veni.202.179); Norway: Research Council of Norway (RCN-314472); Poland: Ministry of Science and Higher Education (IDUB AGH, POB8, D4 no 9722), Polish National Agency for Academic Exchange (PPN/PPO/2020/1/00002/U/00001), Polish National Science Centre (NCN 2021/42/E/ST2/00350, NCN OPUS nr 2022/47/B/ST2/03059, NCN UMO-2019/34/E/ST2/00393, NCN \& H2020 MSCA 945339, UMO-2020/37/B/ST2/01043, UMO-2021/40/C/ST2/00187, UMO-2022/47/O/ST2/00148, UMO-2023/49/B/ST2/04085, UMO-2023/51/B/ST2/00920); Slovenia: Slovenian Research Agency (ARIS grant J1-3010); Spain: Generalitat Valenciana (Artemisa, FEDER, IDIFEDER/2018/048), Ministry of Science and Innovation (MCIN \& NextGenEU PCI2022-135018-2, MICIN \& FEDER PID2021-125273NB, RYC2019-028510-I, RYC2020-030254-I, RYC2021-031273-I, RYC2022-038164-I); Sweden: Carl Trygger Foundation (Carl Trygger Foundation CTS 22:2312), Swedish Research Council (Swedish Research Council 2023-04654, VR 2018-00482, VR 2022-03845, VR 2022-04683, VR 2023-03403, VR grant 2021-03651), Knut and Alice Wallenberg Foundation (KAW 2018.0458, KAW 2019.0447, KAW 2022.0358); Switzerland: Swiss National Science Foundation (SNSF - PCEFP2\_194658); United Kingdom: Leverhulme Trust (Leverhulme Trust RPG-2020-004), Royal Society (NIF-R1-231091); United States of America: U.S. Department of Energy (ECA DE-AC02-76SF00515), Neubauer Family Foundation.

%
%

% End of text imported from the .//acknowledgements/Acknowledgements.tex input file

%
\clearpage
\appendix
\part*{Appendix}
\addcontentsline{toc}{part}{Appendix}
\section{In-situ measurement of the jet energy scale and resolution}
\label{sec:JetResp}

% The next lines are included from the .//tex/appendixA.tex input file

\subsection{Samples used for calibration}

In-situ studies of the jet energy scale and resolution were performed using $334$ \ipb of $13$ TeV $pp$ data taken by ATLAS in 2017 and 2018. Performing calibration studies in this different collision system is necessary for two reasons: UPCs are not energetic enough to produce the $Z$+jet pairs needed to derive the absolute energy scale and alternative MC generators are not available for UPCs in order to test the impact of the MC model on the JES corrections. The \pp data were taken in a period where the typical number of interactions per bunch crossing was $\mu = 2$, allowing the data to be taken using the same detector configuration as the UPC data sample. Two sets of triggers were used to collect the data for these studies: a sample requiring at least one high-\pt\ lepton for studies of $Z$+jet balance and a sample requiring at least one high-\pt\ jet in order to study dijet balance for measures of $\eta$-dependence of the JES and to study the JER.\par

These studies also involved the use of MC samples generated using some combination of \pythia \cite{Sjostrand:2014zea}, \textsc{Herwig++} \cite{Bellm:2015jjp}, \textsc{Powheg} \cite{Alioli:2010xd,Alioli:2008gx}, and \Sherpa \cite{Bothmann:2019yzt}. Samples with the jet cross-section calculated at NLO were produced using either \Sherpa or a combination of \textsc{Powheg} with either \pythia or \textsc{Herwig++} to model the parton shower and hadronization. Each sample was produced with the A14 tune \cite{ATL-PHYS-PUB-2014-021} and NNPDF 2.3 free proton PDFs \cite{Ball:2012cx}.  For simulating $Z$+jet production, two NLO samples were produced. The first sampled used \textsc{Powheg}+\pythia, where \textsc{Powheg} was used to compute the matrix element and \pythia was used to simulate the parton shower and hadronization. The second sample used \Sherpa for both aspects of the event generation. For dijet balance, two samples were used: \textsc{Powheg}+\pythia and \textsc{Powheg}+ \textsc{Herwig++}.\par

\subsection{Jet energy scale}

The strategy used in this measurement for calibrating the jet energy scale follows previous work from ATLAS \cite{JETM-2018-05}, and it involves two steps: studies of the absolute energy scale using $Z$+jet balance and studies of $\eta$-dependent variation of the JES using dijet balance ($\eta$-intercalibration). This procedure works by first establishing the absolute jet energy scale in a reference rapidity interval ($|\eta| < 0.8$) by studying the balance of jets against a well-measured reference object, the $Z$ boson in this case. Then, the much higher-rate process of dijet production is used to calibrate other $\eta$ regions relative to this reference region.

\subsubsection{Absolute energy scale}

Studies of the absolute energy scale begin with a sample of events which are required to have at least two opposite-charge leptons (either \ee or \mumu), where both leptons have $\pt > 20$~\GeV. Additionally, the dilepton invariant mass is selected to be within the $Z$ peak, and events are required to have at least one jet that balances the $Z$ boson in azimuth with no additional jets in the event. After applying these selections, data distributions of $r = \pTjet / \pTz$ are constructed in bins of \pTz. In these studies, a new approach was taken to the fitting procedure, since data taken with low $\mu$ is statistically limited. In this approach, the data \ee and \mumu channel distributions are fitted simultaneously by constructing distributions of $r = R_{\text{Data}/\text{MC}}\pTjet / \pTz$ using the MC simulation, where $R_{\text{Data}/\text{MC}}$ is a re-scaling factor used to calibrate the jet \pt\ in data, in order to account for data-to-MC differences in the absolute JES. This in-situ correction factor is extracted by iterative $\chi^2$ minimization fitting of these re-scaled MC distributions to match the binned data distribution.\par

\begin{figure}[!tb]
\centerline{
\includegraphics[width=0.7\textwidth]{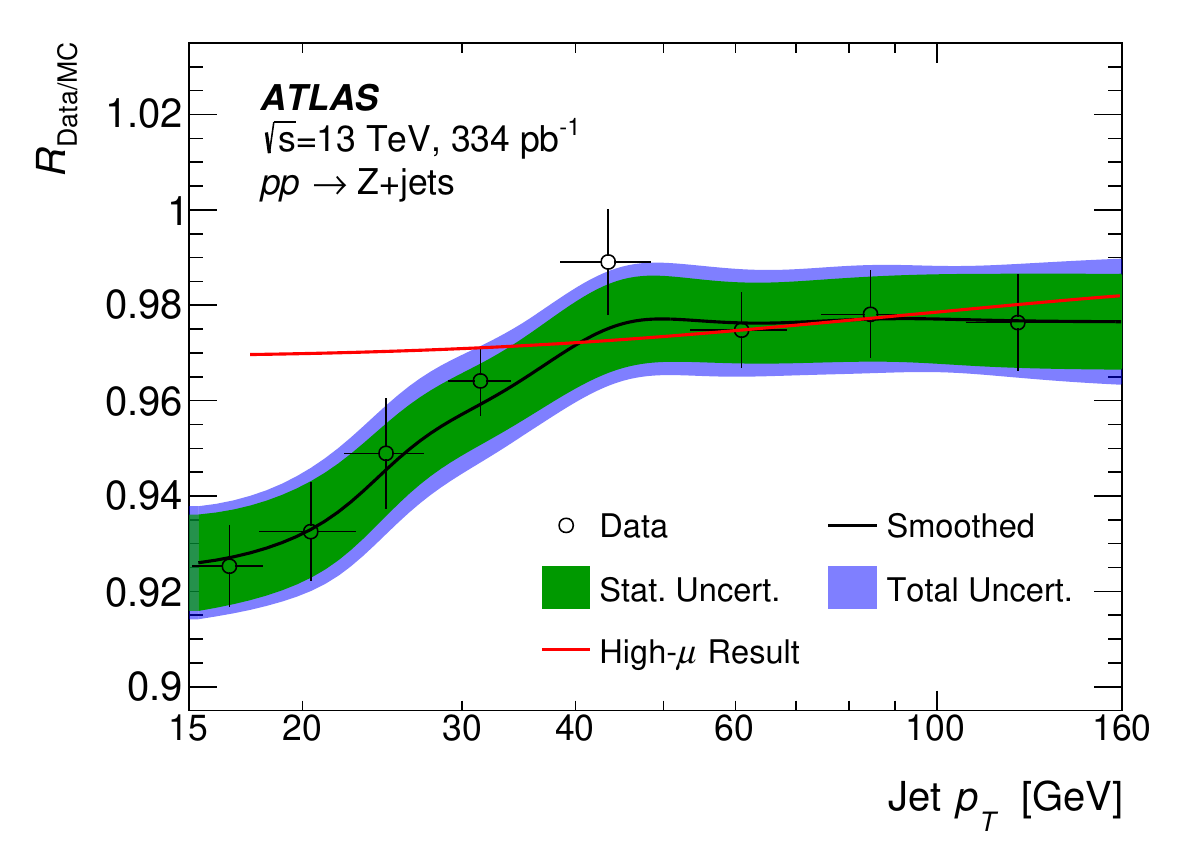}
}
\caption{
In-situ correction to the absolute JES as a function of \pTjet derived from data and MC evaluation of the $Z$+jet balance in low-$\mu$ ($\mu \sim 2$) \pp collisions at 13~\TeV. The binned results are smoothed with a Gaussian kernel smoothing procedure, the result of which is shown as a black line.  The red line shows the result from high-$\mu$ \pp collisions.
}
\label{fig:ZJetFig}
\end{figure}

This fitting procedure is then performed in bins of \pTjet, with statistical uncertainties derived using the bootstrap method \cite{ATL-PHYS-PUB-2021-011}, and the binned results are smoothed using a Gaussian kernel smoothing procedure. The results of this calibration procedure, compared to the same calibration factors derived in $\mu = 40$ \pp collisions, are shown in Figure \ref{fig:ZJetFig}. The shaded bands show the statistical and systematic uncertainties, indicating that the primary differences in the in-situ JES arise at low \pt, likely due to increased sensitivity to low energy topo-clusters most impacted by the different calorimeter conditions. The systematic uncertainties are assigned by comparing the results of two different MC simulation models (\textsc{Powheg}+\pythia and \Sherpa), varying the event selection criteria, and propagating uncertainty on the electron or muon energy scales. An additional uncertainty is assigned for MC modelling of out-of-cone radiation, which may impact the relative balance between $Z$ bosons and jets in these studies.\par

\subsubsection{$\eta$-intercalibration}

The $\eta$-intercalibration uses a sample of events that are selected by requiring two jets, each with $\pTjet > 10$ GeV. Further selections are applied in order to ensure the pair should be balanced in \pt, including that the leading dijet pair be azimuthally balanced and that there are no other jets in the event. Then, using this sample, distributions of $(p_{\text{T}_1} - p_{\text{T}_2})/\langle p_{\text{T}} \rangle$ are constructed in bins of $\eta_1$, $\eta_2$, and $\langle p_{\text{T}}\rangle$, where $\langle p_{\text{T}}\rangle = (p_{\text{T}_1} + p_{\text{T}_2})/2$. In order to derive $\eta$-dependent scale factors from these distributions, the ATLAS matrix method \cite{JETM-2022-01} is employed, where a Gaussian fit to each bin in $\eta_1$ and $\eta_2$ provides a measurement of the product of the relative scale factors in both bins. The results from all bins in $\eta_1$ and $\eta_2$ are combined via a linear regression procedure to extract a scale factor in each $\eta$ and $\langle p_{\text{T}}\rangle$ bin.\par

\begin{figure}[!tb]
\centerline{
\subfloat[]{\includegraphics[width=0.49\textwidth]{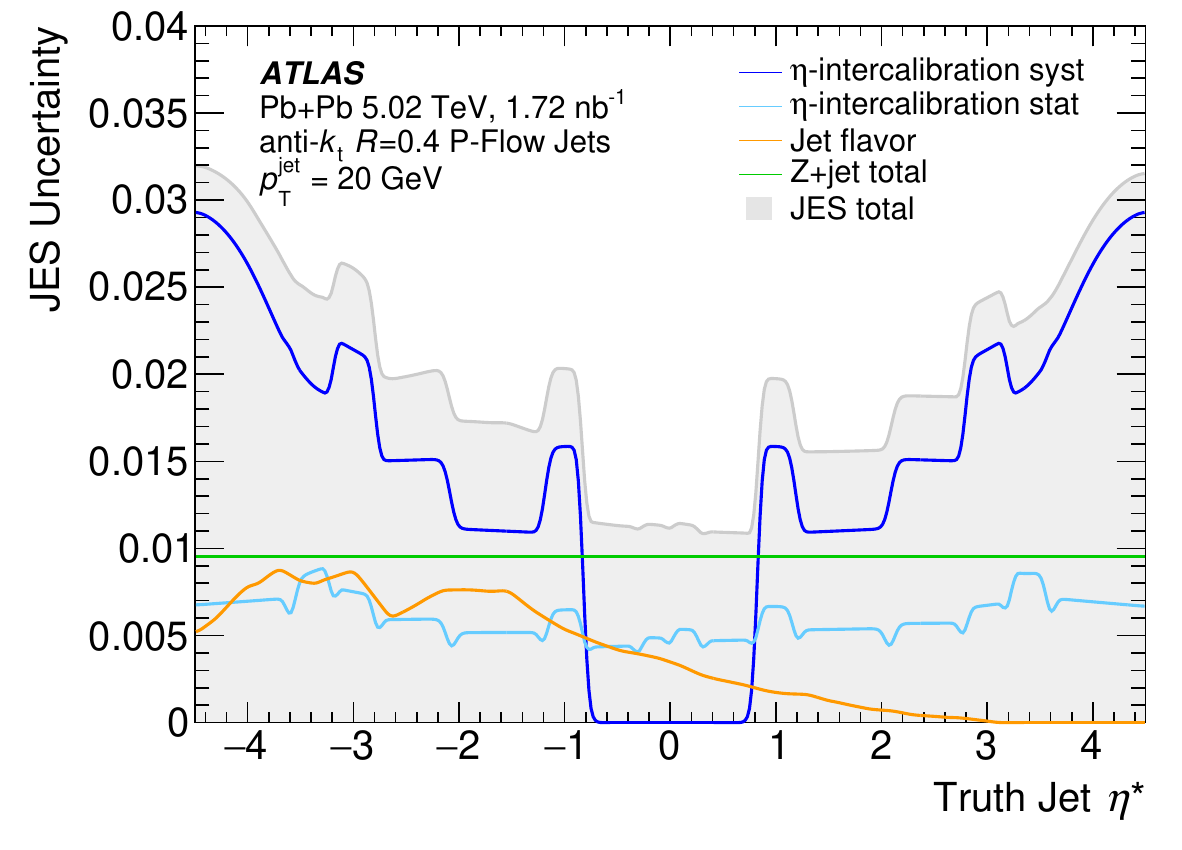}}
\subfloat[]{\includegraphics[width=0.49\textwidth]{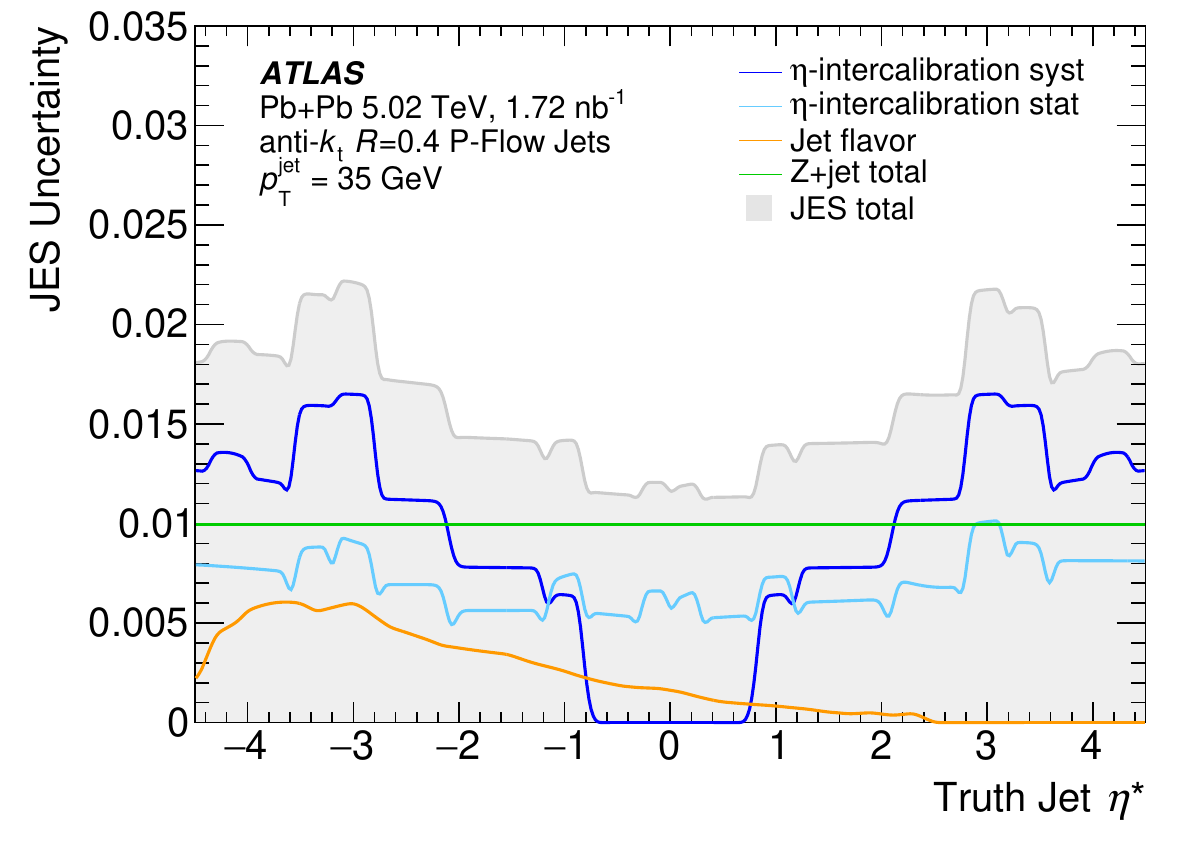}}
}
\caption{
A summary of all sources of JES systematic uncertainties as a function of $\eta^*$ for (a) $\pTjet = 20$~\GeV and (b) $\pTjet = 35$~\GeV.
}
\label{fig:JetSystEtaDep}
\end{figure}

This same procedure is repeated in data and MC simulation, where the ratio of the scale factors between data and MC measures the relevant data-MC difference in jet response, providing the in-situ correction factor. These correction factors are then re-scaled in each \pt\ bin so that the correction in the reference region, where the absolute energy scale is determined by $Z$+jet balance, is unity. Systematic uncertainties in the $\eta$-intercalibration are assessed by varying event selections and the MC generator (\textsc{Powheg}+\pythia vs. \textsc{Powheg}+\textsc{Herwig++}). Statistical uncertainties are determined via the bootstrap method \cite{ATL-PHYS-PUB-2021-011}. The resulting calibration factors and systematic uncertainties are smoothed using a two-dimensional Gaussian kernel smoothing procedure, and an additional uncertainty is assessed for the non-closure of the procedure, using a combination of studies in \gamAjets events and \pp collisions. While the $\eta$-intercalibration uncertainties are symmetrized with respect to $\eta$, the uncertainties related to the jet flavor composition and response depend on $\eta^*$, which is chosen to be positive in the photon-going direction. The total $\eta^*$-dependent systematic uncertainties for two particularly relevant values of \pTjet are summarized in Figure \ref{fig:JetSystEtaDep}. \par

\subsubsection{MC-Derived UPC correction}

\begin{figure}[!tb]
\centerline{
\includegraphics[width=0.7\textwidth]{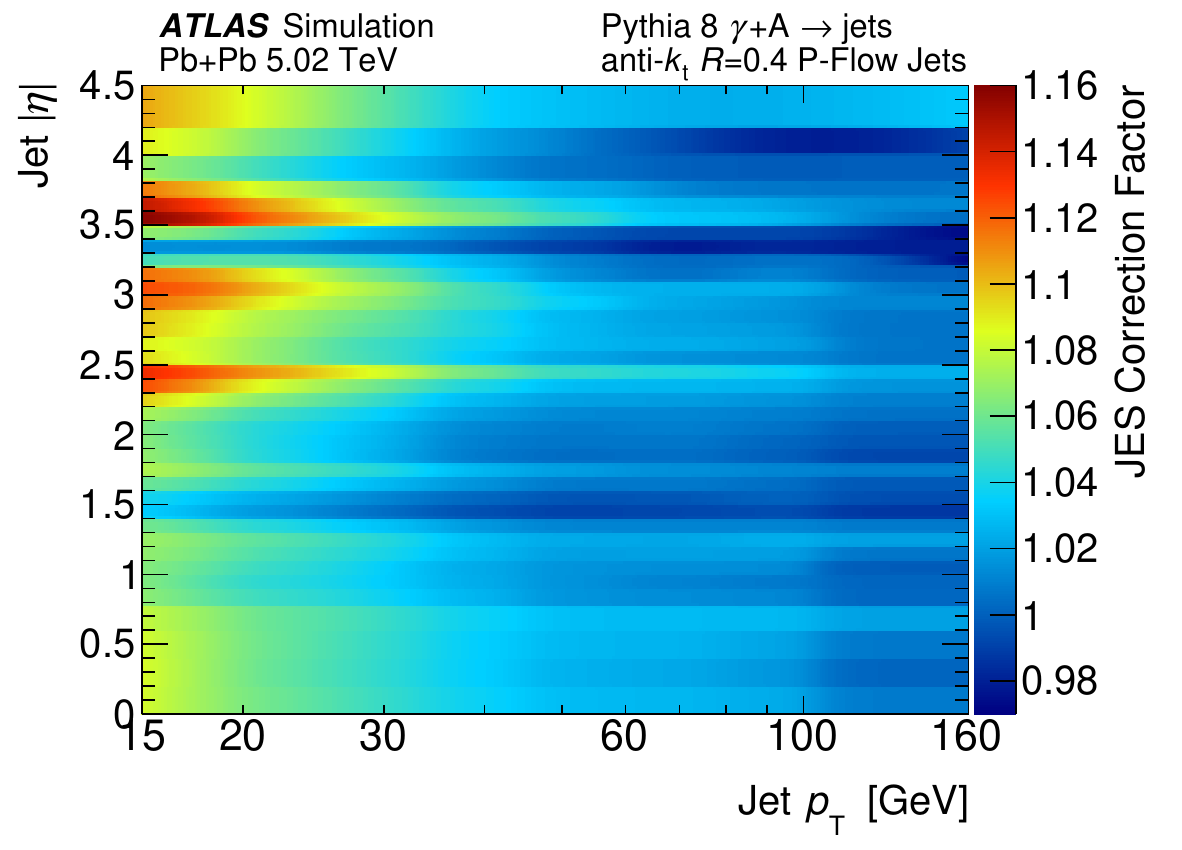}
}
\caption{
Correction factors to the jet energy scale, derived in low-$\mu$ ($\mu \sim 2$) 13~\TeV\pp collisions, used to translate the JES to jets in photonuclear \PbPb events. The $\eta$-dependence of the correction is largely driven by the structure of the ATLAS calorimeter.
}
\label{fig:CorrUPC}
\end{figure}

In order to translate results on the jet energy calibration in $\mu = 2$ \pp to photonuclear jets, an additional correction must be derived from MC simulations to account for differences in the jet energy scale between these two systems. These differences primarily arise from three sources: the different jet flavor compositions of the two samples, the difference in $\mu$ between the two systems, and the different underlying event in \pp and \gamA events. In order to determine a correction accounting for these differences, the calibration derived in $\mu = 2$ \pp collisions is applied in photonuclear \PbPb\ event generated with \pythia, and truth jets are matched to their nearest reconstructed jet within $\Delta R < 0.3$. Then, using matched pairs, $\pTreco / \pTtruth$ distributions are constructed in intervals of \pTtruth\ and
\etaTruth. Then, the mean jet response in each bin is determined by a Gaussian fit to the distribution, and a correction factor is derived from the mean responses via a numerical inversion procedure~\cite{PERF-2011-03}. The results of this procedure are shown in Figure \ref{fig:CorrUPC}.

\subsection{Jet energy resolution}

In addition to correcting for differences between the JES in data and MC, any differences in the jet energy resolution \cite{JETM-2018-05} may also impact this measurement, thus a direct measurement of any difference and its associated systematic uncertainties is essential for a precise measurement of the cross-sections. Unlike the case for the JES, only dijets are necessary to measure the resolution, so these studies are performed directly in UPC jet events. Events are selected with a dijet pair, each with $\pTjet > 10$ GeV. The jets are required to be nearly back-to-back in azimuth and a veto is applied on events with extra jets, in order to ensure the \pt\ balance of the pair. The JER is measured in regions of $\eta$, and for each region, one jet in the pair (the probe jet) is required to be in that region, while the other jet must be in the designated reference region for the probe jet's $\eta$. For jets with $|\eta| < 2.8$, reference regions are the same as the probe regions, but more forward bins use the reference region $1.8 < |\eta| < 2.5$, in order to reduce the statistical uncertainty on the JER measurement. Distributions of $(p^\text{probe}_\text{T} - p^\text{ref}_\text{T})/\langle p_\text{T}\rangle$ are then constructed for all probe/reference jet pairs.\par

Binned measurements of the jet energy resolution are then extracted via a convolution fitting procedure \cite{JETM-2018-05}. First, the asymmetry distribution of particle-level jets is fitted with a Gaussian times exponential distribution, and then this distribution is convolved with a Gaussian, whose mean and standard deviation are fitted to describe the measured distribution. The width of this convolved Gaussian corresponds to the quadrature sum of the probe and reference resolutions. In $\eta$ bins where the same region is used as a reference, the extracted resolution therefore must be divided by $\sqrt{2}$. For bins using an external reference, the resolution of the reference bin is subtracted in quadrature. Then, for each $\eta$ interval, the binned results are fitted as a function of \pt\ using
\begin{equation}
\label{eqn:JERfit}
\sigma (\pt) = \frac{N}{\pt} \oplus \frac{S}{\sqrt{\pt}} \oplus C,
\end{equation}
where the $N$ term, corresponding to the contribution from detector noise, is fixed independently via random cone studies. In those studies, the noise contribution is directly measured by sampling $R = 0.4$ random cones of track or cluster energy. Data events are selected with a trigger requiring only energy in exactly one ZDC, which is dominated by empty events, while a sample of minimum-bias \gamA collisions simulated with \pythia are used to perform the studies in the MC simulation. Differences in the results of these $p_\text{T}$-dependent fits of the JER are treated as a systematic uncertainty, and additional uncertainties are assessed by varying the event selections, the MC generator (\pythia vs. \Sherpa), and random cone noise term. Additional uncertainties are also assessed for the impact of JES uncertainties on this determination and differences between the estimated JER in the MC simulation via this procedure and the truth-matched JER.\par
The results for the JER in certain $\eta$ intervals are shown in the right panel of Figure \ref{fig:JetPerf}. Due to the lack of pile-up in this sample, the noise contribution from Eq.~(\ref{eqn:JERfit}) is negligible in all bins, so the stochastic and constant terms dominate. For this reason, the $\eta$-dependence of the JER is the same as at high-$\mu$, but it grows much less rapidly with 1/\pTjet, allowing for much more precise measurements at low \pTjet. Systematic uncertainties in the in-situ measurement of the JER are typically a bit larger in this sample than in the high-$\mu$ calibration, but the total uncertainty in the measurement is still reduced due to the smaller overall JER values.

% End of text imported from the .//tex/appendixA.tex input file

\clearpage
\section{Fiducial and geometric acceptance definition}
\label{sec:fiducialaccept}

% The next lines are included from the .//tex/appendixB.tex input file
The jet and event selections described in
Table~\ref{tbl:select} limit the kinematic acceptance of this
measurement at the particle-level in both sets of kinematic variables,
$(\yjets,\mjets,\HT)$ and $(\xA,\zg,\HT)$. To determine the true acceptance
of the measurement, each of the intervals defined in Table~\ref{tbl:kinranges}
is subdivided into much finer intervals in each of the three dimensions.
For each of the resulting finer volumes
in \kinsetjet or \kinsetpart, the fraction of MC \gamA\ events
falling within the sub-volume that satisfy the
event selection is determined using particle-level kinematics and selections. The resulting acceptance
functions, $A\kinsetjet$ and $A\kinsetpart$, are used to determine the
geometric acceptance region, which is defined
to contain all sub-divided volumes that have $A > 0.1$. The segmentation of the volume elements
is chosen to be sufficiently fine that varying it has negligible impact
on the resulting procedure. Two effects that impact the differential cross-section
are considered in the following section: the missing acceptance within the accepted regions
and the accepted volume covering only a fraction of the total bin volume.

Using the result of the geometric acceptance determination, the accepted volumes in
$\Delta V\kinsetjet$ and $\Delta V\kinsetpart$ in the nominal
binning (bins $i,j,k$ in each dimension) are calculated according to
\begin{equation}
\Delta V\kinsetjet = \sum_i\sum_j\sum_k \Delta\HT^{i}\Delta\yjets^{j}\Delta\mjets^{k}\Theta\big[ A\kinsetjet - 0.1 \big],
\label{eq:volcompjet}
\end{equation}
\begin{equation}
\Delta V\kinsetpart = \sum_i\sum_j\sum_k \Delta\HT^{i}\Delta\xA^{j}\Delta\zg^{k}\Theta\big[ A\kinsetpart - 0.1 \big],
\label{eq:volcomppart}
\end{equation}
where $\Delta\HT^{i}$, $\Delta\yjets^{j}$, $\Delta\mjets^{k}$, $\Delta\xA^{j}$, and $\Delta\zg^{k}$ are the
widths of sub-bin ($i$, $j$, $k$) for the nominal bin in these variables. This more detailed calculation of the
accepted phase-space volume from Eqs. (\ref{eq:volcompjet}) and (\ref{eq:volcomppart}) is used to normalize the
differential cross-section, as described in Section \ref{sec:crosssecdef}. The accepted volume fraction, $f_\text{vol}$, is
defined for either variable set as
\begin{equation}
f_\text{vol}\kinsetjet \equiv \frac{\Delta V\kinsetjet}{\Delta\HT\Delta\yjets\Delta\mjets},
\end{equation}
\begin{equation}
f_\text{vol}\kinsetpart \equiv \frac{\Delta V\kinsetpart}{\Delta\HT\Delta\xA\Delta\zg}.
\end{equation}
The volume fractions are then used to impose two requirements on the bins that are reported in this measurement.
First, any bin where more than 2.5\% of the total cross-section in that bin falls outside of the geometric acceptance region
is excluded from the reported results. Second, any bin is excluded if $f_\text{vol} < 0.5$.

An additional correction is applied, which accounts for the difference between the measured fiducial cross-section
and the total \gamAjets cross-section. This correction accounts for events that fall into
a given kinematic bin but fail the single-jet rapidity ($\etajet < 4.4$) or jet system mass ($0.9 < \mjets/\HT < 4$)
requirements. These corrections do not attempt to account for the single-jet \pT ($\pTjet > 15$ \GeV) requirement or the
requirement of at least two jets in the event. For any bin with $A < 0.975$, the bin is excluded from the results reported
in the measurement, while bins with $A > 0.975$ are corrected for their partial acceptance. This approach limits
the impact of potential mis-modelling of the fiducial acceptance on this correction.

\begin{figure}[!tb]
\centerline{
\includegraphics[width=0.99\textwidth]{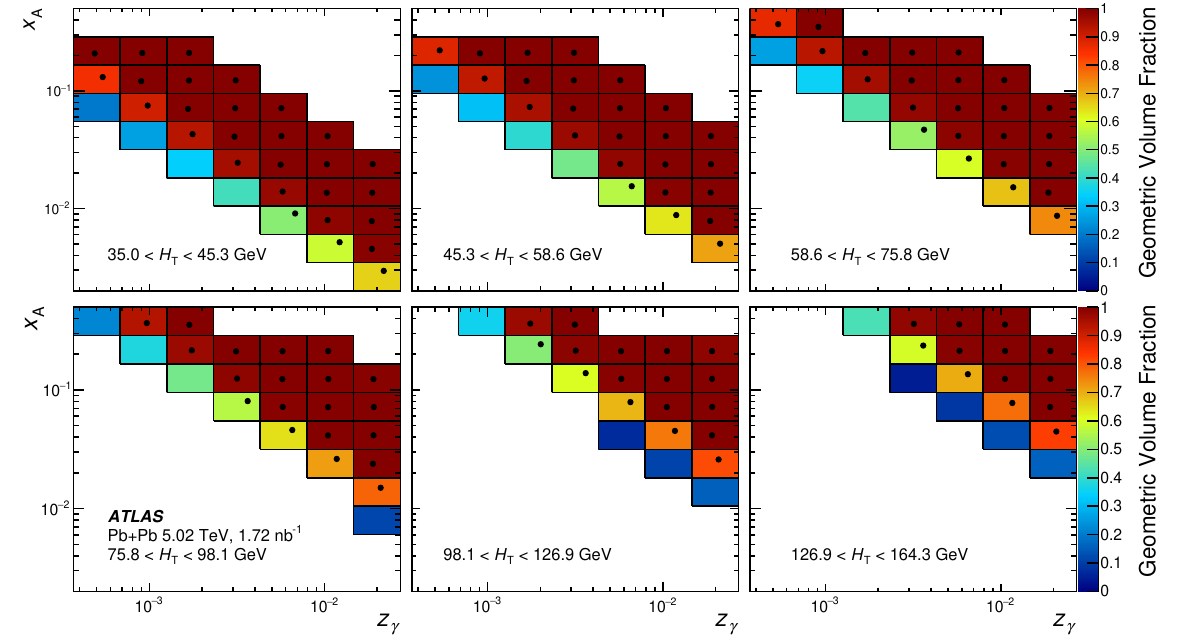}
}
\caption{
The fraction of each bin volume that is included in the geometric acceptance region for the hard-scattering kinematic variables. Re-computed bin means for each bin are shown as markers. Only bins that pass a minimum threshold on the fiducial and geometric acceptance are shown, and the means are only shown for bins with $f_\text{vol} > 50\%$.
}
\label{fig:BinAcceptanceIS}
\end{figure}

\begin{figure}[!tb]
\centerline{
\includegraphics[width=0.99\textwidth]{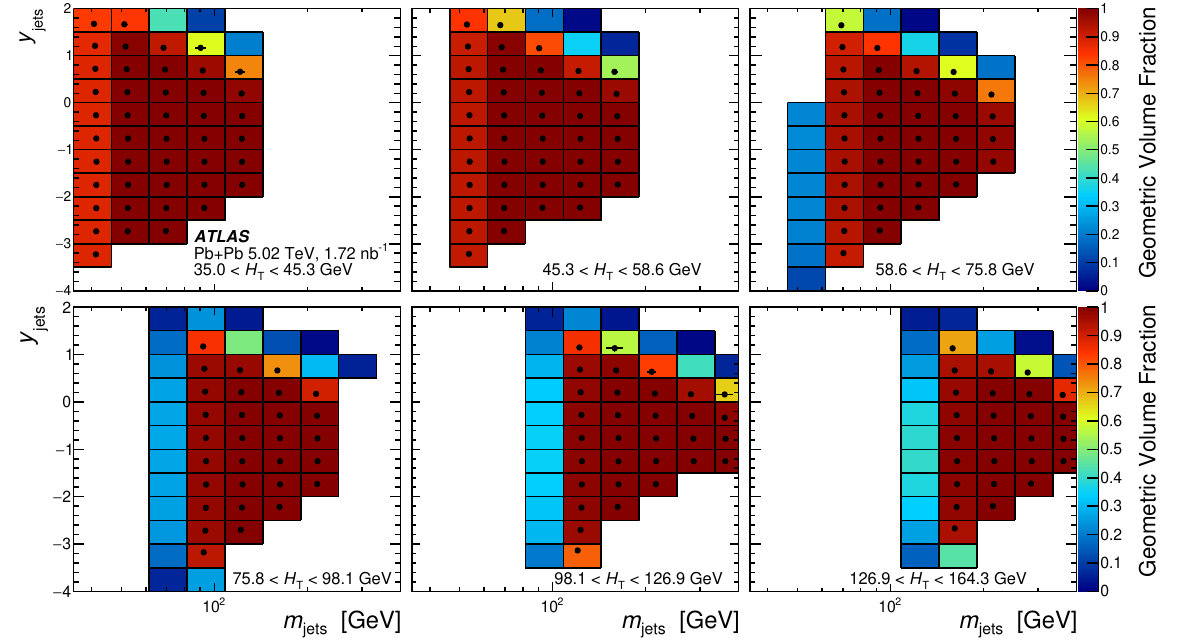}
}
\caption{
The fraction of each bin volume that is included in the geometric acceptance region for the jet system kinematic variables. Re-computed bin means for each bin are shown as markers. Only bins that pass a minimum threshold on the fiducial and geometric acceptance are shown, and the means are only shown for bins with $f_\text{vol} > 50\%$.
}
\label{fig:BinAcceptanceFS}
\end{figure}

The results of the acceptance calculation using the sub-divided volumes
are also used to determine the average values for each of the
kinematic variables within the larger volumes defined in
Table~\ref{tbl:kinranges}. These averages are calculated according to
\begin{equation}
\langle k \rangle = \frac{\sum_i k_i \sigma_i}{\sum_i \sigma_i},
\label{eq:kinemavgacc}
\end{equation}
where $k$ represents one of the kinematic variables, $i$ runs over the
sub-volumes that pass the $A > 0.1$ criterion, $k_i$ represents the
value of $k$ at the middle of the sub-volume, and $\sigma_i$
represents the total cross-section in sub-volume $i$ in \kinsetjet- or \kinsetpart-space.
This calculation assumes that events populate the accepted region in
each sub-volume uniformly. It also relies on the shape of the \pythia
cross-section distribution in order to determine the bin mean values, but
no such dependence is built into the actual reported cross-sections.
Figures \ref{fig:BinAcceptanceIS} and \ref{fig:BinAcceptanceFS} display
all the different calculations and selections described in this section,
showing $f_\text{vol}$ for all bins that pass the requirements on fiducial
acceptance within the geometric acceptance region. For bins with $f_\text{vol} > 0.5$,
the bin means are shown, as calculated by Eq.~(\ref{eq:kinemavgacc}). These
figures demonstrate that most bins have $f_\text{vol} = 1$, with smaller acceptance fractions
occuring near acceptance edges. The $\etajet < 4.4$ requirement limits the acceptance
at high \xA and \yjets, especially at low \HT. The loss of acceptance near
the diagonal acceptance edges in \xA-\zg space arise due to both the mass requirements
and geometric effects from the three-dimensional binning combined with single-jet \pT
requirements.

% End of text imported from the .//tex/appendixB.tex input file

\FloatBarrier
\clearpage
\section{Results for additional $z_\gamma$ intervals}
\label{sec:morePlots}

\begin{figure}[hbp]
\centerline{
\includegraphics[width=0.85\textwidth]{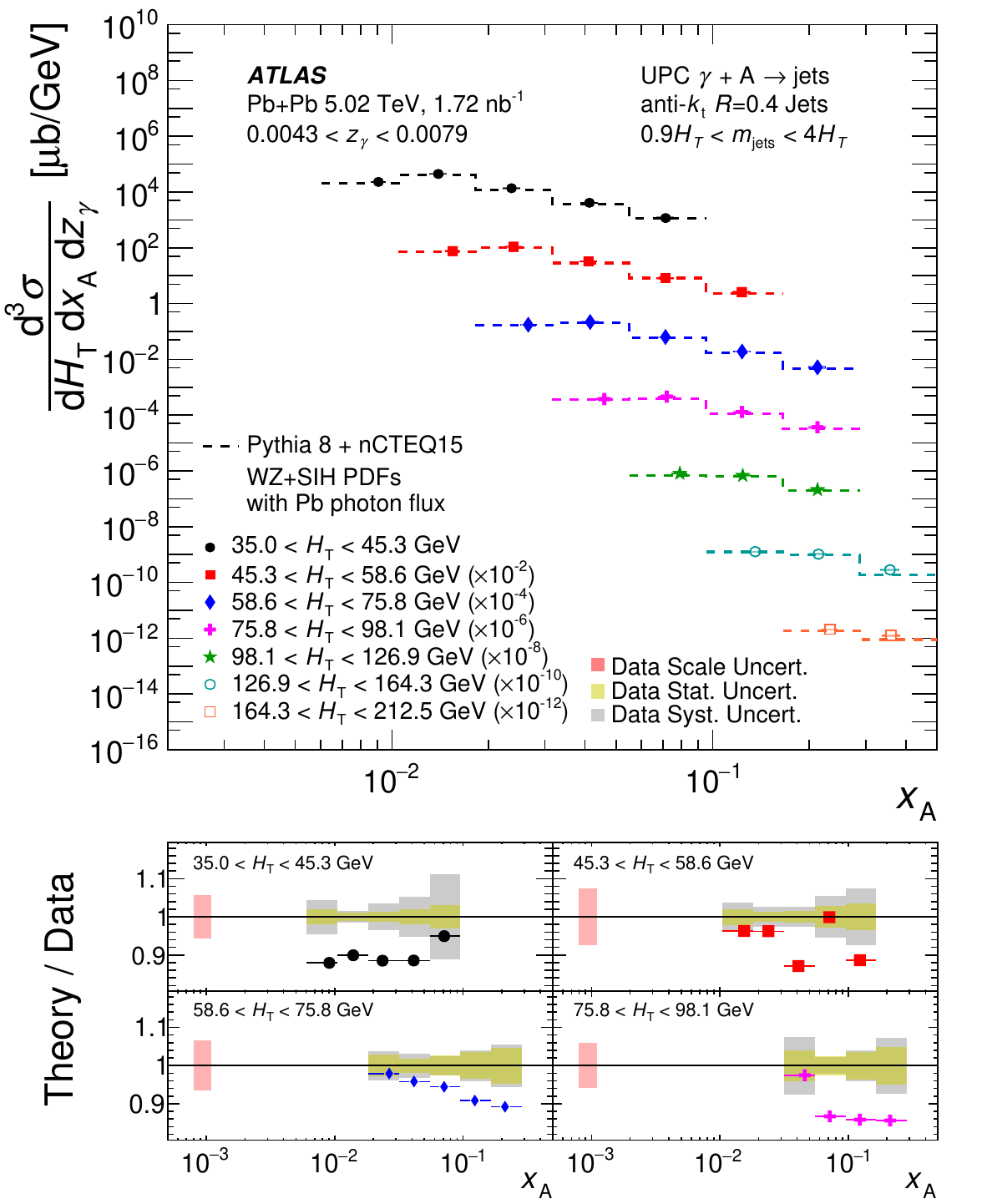}
}
\caption{ Triple-differential cross-sections, \sigtridiff, as a
function of $\xA$ for different bins of $\HT$ for events with emitted
photon energies in the kinematic range $0.0043 < \zg <
0.0079$.  In the upper panel, systematic uncertainties are shown as
shaded boxes, while statistical uncertainties shown as vertical lines
are usually smaller than the size of the markers. A theoretical
comparison is shown to cross-sections computed using \pythia with
nCTEQ15 WZ+SIH PDFs, a photon flux from \textsc{Starlight}, and a \zg-dependent
breakup fraction. The bottom
panels show the ratio between the theory prediction and data for a representative subset of the
bins of \HT. The lorentz-invariant variables \xA, \zg, and \HT are un-ambiguously defined in terms of the particle-level jet kinematics and correspond at leading order to the Bjorken-$x$, emitted photon energy fraction,
and collision $Q^2$, respectively.
The light red bands in the ratio panels are the quadrature sum of scale uncertainties on the cross-section, while the gray band shows the remaining systematic uncertainty. The yellow band  shows the point-to-point statistical uncertainty.  }
\label{fig:Results_xA_TheoryComp_SlicesHT_zGamma8}
\end{figure}

\begin{figure}[p]
\centerline{
\includegraphics[width=0.85\textwidth]{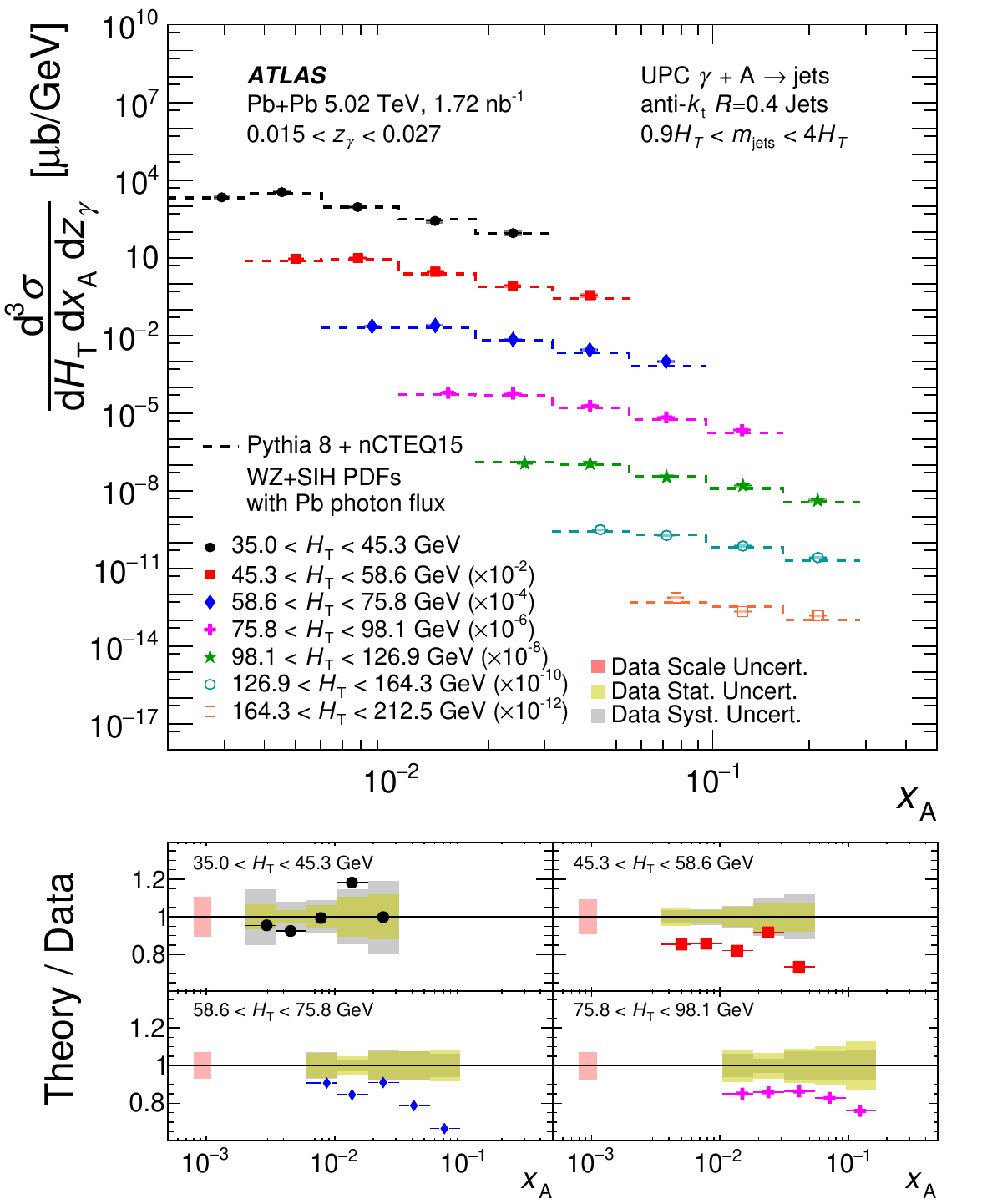}
}
\caption{ Triple-differential cross-sections, \sigtridiff, as a
function of $\xA$ for different bins of $\HT$ for events with emitted
photon energies in the kinematic range $0.015 < \zg <
0.027$.  In the upper panel, systematic uncertainties are shown as
shaded boxes, while statistical uncertainties shown as vertical lines
are usually smaller than the size of the markers. A theoretical
comparison is shown to cross-sections computed using \pythia with
nCTEQ15 WZ+SIH PDFs, a photon flux from \textsc{Starlight}, and a \zg-dependent
breakup fraction. The bottom
panels show the ratio between the theory prediction and data for a representative subset of the
bins of \HT. The lorentz-invariant variables \xA, \zg, and \HT are un-ambiguously defined in terms of the particle-level jet kinematics and correspond at leading order to the Bjorken-$x$, emitted photon energy fraction,
and collision $Q^2$, respectively.
The light red bands in the ratio panels are the quadrature sum of scale uncertainties on the cross-section, while the gray band shows the remaining systematic uncertainty. The yellow band  shows the point-to-point statistical uncertainty.  }
\label{fig:Results_xA_TheoryComp_SlicesHT_zGamma10}
\end{figure}

\begin{figure}[p]
\centerline{
\includegraphics[width=0.85\textwidth]{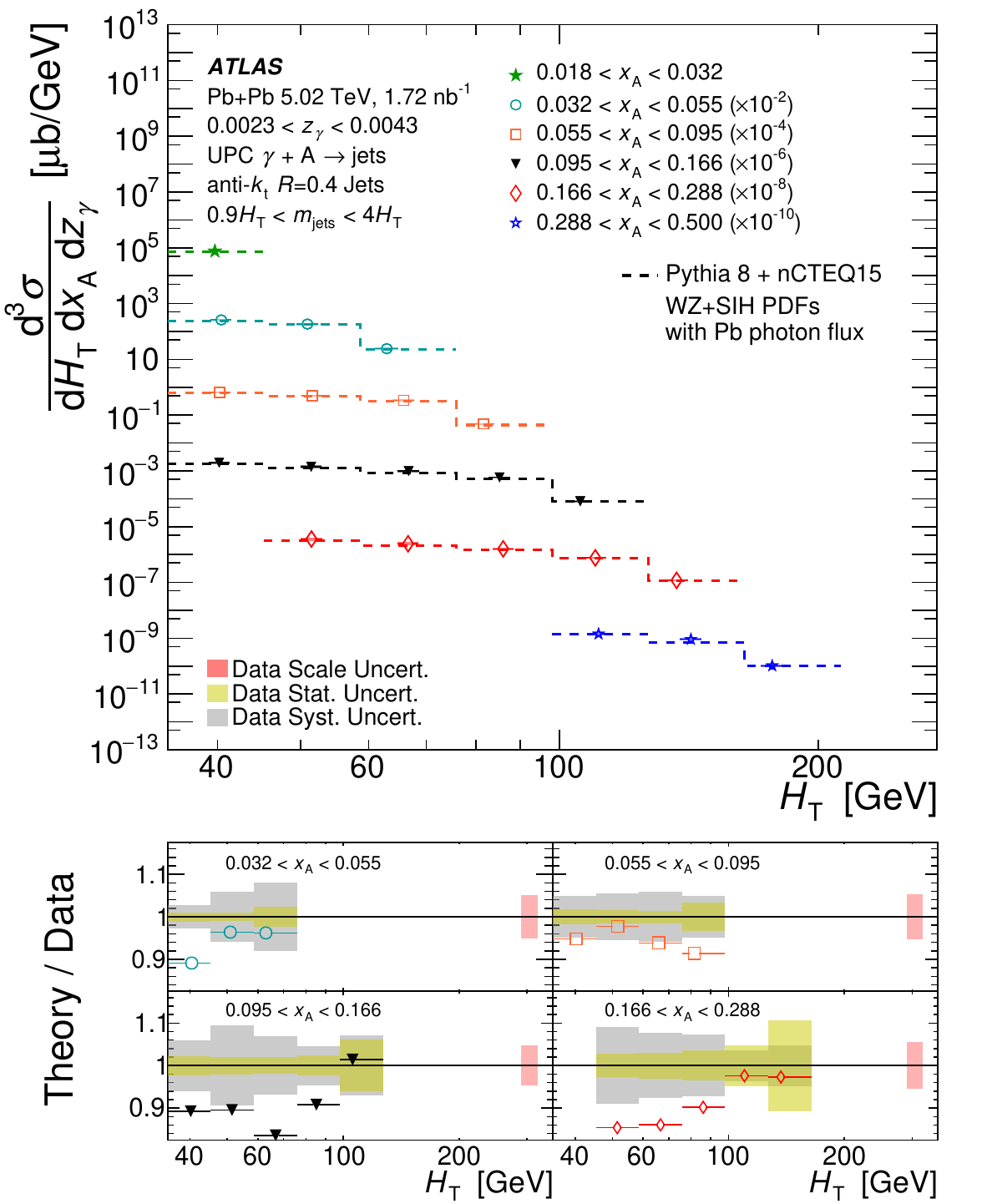}
}
\caption{ Triple-differential cross-sections, \sigtridiff, as a
function of $\HT$ for different bins of $\xA$ for events with emitted
photon energies in the kinematic range $0.0023 < \zg <
0.0043$.  In the upper panel, systematic uncertainties are shown as
shaded boxes, while statistical uncertainties shown as vertical lines
are usually smaller than the size of the markers. A theoretical
comparison is shown to cross-sections computed using \pythia with
nCTEQ15 WZ+SIH PDFs, a photon flux from \textsc{Starlight}, and a \zg-dependent
breakup fraction. The bottom
panels show the ratio between the theory prediction and data for a representative subset of the
bins of \xA. The lorentz-invariant variables \xA, \zg, and \HT are un-ambiguously defined in terms of the particle-level jet kinematics and correspond at leading order to the Bjorken-$x$, emitted photon energy fraction,
and collision $Q^2$, respectively.
The light red bands in the ratio panels are the quadrature sum of scale uncertainties on the cross-section, while the gray band shows the remaining systematic uncertainty. The yellow band  shows the point-to-point statistical uncertainty.  }
\label{fig:Results_HT_TheoryComp_SlicesxA_zGamma7}
\end{figure}

\begin{figure}[!tb]
\centerline{
\includegraphics[width=0.85\textwidth]{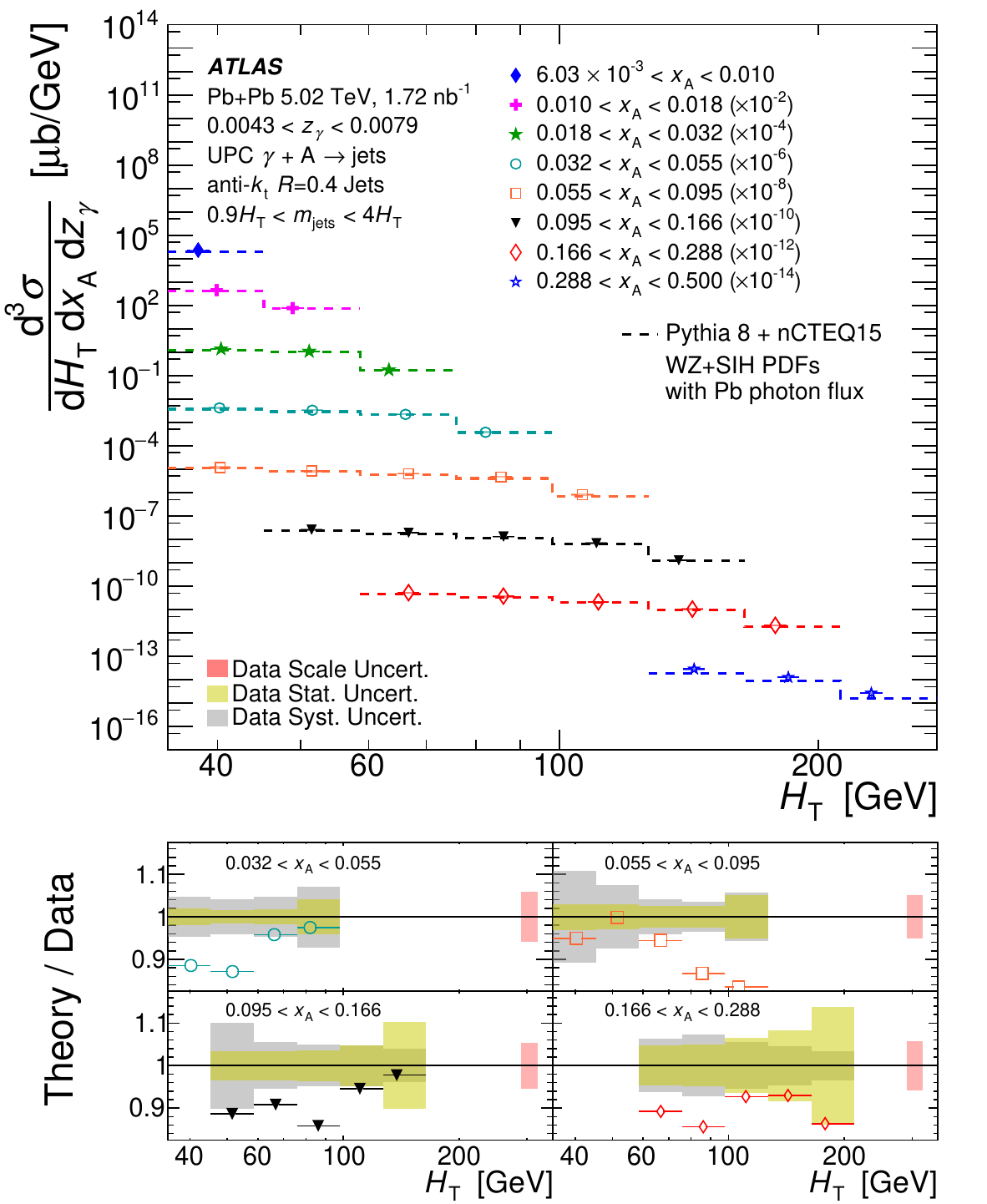}
}
\caption{ Triple-differential cross-sections, \sigtridiff, as a
function of $\HT$ for different bins of $\xA$ for events with emitted
photon energies in the kinematic range $0.0043 < \zg <
0.079$.  In the upper panel, systematic uncertainties are shown as
shaded boxes, while statistical uncertainties shown as vertical lines
are usually smaller than the size of the markers. A theoretical
comparison is shown to cross-sections computed using \pythia with
nCTEQ15 WZ+SIH PDFs, a photon flux from \textsc{Starlight}, and a \zg-dependent
breakup fraction. The bottom
panels show the ratio between the theory prediction and data for a representative subset of the
bins of \xA. The lorentz-invariant variables \xA, \zg, and \HT are un-ambiguously defined in terms of the particle-level jet kinematics and correspond at leading order to the Bjorken-$x$, emitted photon energy fraction,
and collision $Q^2$, respectively.
The light red bands in the ratio panels are the quadrature sum of scale uncertainties on the cross-section, while the gray band shows the remaining systematic uncertainty. The yellow band  shows the point-to-point statistical uncertainty.  }
\label{fig:Results_HT_TheoryComp_SlicesxA_zGamma8}
\end{figure}

\begin{figure}[!tb]
\centerline{
\includegraphics[width=0.85\textwidth]{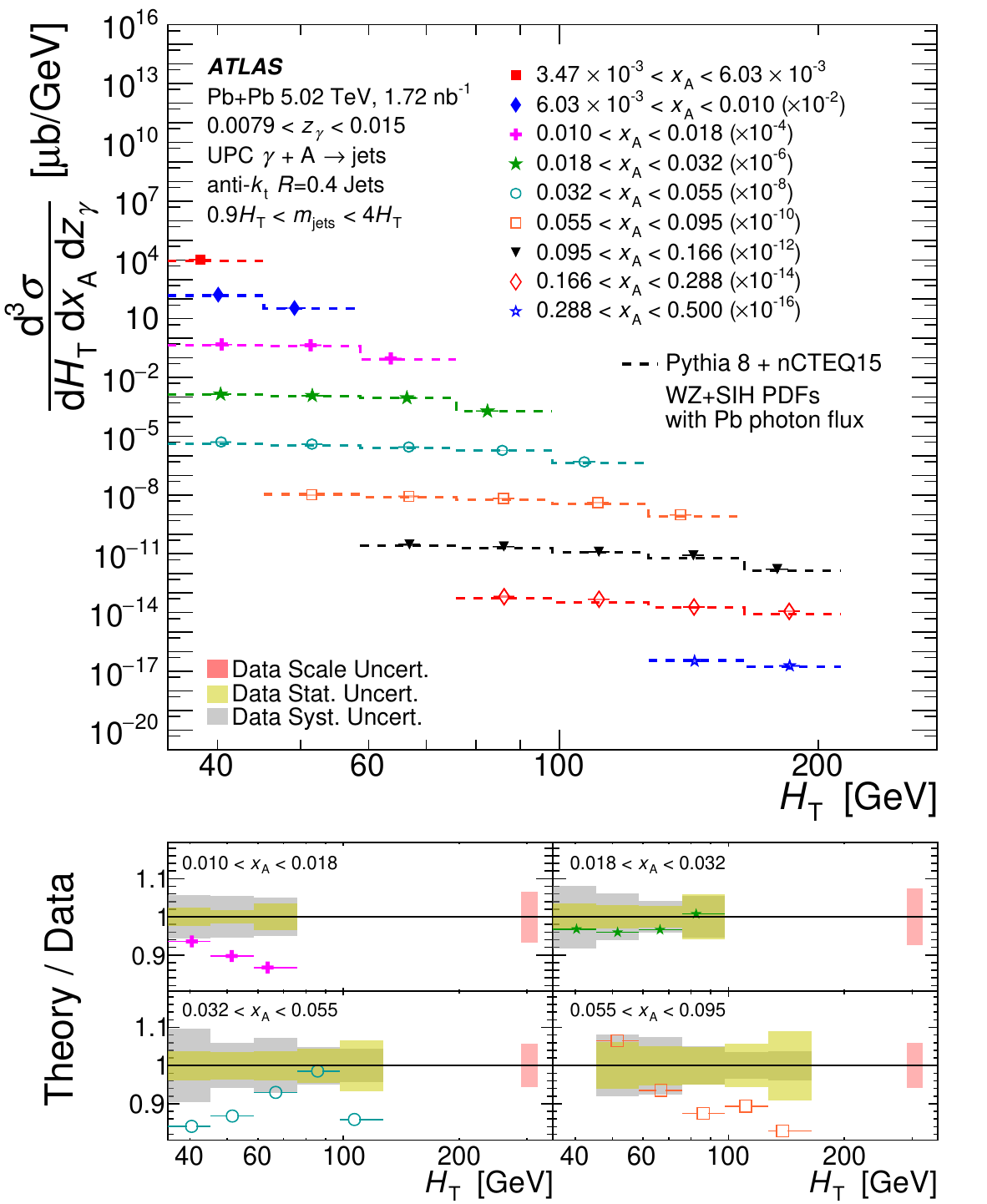}
}
\caption{ Triple-differential cross-sections, \sigtridiff, as a
function of $\HT$ for different bins of $\xA$ for events with emitted
photon energies in the kinematic range $0.0079 < \zg <
0.015$.  In the upper panel, systematic uncertainties are shown as
shaded boxes, while statistical uncertainties shown as vertical lines
are usually smaller than the size of the markers. A theoretical
comparison is shown to cross-sections computed using \pythia with
nCTEQ15 WZ+SIH PDFs, a photon flux from \textsc{Starlight}, and a \zg-dependent
breakup fraction. The bottom
panels show the ratio between the theory prediction and data for a representative subset of the
bins of \xA. The lorentz-invariant variables \xA, \zg, and \HT are un-ambiguously defined in terms of the particle-level jet kinematics and correspond at leading order to the Bjorken-$x$, emitted photon energy fraction,
and collision $Q^2$, respectively.
The light red bands in the ratio panels are the quadrature sum of scale uncertainties on the cross-section, while the gray band shows the remaining systematic uncertainty. The yellow band  shows the point-to-point statistical uncertainty.  }
\label{fig:Results_HT_TheoryComp_SlicesxA_zGamma9}
\end{figure}

\begin{figure}[!tb]
\centerline{
\includegraphics[width=0.85\textwidth]{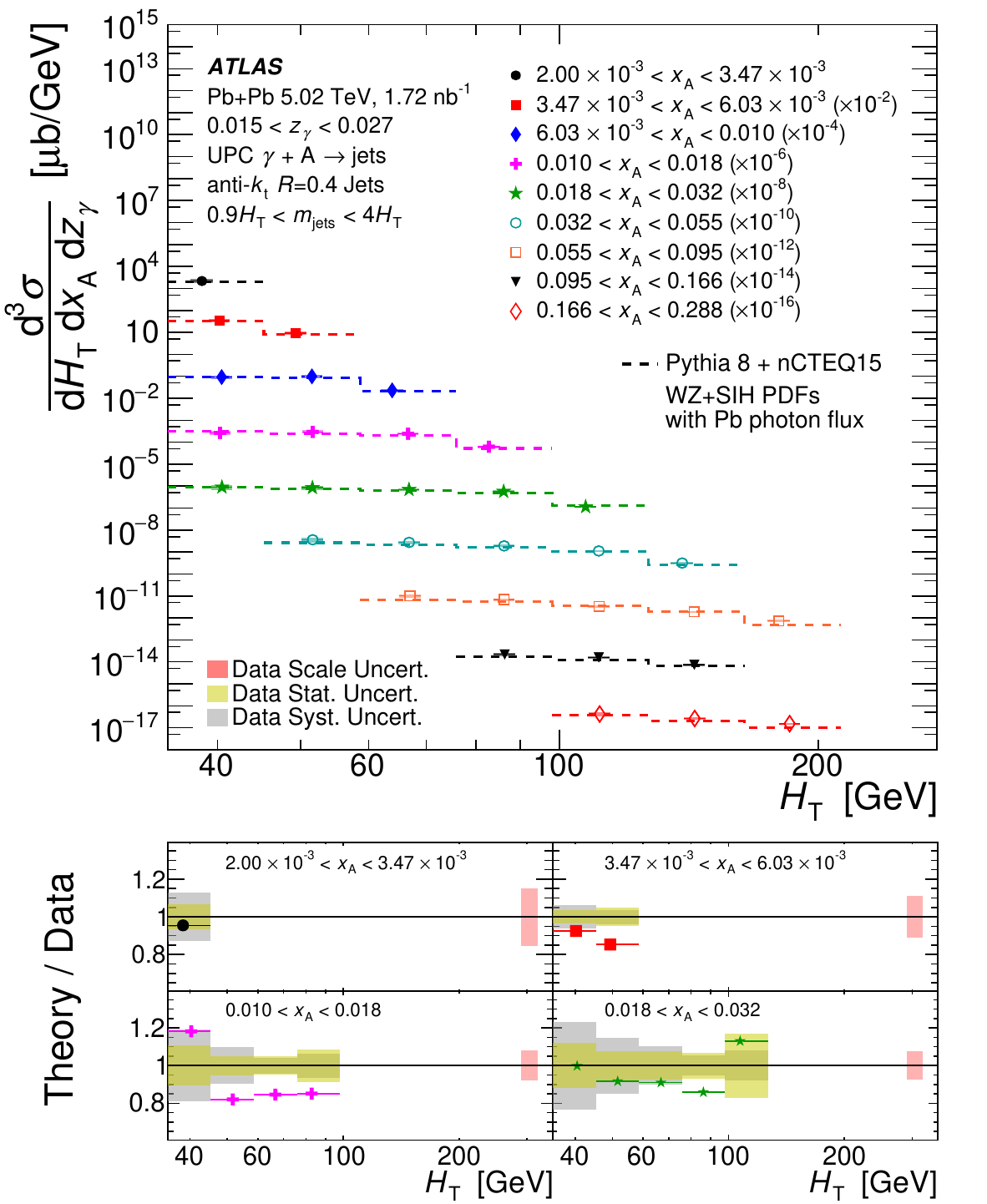}
}
\caption{ Triple-differential cross-sections, \sigtridiff, as a
function of $\HT$ for different bins of $\xA$ for events with emitted
photon energies in the kinematic range $0.015 < \zg <
0.027$.  In the upper panel, systematic uncertainties are shown as
shaded boxes, while statistical uncertainties shown as vertical lines
are usually smaller than the size of the markers. A theoretical
comparison is shown to cross-sections computed using \pythia with
nCTEQ15 WZ+SIH PDFs, a photon flux from \textsc{Starlight}, and a \zg-dependent
breakup fraction. The bottom
panels show the ratio between the theory prediction and data for a representative subset of the
bins of \xA. The lorentz-invariant variables \xA, \zg, and \HT are un-ambiguously defined in terms of the particle-level jet kinematics and correspond at leading order to the Bjorken-$x$, emitted photon energy fraction,
and collision $Q^2$, respectively.
The light red bands in the ratio panels are the quadrature sum of scale uncertainties on the cross-section, while the gray band shows the remaining systematic uncertainty. The yellow band  shows the point-to-point statistical uncertainty.  }
\label{fig:Results_HT_TheoryComp_SlicesxA_zGamma10}
\end{figure}

% The next lines are included from the .//tex/appendixC.tex input file
Several additional figures demonstrating the measured cross-sections
described in Section \ref{sec:Xsec3D} are shown here. These figures
include the \xA dependence of the cross-section in two additional \zg
intervals in Figures \ref{fig:Results_xA_TheoryComp_SlicesHT_zGamma8}
and \ref{fig:Results_xA_TheoryComp_SlicesHT_zGamma10}, as well as the \HT dependence
for four \zg intervals in Figures \ref{fig:Results_HT_TheoryComp_SlicesxA_zGamma7},
\ref{fig:Results_HT_TheoryComp_SlicesxA_zGamma8}, \ref{fig:Results_HT_TheoryComp_SlicesxA_zGamma9},
and \ref{fig:Results_HT_TheoryComp_SlicesxA_zGamma10}. These additional intervals in \zg
allow for a more robust separation of the correlated and un-correlated uncertainty components,
helping to demonstrate the size of point-to-point uncertainties most relevant for constraining
nPDF effects.

% End of text imported from the .//tex/appendixC.tex input file

\FloatBarrier
\clearpage
\printbibliography
\clearpage
% ATLAS Collaboration author list
% Reference date of HION-2022-15 is 2024-07-01
% Author list last updated on date 27-JAN-25
% Data extracted on 27-Jan-2025 for paper reference HION-2022-15
% at 9:59am
 
\begin{flushleft}
\hypersetup{urlcolor=black}
{\Large The ATLAS Collaboration}

\bigskip

\AtlasOrcid[0000-0002-6665-4934]{G.~Aad}$^\textrm{\scriptsize 105}$,
\AtlasOrcid[0000-0001-7616-1554]{E.~Aakvaag}$^\textrm{\scriptsize 17}$,
\AtlasOrcid[0000-0002-5888-2734]{B.~Abbott}$^\textrm{\scriptsize 124}$,
\AtlasOrcid[0000-0002-0287-5869]{S.~Abdelhameed}$^\textrm{\scriptsize 120a}$,
\AtlasOrcid[0000-0002-1002-1652]{K.~Abeling}$^\textrm{\scriptsize 57}$,
\AtlasOrcid[0000-0001-5763-2760]{N.J.~Abicht}$^\textrm{\scriptsize 51}$,
\AtlasOrcid[0000-0002-8496-9294]{S.H.~Abidi}$^\textrm{\scriptsize 30}$,
\AtlasOrcid[0009-0003-6578-220X]{M.~Aboelela}$^\textrm{\scriptsize 46}$,
\AtlasOrcid[0000-0002-9987-2292]{A.~Aboulhorma}$^\textrm{\scriptsize 36e}$,
\AtlasOrcid[0000-0001-5329-6640]{H.~Abramowicz}$^\textrm{\scriptsize 156}$
% ...
\bigskip
\\

% ...
$^{17}$Department for Physics and Technology, University of Bergen, Bergen; Norway.\\
% ...
$^{30}$Physics Department, Brookhaven National Laboratory, Upton NY; United States of America.\\
% ...
$^{36}$$^{(a)}$Facult\'e des Sciences Ain Chock, Universit\'e Hassan II de Casablanca;$^{(b)}$Facult\'{e} des Sciences, Universit\'{e} Ibn-Tofail, K\'{e}nitra;$^{(c)}$Facult\'e des Sciences Semlalia, Universit\'e Cadi Ayyad, LPHEA-Marrakech;$^{(d)}$LPMR, Facult\'e des Sciences, Universit\'e Mohamed Premier, Oujda;$^{(e)}$Facult\'e des sciences, Universit\'e Mohammed V, Rabat;$^{(f)}$Institute of Applied Physics, Mohammed VI Polytechnic University, Ben Guerir; Morocco.\\
% ...
$^{46}$Physics Department, Southern Methodist University, Dallas TX; United States of America.\\
% ...
$^{51}$Fakult\"{a}t Physik , Technische Universit{\"a}t Dortmund, Dortmund; Germany.\\
% ...
$^{57}$II. Physikalisches Institut, Georg-August-Universit\"{a}t G\"ottingen, G\"ottingen; Germany.\\
% ...
$^{105}$CPPM, Aix-Marseille Universit\'e, CNRS/IN2P3, Marseille; France.\\
% ...
$^{120}$$^{(a)}$New York University Abu Dhabi, Abu Dhabi;$^{(b)}$United Arab Emirates University, Al Ain; United Arab Emirates.\\
% ...
$^{124}$Homer L. Dodge Department of Physics and Astronomy, University of Oklahoma, Norman OK; United States of America.\\
% ...
$^{156}$Raymond and Beverly Sackler School of Physics and Astronomy, Tel Aviv University, Tel Aviv; Israel.\\
% ...
$^{a}$ Also Affiliated with an institute covered by a cooperation agreement with CERN.\\
$^{b}$ Also at An-Najah National University, Nablus; Palestine.\\
$^{c}$ Also at Borough of Manhattan Community College, City University of New York, New York NY; United States of America.\\
$^{d}$ Also at Center for High Energy Physics, Peking University; China.\\
$^{e}$ Also at Center for Interdisciplinary Research and Innovation (CIRI-AUTH), Thessaloniki; Greece.\\
$^{f}$ Also at CERN, Geneva; Switzerland.\\
% ...

\end{flushleft}

% Created with Glance <Atlas.Glance@cern.ch>

%

%
%
%
%

%
%
%
%

\end{document}